\documentclass[
structabstract
]{aa}

\usepackage[fleqn]{amsmath}
\usepackage{graphicx}
\usepackage[]{subfigure}
\usepackage{txfonts}
\usepackage{natbib}
\usepackage[colorlinks=true,citecolor=black]{hyperref}

\usepackage{color}

\newcommand{\vONE}[1]{#1}

\newcommand{\Msol}{\mbox{M}_\odot}
\newcommand{\Lsol}{\mbox{L}_\odot}
\newcommand{\Rsol}{\mbox{R}_\odot}
\newcommand{\Msolyr}{\mbox{M}_\odot \mbox{ yr}^{-1}}
\newcommand{\gccm}{\mbox{g cm}^{-3}}
\newcommand{\K}{\mbox{K}}
\newcommand{\kK}{\mbox{kK}}
\newcommand{\yr}{\mbox{yr}}
\newcommand{\kyr}{\mbox{kyr}}
\newcommand{\pc}{\mbox{pc}}
\newcommand{\au}{\mbox{au}}
\newcommand{\kms}{\mbox{km s}^{-1}}
\newcommand{\eV}{\mbox{eV}}

\title{First hydrodynamics simulations of \\radiation forces and photoionization feedback\\ in massive star formation}
\titlerunning{Radiation forces and photoionization feedback in massive star formation}

\author{ 
R.~Kuiper\inst{1} 
\and 
T.~Hosokawa\inst{2} 
}
\institute{
Institute of Astronomy and Astrophysics,
University of T\"ubingen,
Auf der Morgenstelle 10, 
D-72076 T\"ubingen, 
Germany
\\
\email{rolf.kuiper@uni-tuebingen.de}
\and
Department of Physics, 
Kyoto University, 
Kyoto 606-8502, 
Japan
}

\abstract{}
{
We present the first simulations of the formation and feedback of massive stars which account for radiation forces as well as photoionization feedback (along with protostellar outflows).
In two different accretion scenarios modeled, we determine the relative strength of these feedback components and derive the size of the reservoir from which the forming stars gained their masses.
}
{
We performed direct hydrodynamics simulations of the gravitational collapse of high-density mass reservoirs toward the formation of massive stars including self-gravity, stellar evolution, protostellar outflows, continuum radiation transport, photoionization, and the potential impact of ram pressure from large-scale gravitational infall.
For direct comparison, we executed these simulations with and without the individual feedback components.
}
{
Protostellar outflows alone limit the stellar mass growth only in an accretion scenario with a finite mass reservoir;
when including accretion and ram pressure from large scales ($> 0.1~\pc$), protostellar outflows do not limit stellar mass growth at all.
Photoionization and HII regions dominate the feedback ladder only at later times, after the star has already contracted down to the zero-age main sequence, and only on large scales.
Specifically, photoionization yields a broadening of the bipolar outflow cavities
and a reduction of the gravitational infall momentum by about 50\%,
but does not limit the stellar mass accretion.
On the other hand, we find radiation forces restrain the gravitational infall toward the circumstellar disk, impact the gravito-centrifugal equilibrium at the outer edge of the disk, and eventually shut down stellar accretion completely.
The most massive star formed in the simulations accreted $95~\Msol$ before disk destruction; this mass was drawn-in from an accretion reservoir of $\approx 240~\Msol$ and $\approx 0.24~\pc$ in radius.
}
{
In the regime of very massive stars, the final mass of these stars is controlled by their own radiation force feedback.
}


\date{\today}

\begin{document}
\maketitle

%%%%%%%%%%%
%%                       %%
%%  Introduction  %%
%%                       %%
%%%%%%%%%%%
\section{Introduction}
\label{sect:introduction}
Radiative feedback in terms of radiation forces and photoionization is expected to play a major role in the formation of (the most) massive stars in the present-day universe \citep{1971A&A....13..190L}.
During the lifetime of a massive (proto)star, these two feedback components strongly oppose gravity, and hence impact infall and accretion flows on a variety of spatial scales.
The details of these complex interactions in terms of their relative importance in determining the final stellar mass remain unclear.
For comprehensive and/or recent reviews on massive star formation and stellar feedback, we refer the reader to \citet{2007ARA&A..45..481Z}, \citet{2007ARA&A..45..565M}, and \citet{2014prpl.conf..149T}.

% "While this is progress, a few caveats are in order. First, no code yet includes all of these physical processes: e.g., ORION includes MHD, dust-reprocessed radiation and out- flows, but not ionizing radiation, while FLASH has MHD and ionizing radiation, but not outflows or dust-reprocessed radiation." (Tan et al. 2014, PPVI)
Here, we present the first simulations of the formation of massive stars, including both radiative feedback components, namely radiation forces from direct and dust-reprocessed radiation as well as photoionization.
So far, numerical simulations of the feedback during high-mass star formation have been restricted to include either 
only radiation forces but no photoionization 
\citep{
Yorke:1977vk, 
Yorke:2002dn, 
Krumholz:2007hs, 
Krumholz:2009ea,
Krumholz:2010fi,
2010ApJ...722.1556K, % Kuiper - 2D
2011ApJ...732...20K, % Kuiper - 3D
Cunningham:2011dy, % including outflows
2011ApJ...742L...9C, % including MHD, but covers only a tiny time evolution
2012A&A...537A.122K, % Kuiper - cavities
2013ApJ...763..104K, % Kuiper - gas disks
2013ApJ...772...61K, % Kuiper - stellar and host core evolution
Myers:2013jv, % including MHD, but no jets launched (most likely due to resolution)
Harries:2014bg,
2015ApJ...800...86K, % Kuiper - including outflows
2016ApJ...823...28K, % Klassen et al. 
2016MNRAS.463.2553R, % including a ray-tracing method for the stellar feedback (opacity switched to zero in low-density cavities?)
2016ApJ...832...40K, % Kuiper - including outflows
2017MNRAS.471.4111H}
or photoionization only but no radiation forces
\citep{
Yorke:1996wd,
Richling:1997tw,
Richling:1998uj,
Richling:2000bu,
2005MNRAS.358..291D,
2010ApJ...715.1302V,
Peters:2010bw,
Peters:2010bz,
Peters:2011hi, % including MHD
Dale:2011ct,
2012MNRAS.427..625W,
2012MNRAS.427.2852D,
2013MNRAS.435.1701C,
2013MNRAS.431.1062D,
2014MNRAS.442..694D,
Peters:2014bm, %including collective outflows
2016MNRAS.461.2953H,
2017MNRAS.466.3293P,
2017MNRAS.467.1067D,
2017ApJ...834...40H,
2017MNRAS.470.3346H}.
We note that the latter list of ionization-hydrodynamics simulations is restricted to massive star formation only (individual stars or star clusters), hence, does not include the huge literature on interaction of expanding HII regions with the large-scale turbulent ISM, photoionization feedback on star-forming regions from external sources, or triggering of star formation.
\vONE{
On larger scales, the feedback by radiation forces of young star clusters on their parent giant molecular cloud was investigated as well \citep[see][for recent studies]{2015ApJ...809..187S, 2016ApJ...829..130R, Raskutti:2017kv}.
}
For completeness, we would like to mention that \citet{2017MNRAS.466L.123I} and \citet{2018MNRAS.474.1935I} report on one-dimensional hydrodynamics simulations including radiation forces and photoionization to study the expansion of an HII region around an a priorly fixed stellar cluster, hence, not associated to the formation of individual massive stars.
For a broader review of feedback modeling in star formation simulations we refer the reader to the review by \citet{2015NewAR..68....1D}.

With respect to the potential feedback efficiency against gravitational infall and disk accretion toward a forming high-mass star, the studies of photoionization feedback agree on the outcome that the thermal pressure of the HII region is not able to halt accretion (see also \citet{Keto:2003kk} for a semi-analytical model of accretion through ionization fronts).
Nonetheless, photoevaporation will certainly affect the atmospheric layers of the accretion disk \citep{Yorke:1996wd,Richling:1997tw,Richling:1998uj,Richling:2000bu}.
Additionally, \citet{2017MNRAS.470.3346H} report on a clear reduction of the efficiency of star cluster formation in the presence of photoionization feedback.

With respect to radiation forces opposing gravity, the different studies mentioned above agree on the outcome that non-spherical accretion is a necessity to circumvent the so-called radiation pressure barrier in the formation of very massive stars \citep{Yorke:2002dn, Krumholz:2009ea, 2010ApJ...722.1556K, 2011ApJ...732...20K, 2013ApJ...763..104K, 2016ApJ...823...28K}.
In strictly spherical accretion, the maximum mass of the forming star is limited to $\approx 40~\Msol$ \citep{Kahn:1974tk, Yorke:1977vk}.
In the simulations of non-spherical disk accretion it was demonstrated that due to the optically thick circumstellar disk the radiation field becomes anisotropic with most of the radiation being channeled toward the polar regions and little acting in the radial direction against the disk accretion flow.
These simulation results supported the early non-hydrodynamical studies of non-spherical accretion onto forming high-mass stars by \citet{Nakano:1989ce} and \citet{Jijina:1996ic}.
We would like to point out, that from all simulations including radiation forces, up to now only \citet{2010ApJ...722.1556K} has formed stars clearly above (up to $137~\Msol$) the spherical accretion threshold of $\approx 40~\Msol$.
Hence, in these simulations it was for the first time explicitly shown that non-spherical accretion allows the stars to accrete beyond the radiation pressure barrier.
Technically, this could be achieved by solving for the long-term evolution of the star and disk system.

Following \citet{Yorke:1999fo}, we call the anisotropic radiation field caused by the optical thickness of the accretion disk the ``disk's flashlight effect''.
This disk's flashlight effect can be enhanced by the formation of low-density (optically thin) bipolar outflow cavities \citep{Krumholz:2005ds, Cunningham:2011dy, 2015ApJ...800...86K, 2016ApJ...832...40K}.
\vONE{
For a recent discovery of a parsec-scale optical jet from a massive young star in the Large Magellanic Cloud, please see \citet{2018Natur.554..334M}.
This observational detection of a strongly collimated jet supports the idea that luminous, massive stars form via disk accretion \citep{2010ApJ...722.1556K, 2011ApJ...732...20K}, in accordance with their lower mass siblings.
}

Here, we report on the results of the first hydrodynamical simulations of a forming high-mass (proto)star, which includes both kinds of radiative feedback effects -- radiation forces and photoionization.
The aim of this study is to derive the relative importance of both feedback effects on the various spatial scales of gravitational infall and accretion.
The relative importance of the feedback components is determined in a series of simulations that includes
a run with both feedback effects,
a run with photoionization only,
a run with radiation forces only, and
a run with neither of these two feedback components switched on.
In the latter run only centrifugal forces, protostellar outflows, and thermal pressure are opposing gravity.

The analysis is divided into characteristic spatial scales $L$, which are chosen to be 
$L < 0.01~\pc \approx 2000~\au$ (labeled as disk scales),
$0.01~\pc < L < 0.1~\pc$ (labeled as small scales or core scales), and 
$0.1~\pc < L < 1~\pc$ (labeled as large scales or cloud scales).
To isolate the latter two spatial scales and to check the impact of ram pressure from large-scale accretion flows, we model two different kinds of initial conditions or accretion scenarios respectively:
One of the initial conditions is given by a finite mass reservoir of $100~\Msol$ in a sphere of $0.1~\pc$ radius, resembling an isolated pre-stellar core.
The other initial condition is given by a virtually infinite mass reservoir of $1000~\Msol$ in a sphere of $1~\pc$ radius, resembling a pre-stellar core, which is fed by gravitational infall from large scales.

The manuscript is organized as follows:
In Sect.~\ref{sect:methods}, we give an overview of the numerical framework used for the simulations and give details of the numerical grid and boundary conditions.
In Sect.~\ref{sect:simulations}, we elaborate on the physical initial conditions of the simulations and the physics included.
In Sect.~\ref{sect:results}, we present and discuss the results of these simulation series in terms of the 
basic evolution of the system (Sect.~\ref{sect:results_Basics}),
the evolution of the circumstellar accretion disk (Sect.~\ref{sect:results_Disk}),
the evolution of the host star (Sect.~\ref{sect:results_Star})
the impact of the feedback components on the final stellar mass (Sect.~\ref{sect:results_StarFormation}),  
the evolution of the HII regions as well as feedback on large scales (Sect.~\ref{sect:results_HII}), and
the broadening of the bipolar outflow cavities as a consequence of the multiple feedback components and spatial scales involved (Sect.~\ref{sect:results_Outflow}).
In Sect.~\ref{sect:limitations}, we discuss the limitations of the current models and offer an outlook on future directions.
In Sect.~\ref{sect:summary}, we conclude with a brief summary of our findings.

%%%%%%%%%%%%%%%
%%                                    %%
%%  Simulation Methods  %%
%%                                    %%
%%%%%%%%%%%%%%%
\section{Methods}
\label{sect:methods}

\begin{table*}[tbp]
\caption{ 
Overview of simulations.
The columns from left to right denote
the run label,
the mass $M_\mathrm{res}$ of the initial mass reservoir,
the size $R_\mathrm{res}$ of the initial mass reservoir,
the feedback effects included in the simulations (outflows, radiation forces, and photoionization),
some quantitative results, namely
the final mass $M_\mathrm{*}^\mathrm{final}$ of the central star at the end of its accretion phase (if still accreting at the end of the simulation, lower limits are given), 
the duration of the accretion phase (if still accreting at the end of the simulation, lower limits are given), 
and the associated size of the accretion region, i.e.~the spatial scales from which the star has drawn its mass.
The final column gives the line style used for the corresponding simulation data in the analyses figures throughout the manuscript.
%The small-scale initial conditions include $M_\mathrm{res} = 100 \mbox{ M}_\odot$ and large-scale initial conditions include $M_\mathrm{res} = 1000 \mbox{ M}_\odot$.
The label PO (protostellar outflow) refers to the feedback of protostellar outflow \& jets.
The label RAD (radiation) refers to simulations that account for radiation forces.
The label ION (ionization) refers to taking into account the effect of the stellar photoionizing flux and the subsequent formation and expansion of an HII region around the high-mass star.
}
\label{tab:sims}
\centering
\begin{tabular}{l | c c | c c c | c  c c | r}
\hline\hline
Run Label & 
$M_\mathrm{res}$~[$\Msol$] & 
$R_\mathrm{res}$~[pc] & 
Outflow & 
Radiation & 
Ionization &
$M_\mathrm{*}^\mathrm{final}~[\mbox{M}_\odot]$ &
$t_\mathrm{acc}^\mathrm{tot}~[\mbox{kyr}]$ &
$R_\mathrm{acc}^\mathrm{max}~[\mbox{pc}]$ &
Line Style
\\
\hline
0.1pc-PO				& ~~100	& 0.1 & Yes & No	& No		& $66$		&$> 300$	&$R_\mathrm{res}$	& Gray Solid	\\
0.1pc-PO-ION			& ~~100	& 0.1 & Yes & No	& Yes	& $66$		&$220$	&$R_\mathrm{res}$	& Blue Solid	\\
0.1pc-PO-RAD			& ~~100	& 0.1 & Yes & Yes	& No		& $51$		&$100$	&$R_\mathrm{res}$	& Red Solid	\\
0.1pc-PO-RAD-ION		& ~~100	& 0.1 & Yes & Yes 	& Yes	& $59$		&$115$	&$R_\mathrm{res}$	& Black Solid	\\
\hline
1.0pc-PO				& 1000	& 1.0 & Yes & No	& No		& $> 300$		&$> 600$	&$R_\mathrm{res}$ & Gray Dashed	\\
1.0pc-PO-ION			& 1000	& 1.0 & Yes & No	& Yes	& $> 277$		&$> 524$	&$R_\mathrm{res}$	& Blue Dashed	\\
1.0pc-PO-RAD			& 1000	& 1.0 & Yes & Yes	& No		& $79$		&$140$	&$0.27$			& Red Dashed	\\
1.0pc-PO-RAD-ION		& 1000	& 1.0 & Yes & Yes	& Yes	& $95$		&$126$	&$0.24$			& Black Dashed	
\end{tabular}
\end{table*}

\subsection{Included physics}
We computed the evolution of a gravitationally collapsing mass reservoir and the consequential formation of a massive (proto)star and its circumstellar disk including its feedback onto the natal environment.
The overall numerical framework used to model these physics was presented in \citet{MakemakeSedna}.

We solved the hydrodynamical evolution of the system in axial symmetry including 
an $\alpha$-shear viscosity of the forming accretion disk,
the self-gravity of the gas as well as the gravity of the forming star,
outflow feedback from the protostar,
continuum radiation feedback of the star as well as the absorption and (re-)emission by the dust and gas, and
the photoionization feedback of the star.

For the viscous hydrodynamics, we utilized the open source magneto-hydrodynamics code \emph{Pluto} introduced in \citet{2007ApJS..170..228M}  and \citet{2012ApJS..198....7M}.
The self-gravity module (named \emph{Haumea}) and
the stellar evolution module
are described in \citet{2010ApJ...722.1556K}.
\vONE{
The $\alpha$-shear viscosity description was configured to model the angular momentum transport in the forming accretion disk via gravitational torques, please see \citet{2011ApJ...732...20K} for a direct comparison of a suite of axially symmetric viscous disk formation simulations with a corresponding three-dimensional simulation.
}
%we use constant values of $\alpha = 1$, $H/R = 0.1$, and set the angular velocity from the condition of gravito-centrifugal equilibrium at each radius, including the effect of self-gravity of the disk.
The protostellar outflow feedback module is identical to \citet{2015ApJ...800...86K} and \citet{2016ApJ...832...40K}; 
the outflow was injected into the computational domain at $t = 4~\mbox{kyr}$ with an ejection-to-accretion efficiency of 10\%, a flattening parameter of $0.01$, and a velocity of three times the local escape velocity with respect to the current protostellar mass.
Details of the continuum radiation transport module (named \emph{Makemake}) and photoionization module (named \emph{Sedna}) are given in \citet{MakemakeSedna}, which presents a major code update to the previous radiation transport module introduced in \citet{2010A&A...511A..81K} and \citet{2013A&A...555A...7K}.
This numerical star formation framework was also used to participate in the first study of the StarBench initiative \citep{2015MNRAS.453.1324B}, a code benchmark of the D-type expansion of HII regions.

For the evolution of the central (proto)star, we followed the evolutionary tracks by \citet{Hosokawa:2009eu} (please see \citet{2010ApJ...722.1556K} and \citet{2013ApJ...772...61K} for details).
To properly compute the ionizing photons emitted by the evolving star, we have implemented a stellar atmosphere model based on \citet{Kurucz:1979kj} (see \citet{MakemakeSedna} for details).
The dust and radiation temperatures were computed using the updated two-temperature hybrid continuum radiation transport scheme. 
The gas temperature of a fully ionized medium was set to $8000 \mbox{ K}$.
The gas temperature of a fully neutral medium was set equal to the dust temperature.
The gas temperature of a partially ionized medium was computed as a linear regression between these two cases.
The photoionization was computed assuming the on-the-spot approximation for direct recombination of free electrons into hydrogen's ground state.
We used a recombination cross section of $\sigma_\mathrm{rec} = 5.53 \times 10^{-18} \mbox{ cm}^2$, and analytic fit formulae for the recombination rates and collisional ionization rates as described in \citet{MakemakeSedna}.

We used the dust opacities from \citet{Laor:1993bv}.
During ray-tracing, the spectrum of the star was split into 65 frequency bins.
The FUV regime ($6 \mbox{ eV} \le h\nu < 13.6 \mbox{ eV}$) and EUV regime ($h\nu \ge 13.6 \mbox{ eV}$) were each treated as a single bin with dust opacity $\kappa_\mathrm{dust} = 4\times10^4 \mbox{ cm}^2 \mbox{ g}^{-1}$ and $2\times10^4 \mbox{ cm}^2 \mbox{ g}^{-1}$, respectively.
This choice is based on the dust opacities from \citet{Laor:1993bv} and \citet{Ossenkopf:1994tq}.
%, see Fig.~\ref{fig:Opacities}.
%\begin{figure}[htb!]
%\includegraphics[width=0.49\textwidth]{./images/OpacitiesInset}
%\caption{
%Comparison of dust opacities from \citet{Laor:1993bv}, \citet{Draine:1984jw}, and \citet{Ossenkopf:1994tq}.
%The inset highlights the opacities in the FUV/EUV regime, vertical lines in the inset correspond to the FUV and EUV transitions at $6 \mbox{ eV}$ and $13.6 \mbox{ eV}$, horizontal lines in the inset correspond to the chosen constant dust opacities within these bins.
%}
%\label{fig:Opacities}
%\end{figure}
For the gaseous medium, we used an opacity of $\kappa_\mathrm{gas} = 0.01 \mbox{ cm}^2 \mbox{ g}^{-1}$.
We updated the dust evaporation and sublimation module to consider a time-dependent evolution of the dust, see \citet{Bhandare} for details.

\subsection{Numerical grid}
We solved the evolution of the system on a numerical grid in spherical coordinates, assuming axial symmetry and midplane symmetry of the system.
The inner rim of the computational domain was fixed to $R_\mathrm{min} = 3 \mbox{ au}$, but please see Sect.~\ref{sect:Rmin} for a simulation series dedicated to checking for the proper size of the central sink cell.
Depending on the initial conditions, the grid extends out to $R_\mathrm{max} = R_\mathrm{res} = 0.1 \mbox{ or } 1.0 \mbox{ pc}$.
In the polar direction, the grid extends from $\theta_\mathrm{min} = 0$ to $\theta_\mathrm{max} = \pi/2$, assuming midplane symmetry for the evolution of the lower hemisphere.
In the azimuthal dimension, we further assumed axial symmetry of the system, hence, solved the evolution on a two-dimensional grid.
The number of grid cells in our default configuration is $N_\mathrm{r} \mbox{ x } N_\mathrm{\theta} = 90 \mbox{ x } 16$ for the small mass reservoir and $114 \mbox{ x } 16$ for the large mass reservoir.
In the radial direction, the grid cells increase in size proportional to the spherical radius, a so-called log-grid.
In the polar direction, the grid cells are uniformly spaced in the polar angle.

Due to the radially logarithmic scaling of the spatial grid size, this fairly small number of grid cells already gives a very high spatial resolution of the stellar surroundings, namely the forming accretion disk.
In the default resolution given above, the smallest grid cells have a spatial extent of $0.33 \mbox{ au}$, but please see Sect.~\ref{sect:resolution} for a dedicated resolution study.
The relatively high computational expense of these simulations arises from the multi-physical processes modeled as well as the combination of very high spatial resolution of the stellar surrounding ($\Delta x_\mathrm{min} \approx 0.33~\au$) \vONE{and the energetic feedback in those regions along} with the long-term evolution investigated.

\subsection{Boundary conditions}
The solvers for hydrodynamics, self-gravity, continuum radiation transport, and ionization require proper boundary conditions at the edges of the computational domain.
For the hydrodynamics, we used a semi-permeable wall at the radially outer boundary $R_\mathrm{max}$, i.e.~we allowed gas to leave (as a result of the feedback) but not to enter the computational domain.
At the inner rim $R_\mathrm{min}$, we used basically the same semi-permeable wall, i.e.~we allowed gas to leave the computational domain (as stellar accretion), but not to enter the domain.
This boundary condition was superimposed with the protostellar outflow injection module, the resulting mass flux at the interface was solved assuming momentum conservation (see \citet{2015ApJ...800...86K} and \citet{2016ApJ...832...40K} for details).
No hydrodynamic or radiative flux was allowed across the polar axis boundary or disk midplane.

For the ray-tracing of stellar continuum radiation and photoionization, the stellar emission was followed along the radial coordinate from the sink cell's radius $R_\mathrm{min}$ up to the outer boundary $R_\mathrm{max}$ of the computational domain; of course the remaining radiative flux at $R_\mathrm{max}$ simply leaves the domain.
The thermal continuum (re-)emission of the dust and gas was treated in the flux-limited diffusion approximation.
At the outer boundary, we applied a free streaming condition in the optically thin limit, i.e.~a zero gradient in the radiative flux.
%The inner boundary was set to the initial temperature of $10 \mbox{ K}$; in the hybrid scheme, the stellar irradiation is computed using an independent ray-tracing step and, hence, does not contribute to the thermal continuum radiation energy computed within the flux-limited diffusion solver.

\section{Physical initial conditions and feedback physics}
\label{sect:simulations}
The gas density was initialized as a spherically symmetric distribution, which decreases quadratically toward larger spherical radii: $\rho_\mathrm{gas} \propto r^{-2}$.
The gas velocity was initialized as a pure axially symmetric rotation ($v_\mathrm{r} = v_\mathrm{\theta} = 0$), which decreases linearly toward larger cylindrical radii: $\Omega_\mathrm{gas} \propto R^{-1}$.
The initial temperatures were set to a uniform value of $T_\mathrm{gas} = T_\mathrm{dust} = T_\mathrm{rad} = 10 \mbox{ K}$.

We initialized the mass reservoir out to a maximum size of either $R_\mathrm{res} = 0.1 \mbox{ pc}$ or $R_\mathrm{res} = 1.0 \mbox{ pc}$.
The smaller mass reservoir was initialized such that $M_\mathrm{res} = 100~\Msol$ and the large-scale mass reservoir to $M_\mathrm{res} = 1000~\Msol$.
The corresponding free-fall timescale of the reservoirs is $t_\mathrm{ff} = 52.4~\kyr$ and $524~\kyr$, respectively.
The normalization of the rotation was fixed such that the rotational to gravitational energy ratio is $E_\mathrm{rot} / E_\mathrm{grav} = 2\%$.
Note that the choice to make the angular momentum power-law index half the density power-law index results in the rotational to gravitational energy ratio being equal in the small scale and large scale configurations.
In fact, the large-scale mass reservoir is simply an extension of the small-scale reservoir, i.e.~the inner $0.1 \mbox{ pc}$ region of both setups are identical.

Not including the convergence tests (see Sect.~\ref{sect:convergence}), we present the analyses of eight simulations performed in total.
First, we distinguish two kinds of initial mass reservoirs: a $100 \mbox{ M}_\odot$ reservoir of radius $R_\mathrm{res} = 0.1 \mbox{ pc}$ and a $1000 \mbox{ M}_\odot$ reservoir of radius $R_\mathrm{res} = 1.0 \mbox{ pc}$.
In the following, we will refer to these configurations as small-scale and large-scale mass reservoirs.

The underlying aim of the variation of the size (and mass) of the initial reservoir is to study the resulting feedback and star formation efficiency in two very different accretion scenarios:
The small-scale reservoir represents a finite mass reservoir for the forming massive star, and, hence resembles an isolated pre-stellar high-mass core.
The large-scale mass reservoir was chosen to be large enough in size that its free-fall time is larger than the accretion time of the forming star. 
Hence, the large-scale mass reservoir represents a virtually infinite mass reservoir for the forming massive star, and hence resembles for example a pre-stellar core embedded in a larger scale cloud or filament, leading to a sustained supply of mass accretion from these large scales.
Moreover, by direct comparison of the simulations with and without the large-scale mass reservoir, we can examine the impact of the ram pressure from large-scale accretion flows on the stellar envelope, the bipolar cavity and the accretion disk.

Additionally, in both configurations, we varied the feedback physics included by artificially switching off the radiative force terms and/or the photoionization solver (see Table~\ref{tab:sims} for an overview of simulations presented).
This approach allows us to investigate the impact of these feedback effects, and their potential interplay, in a direct comparison of the resulting evolution of the collapsing cloud as well as the stellar, disk, and envelope properties.

%%%%%%%%%%%%%%%%
%%                                       %%
%%  Results \& Discussion  %%
%%                                       %%
%%%%%%%%%%%%%%%%
\section{Results and discussion}
\label{sect:results}

% TOC of Results Section:
In this section, we present and discuss the results of the simulations by breaking them down into subsections focused on specific aspects of star formation and feedback.
First, we start with an overview of the overall evolution of the system, which we divide into characteristic epochs of specific dominant forces and feedback mechanisms (Sect.~\ref{sect:results_Basics}).
Afterwards, we analyze
the evolution of the circumstellar accretion disk (Sect.~\ref{sect:results_Disk}),
the evolution of the massive (proto)star (Sect.~\ref{sect:results_Star}),
the feedback efficiency of the radiation forces and photoionization in terms of the final stellar mass and the associated accretion timescale (Sect.~\ref{sect:results_StarFormation}), 
the expansion of HII regions with and without the effect of radiation pressure as well as photoionization feedback on large scales (Sect.~\ref{sect:results_HII}), and
the evolution of the bipolar outflow cavities (Sect.~\ref{sect:results_Outflow}).

\begin{figure}[thbp]
\centering
\includegraphics[width=0.35\textwidth]{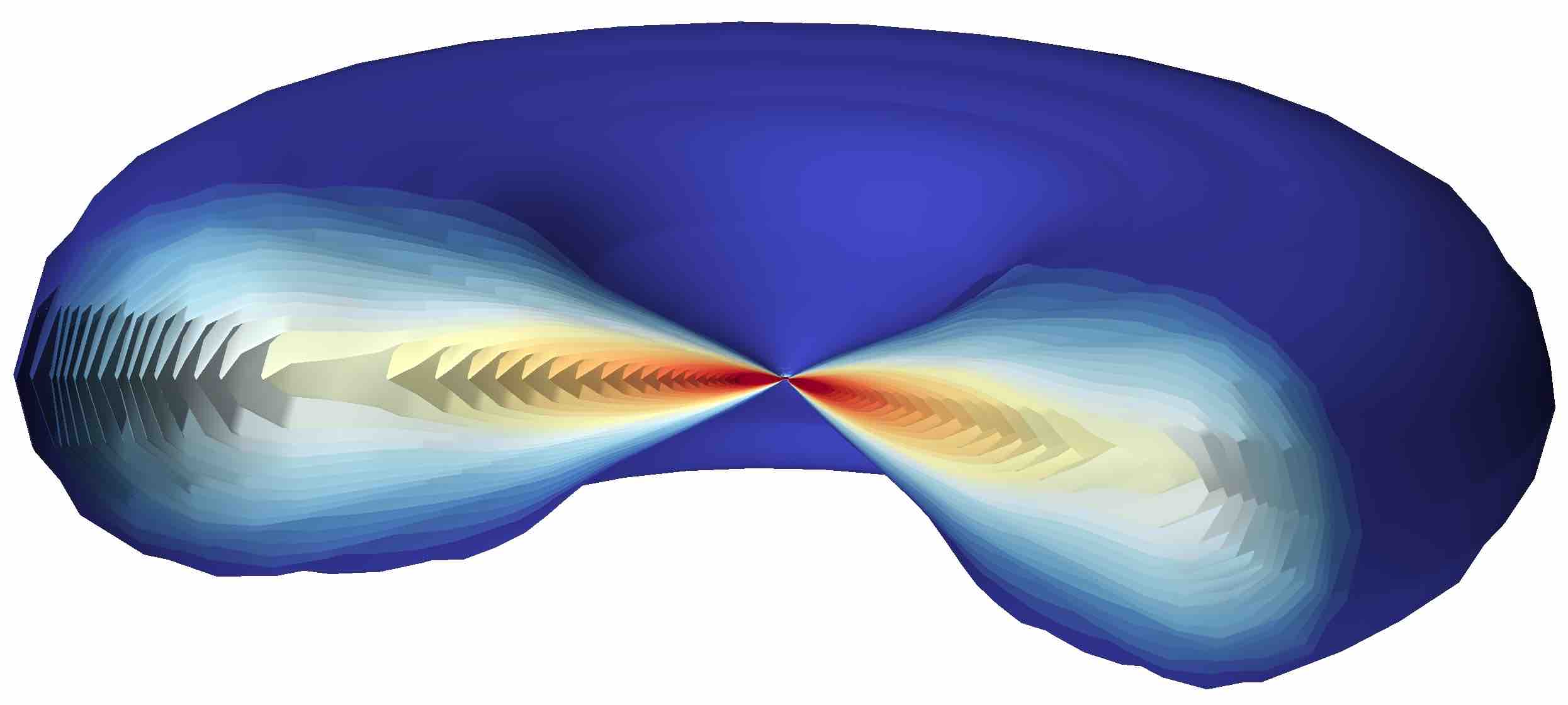}\\
$\log\left(\rho_\mathrm{gas}~[\gccm]\right)$\\
\includegraphics[width=0.49\textwidth]{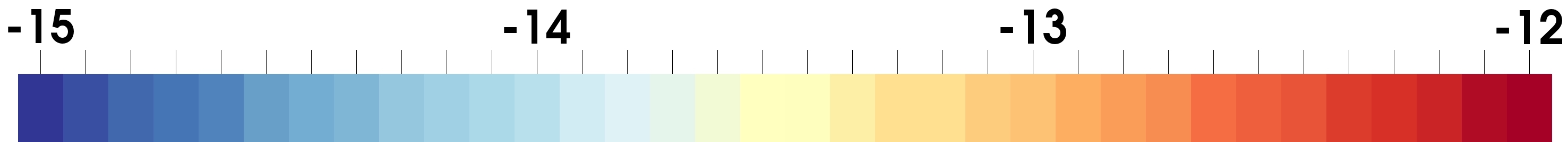}
\caption{
3D iso-contour rendering of the circumstellar accretion disk at $t = 10~\kyr$.
The image covers $\approx 1000~\au$ in width.
}
\label{fig:visualization_disk}
\end{figure}

\begin{figure}[ht!]
\centering
\includegraphics[width=0.35\textwidth]{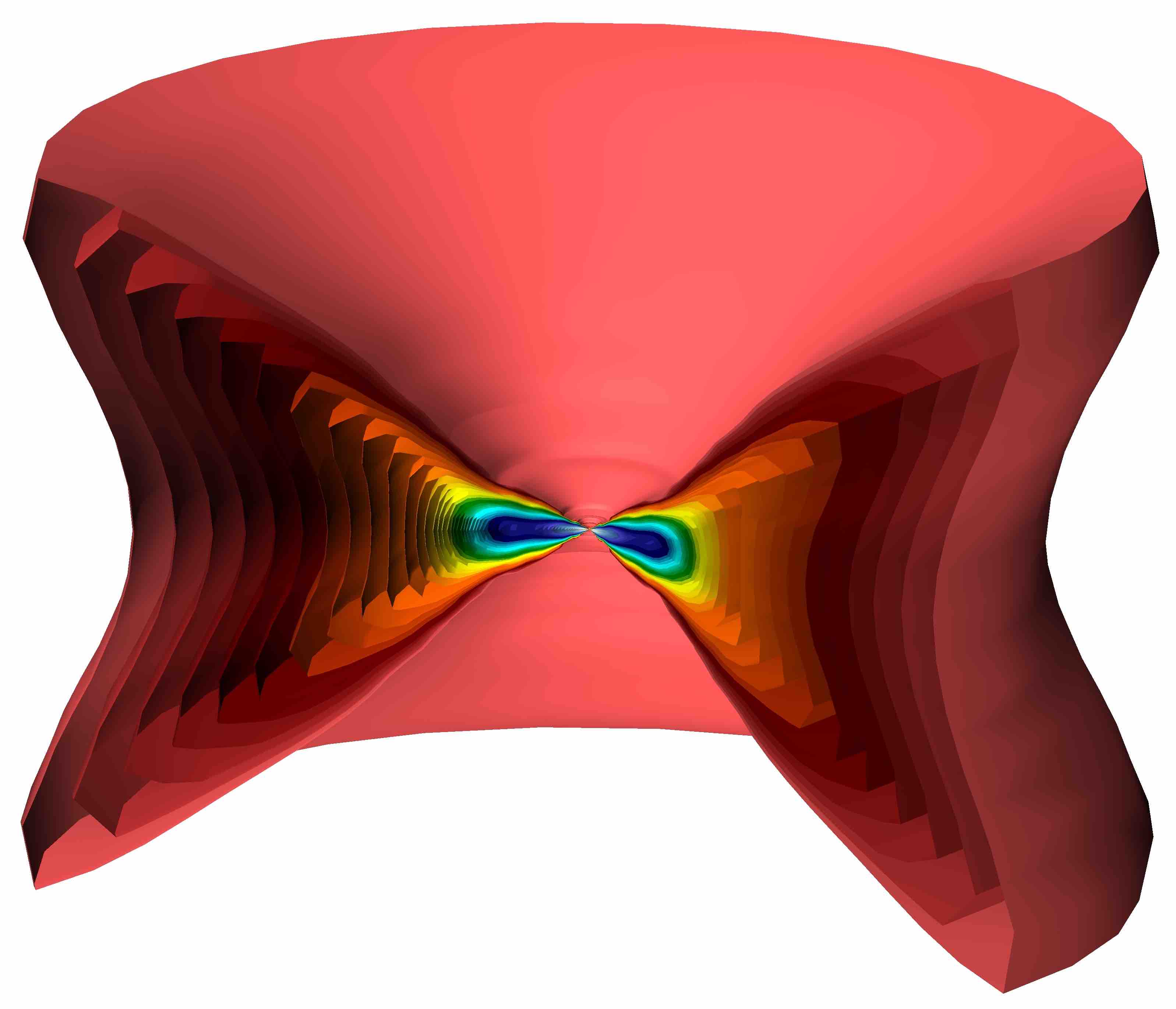}\\
$\log\left(\rho_\mathrm{gas}~[\gccm]\right)$\\
\includegraphics[width=0.49\textwidth]{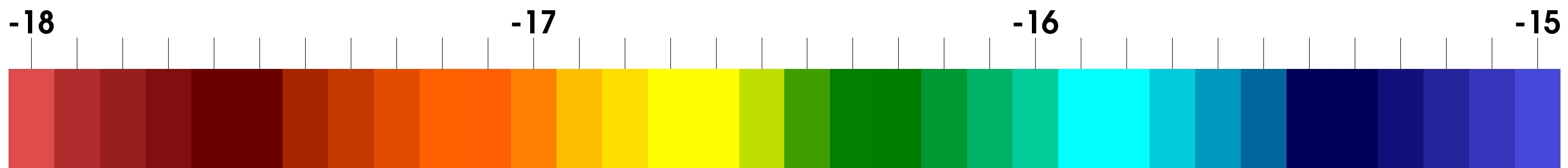}
\caption{
3D iso-contour rendering of the disk plus the surrounding torus at $t = 70~\kyr$.
The disk density is color-coded as in Fig.~\ref{fig:visualization_disk}.
The image covers $\approx 0.1~\pc$ in height.
}
\label{fig:visualization_torus}
\end{figure}

\begin{figure}[tbhp]
\centering
\rotatebox{90}{\hspace{12mm}$v_\mathrm{r}~[\kms]$}
\includegraphics[width=0.055\textwidth]{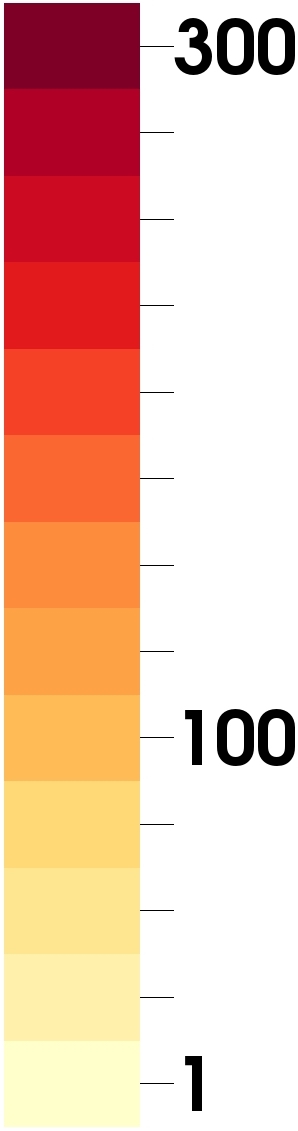}
\includegraphics[height=0.49\paperheight]{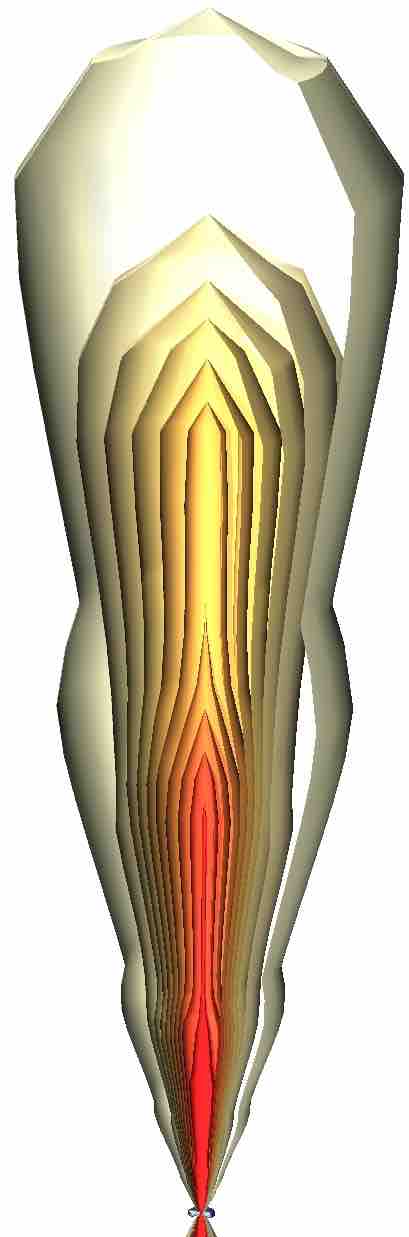}
\caption{
3D iso-contour rendering of the disk and the large-scale outflow at $t = 10~\kyr$.
The disk density is color-coded as in Fig.~\ref{fig:visualization_disk}.
The image covers $\approx 0.25~\pc$ in height.
}
\label{fig:visualization_outflow}
\end{figure}

\begin{figure}[tbhp]
\centering
\includegraphics[height=0.49\paperheight]{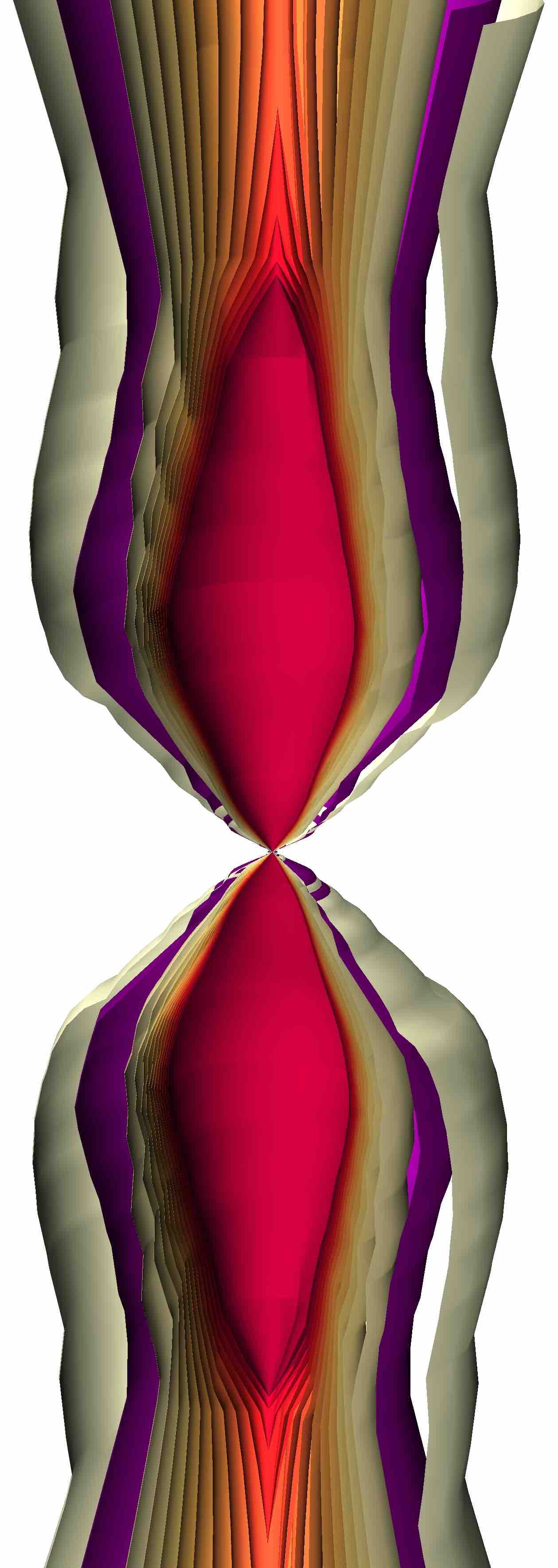}
\caption{
3D iso-contour rendering of the disk plus outflow plus HII region at $t = 90~\kyr$.
The extent of the ionized HII region is shown as a purple contour.
The outflow velocity is color-coded as in Fig.~\ref{fig:visualization_outflow}.
The image covers $\approx 1.8~\pc$ in height.
}
\label{fig:visualization_HII}
\end{figure}

\subsection{Dominant forces and basic dynamical and thermodynamical evolution}
\label{sect:results_Basics}
In this section, we focus on the fiducial case including all kinds of feedback, i.e., run ``1pc-PO-RAD-ION''.
We divide the evolution of the system into five evolutionary phases, based on the dominating forces or dynamics during these epochs respectively:
1) global gravitational collapse,
2) circumstellar disk formation,
3) protostellar outflows and jets,
4) radiation pressure feedback,
5) ionization feedback and HII region formation.
Although these epochs strongly overlap in time rather than being distinct, 
in each of the phases, another force becomes as strong as or stronger than gravity, hence, hinders the gravitational collapse of the mass reservoir.
While in phase 1), gravity is the only important force, phase 2) is characterized by the increase of centrifugal forces on small scales.
In phases 3) - 5), feedback mechanisms, i.e.~forces that exceed the gravitational attraction of the forming star, come into play, namely
the injected momentum from the jet and outflow,
the radiation forces from the stellar irradiation, and
the thermal pressure from the hot, ionized gas.

\begin{figure*}
\centering
\begin{subfigure}[Gravitational Collapse \label{fig:PhaseDiagrams-GravitationalCollapse}]{
\includegraphics[width=0.3\textwidth]{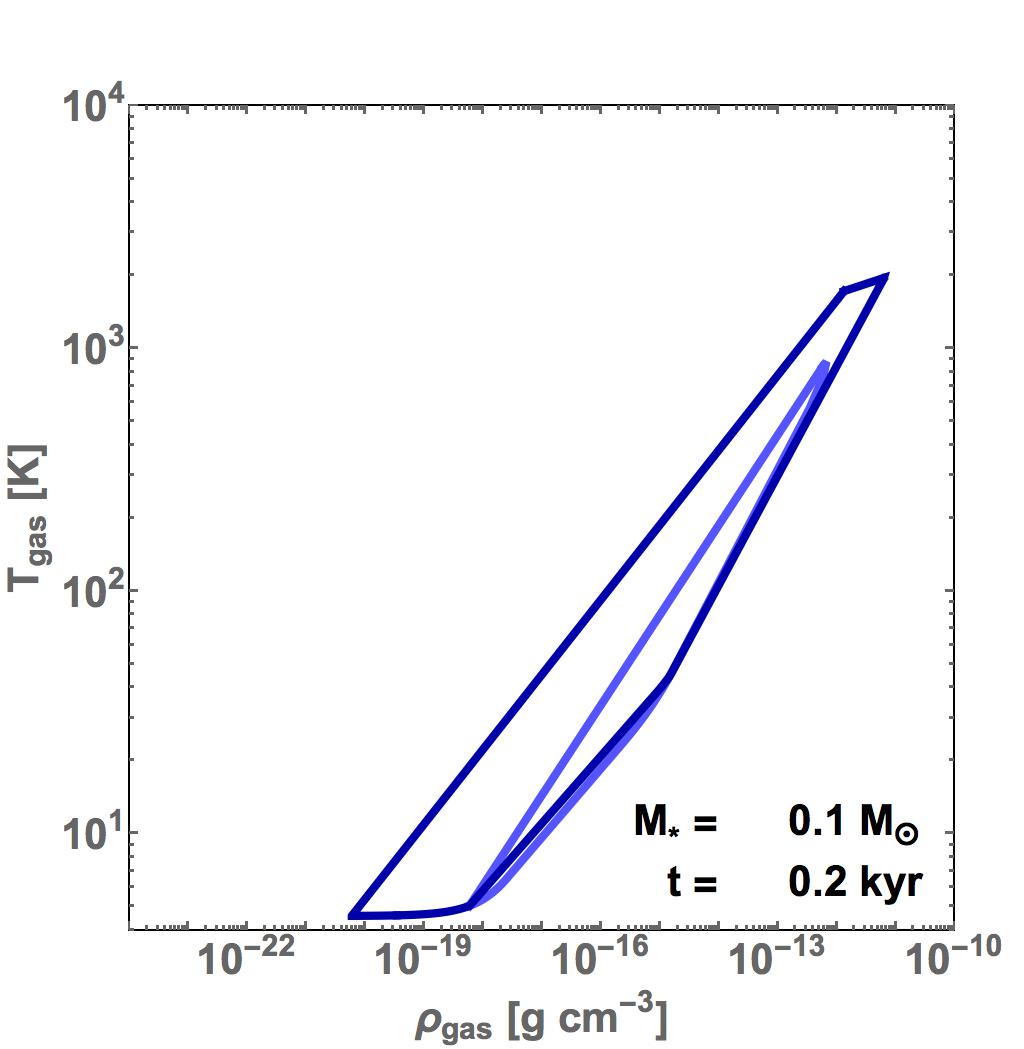}
}
\end{subfigure}
\begin{subfigure}[Torus Formation \label{fig:PhaseDiagrams-TorusFormation}]{
\includegraphics[width=0.3\textwidth]{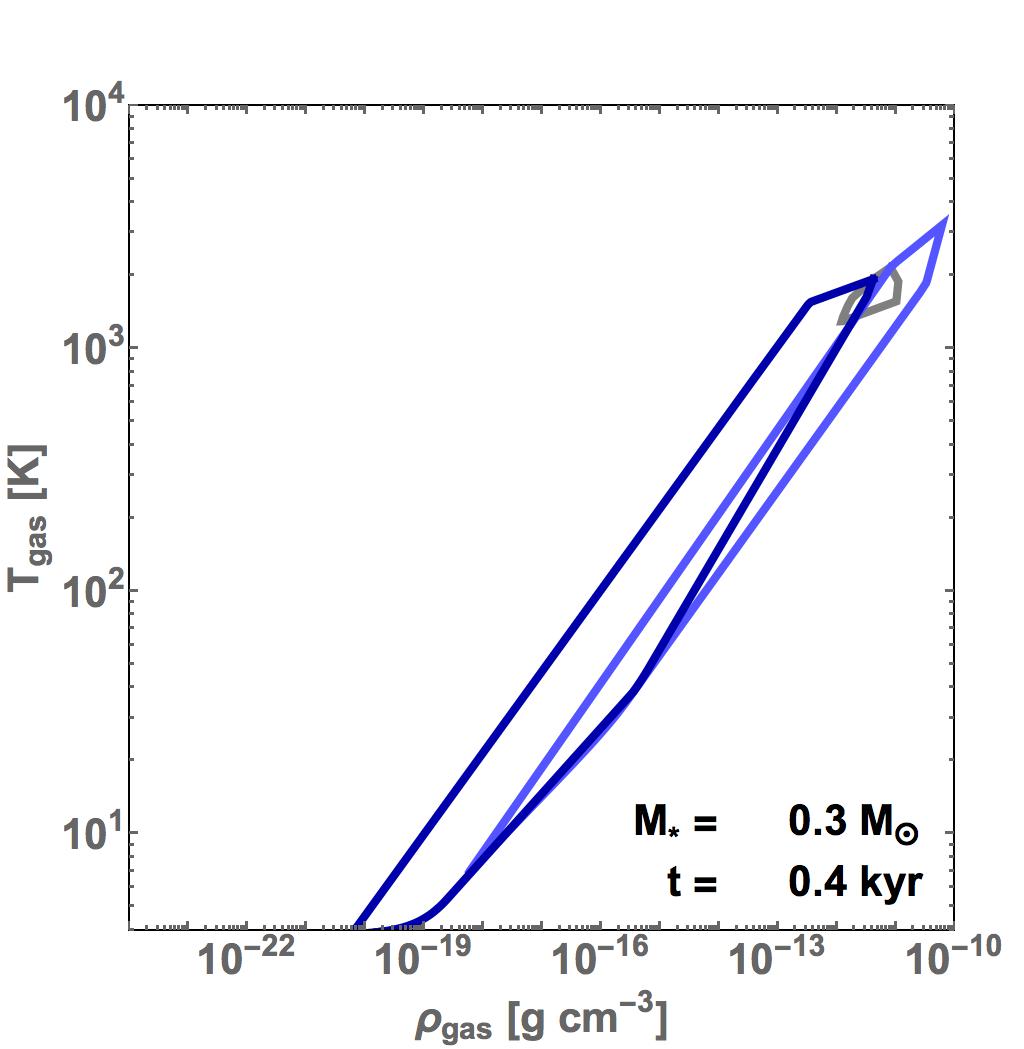}
}
\end{subfigure}
\begin{subfigure}[Disk Formation \label{fig:PhaseDiagrams-DiskFormation}]{
\includegraphics[width=0.3\textwidth]{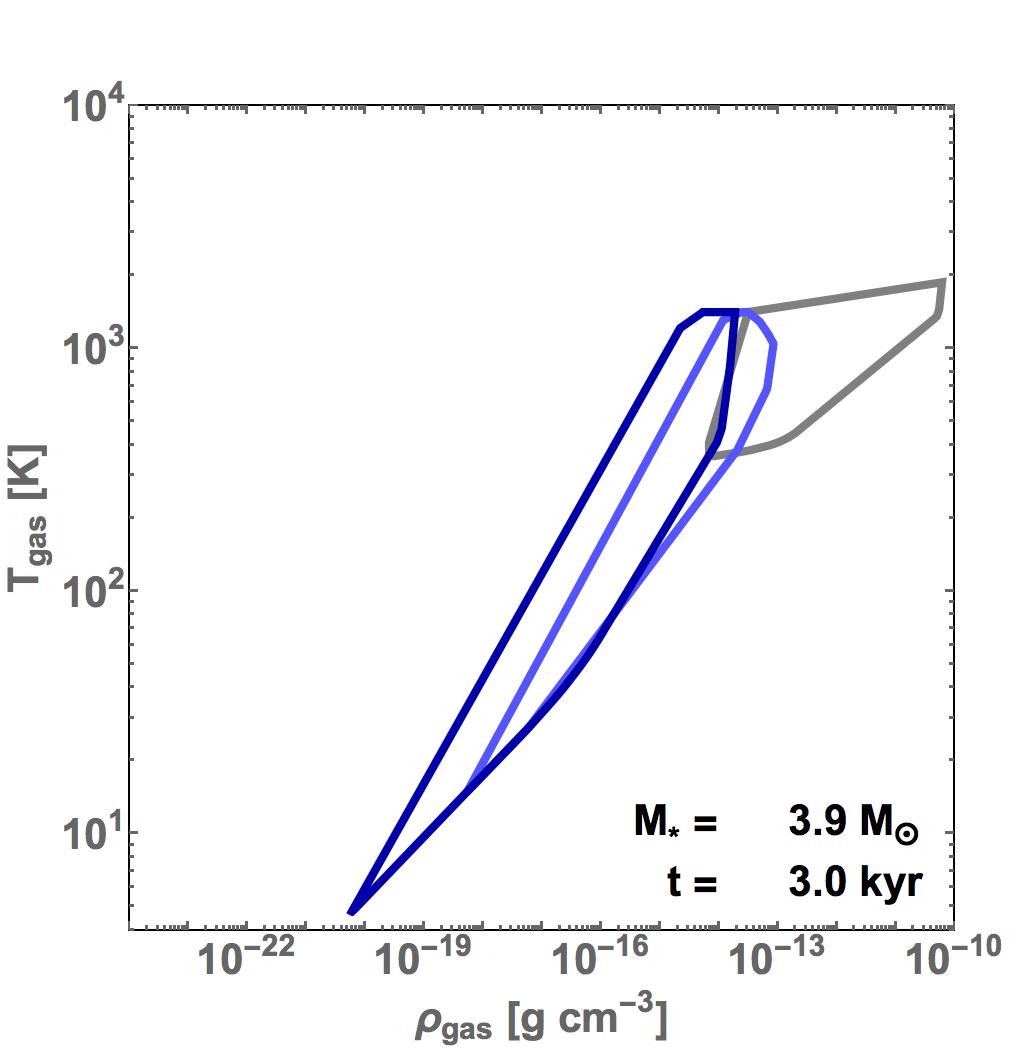}
}
\end{subfigure}
\begin{subfigure}[Jet \& Protostellar Outflow Feedback \label{fig:PhaseDiagrams-OutflowFeedback}]{
\includegraphics[width=0.3\textwidth]{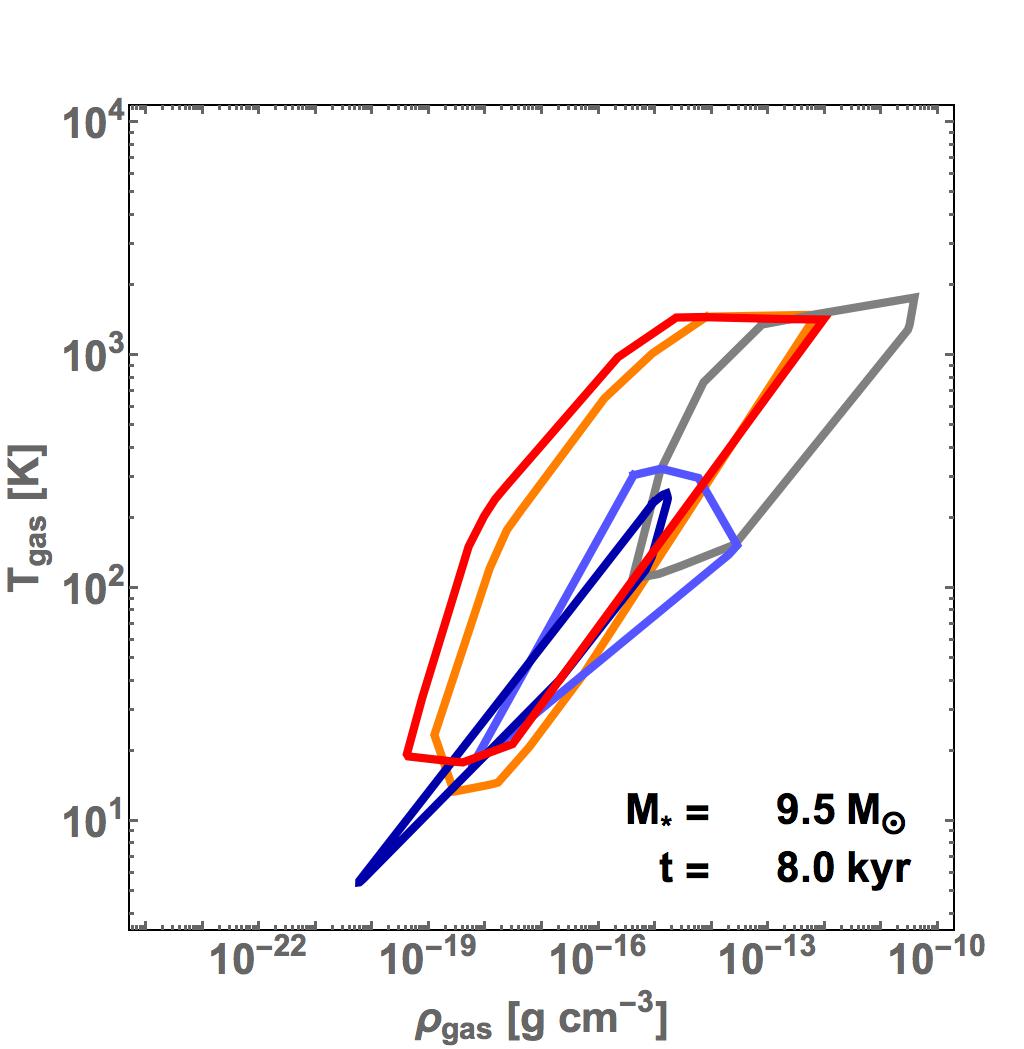}
}
\end{subfigure}
\begin{subfigure}[Radiation Force Feedback \label{fig:PhaseDiagrams-RadiationForceFeedback}]{
\includegraphics[width=0.3\textwidth]{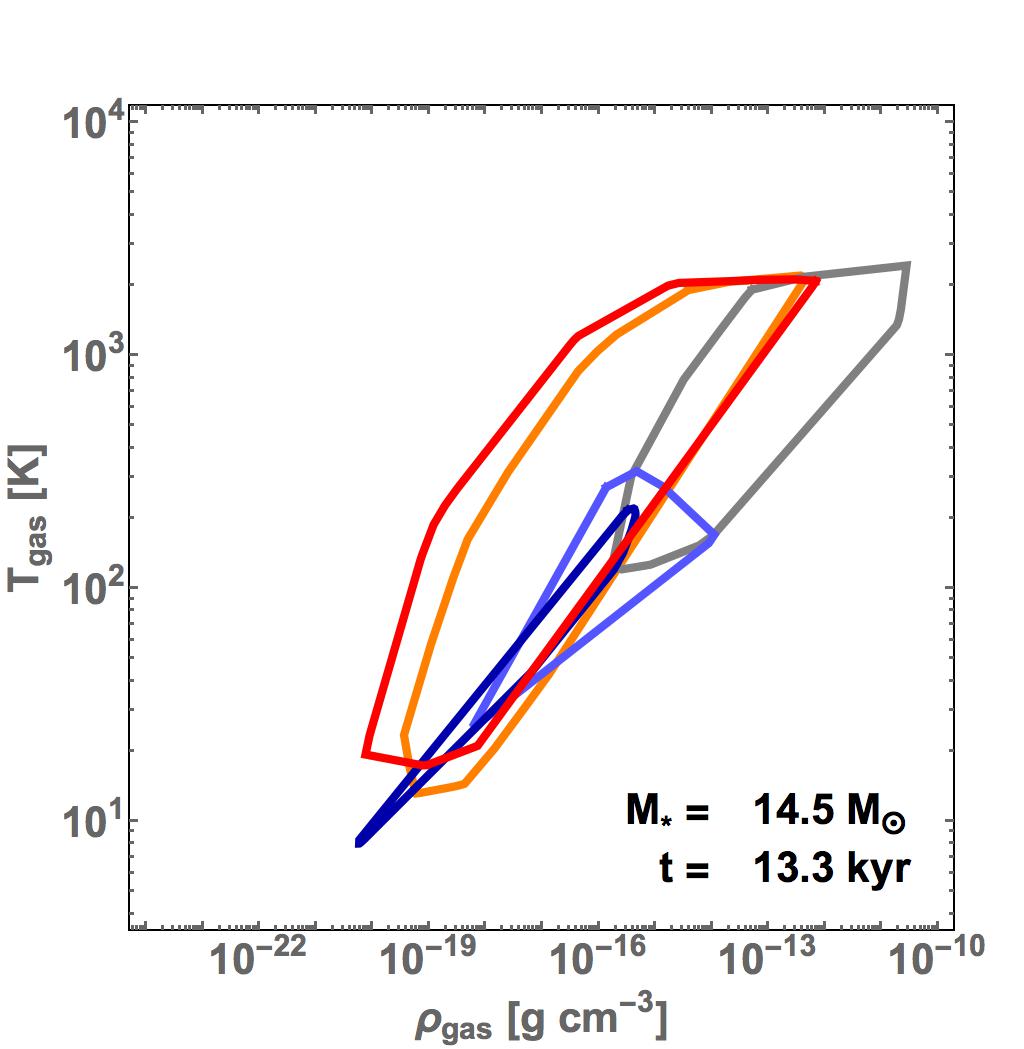}
}
\end{subfigure}
\begin{subfigure}[Photoionization Feedback \label{fig:PhaseDiagrams-PhotoionizationFeedback}]{
\includegraphics[width=0.3\textwidth]{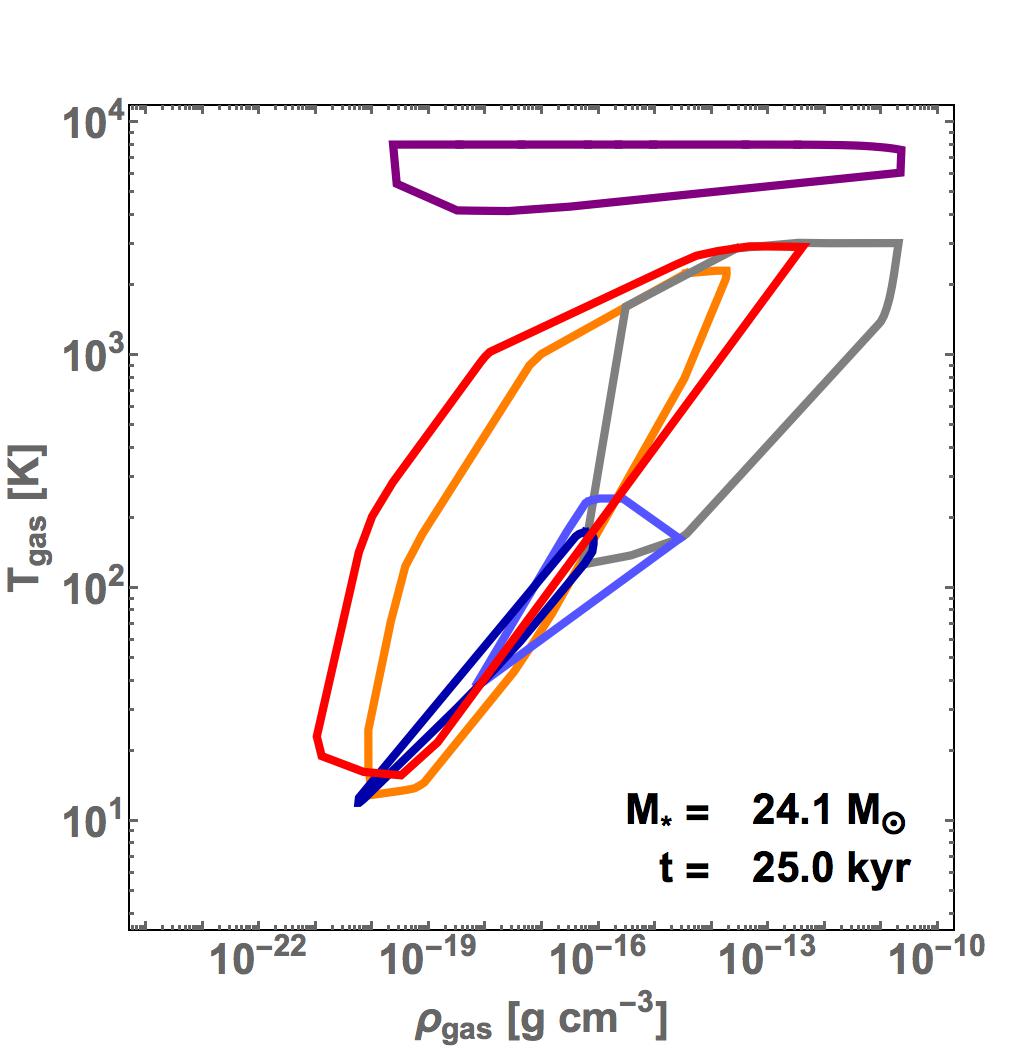}
}
\end{subfigure}
\begin{subfigure}[HII Region Expansion \label{fig:PhaseDiagrams-HIIRegionExpansion}]{
\includegraphics[width=0.3\textwidth]{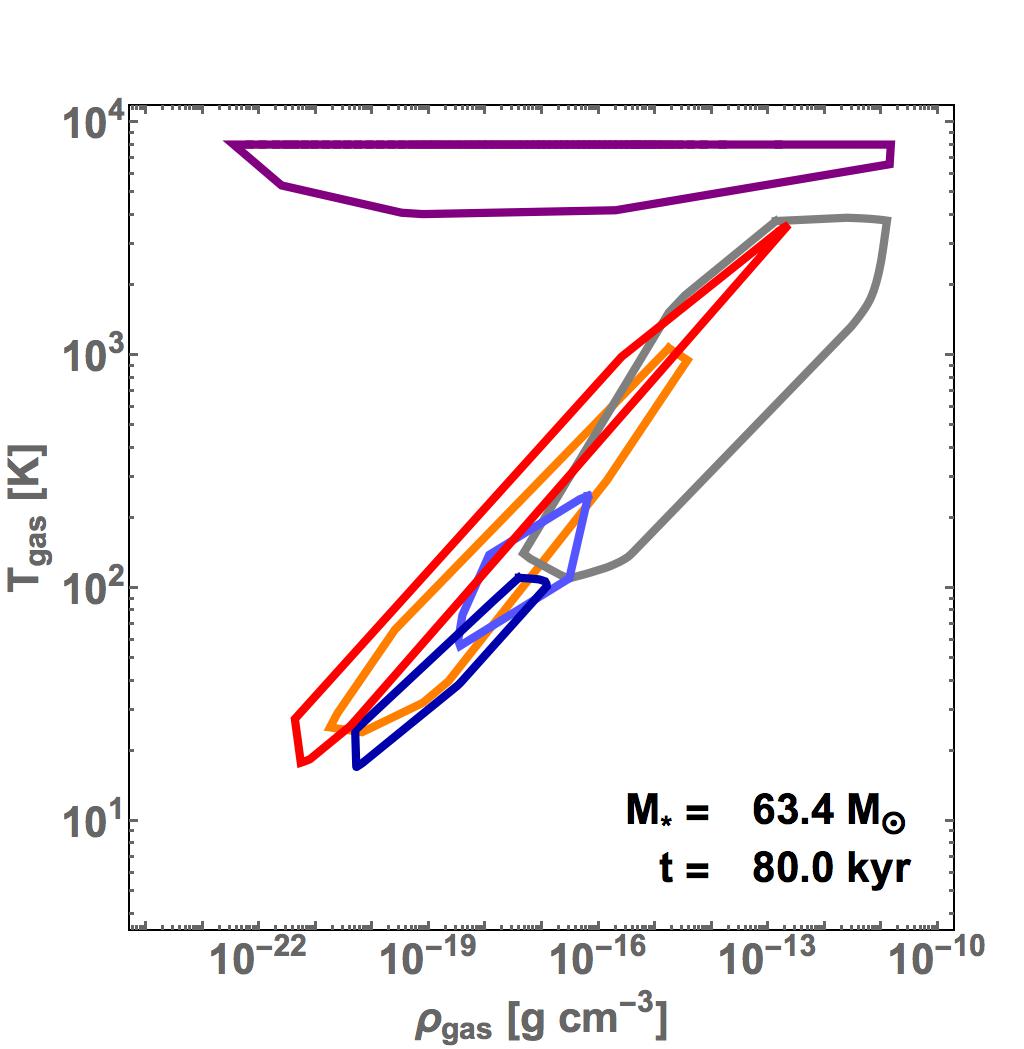}
}
\end{subfigure}
\begin{subfigure}[Disk Destruction \label{fig:PhaseDiagrams-DiskDestruction}]{
\includegraphics[width=0.3\textwidth]{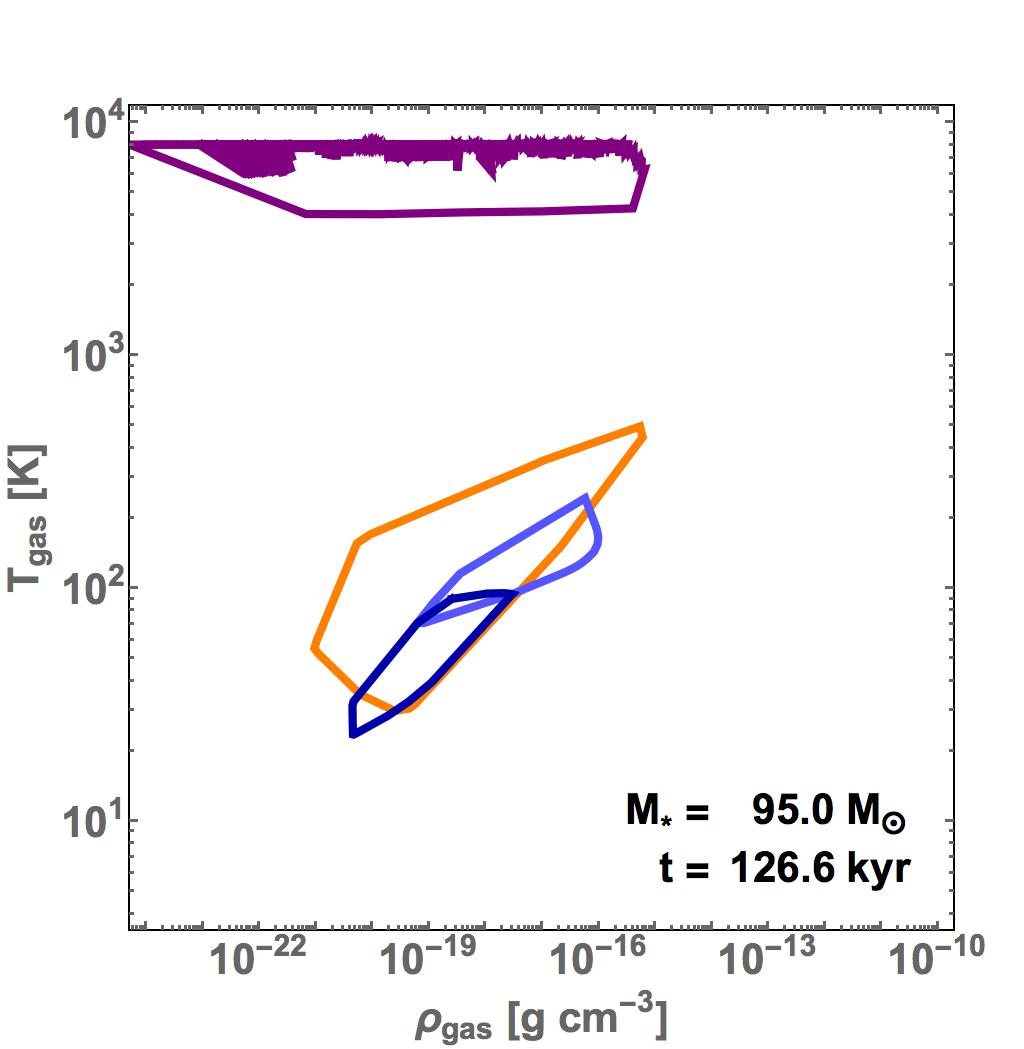}
}
\end{subfigure}
\begin{subfigure}[Legend of color-coding]{
\begin{tabular}{l l l}
\vspace{-60mm}\\
Region & Selection Criterion & Color Code \\
\hline
Infall 	& $v_\mathrm{r} < 0$ and $|v_\mathrm{r}| > |v_\mathrm{\phi}|$ 	& Dark Blue 	\\
Torus 	& $v_\mathrm{r} < 0$ and $|v_\mathrm{r}| < |v_\mathrm{\phi}|$ 	& Light Blue 	\\
Disk 		& $v_\mathrm{\phi} \approx v_\mathrm{Kepler}$ 			& Gray 		\\
Outflow 	& $v_\mathrm{r} > 1~\kms$ 							& Orange 		\\
Jet 		& $v_\mathrm{r} > 100~\kms$ 							& Red 		\\
HII 		& $x > 0.5$ 										& Purple
\end{tabular}
}
\end{subfigure}
\caption{
Phase diagrams (gas temperature -- gas density plane) for the different epochs of the evolution of the stellar surrounding.
The time after the onset of the initial global gravitational collapse and the current mass of the (proto)star is given at the bottom right corner of each panel, the associated label is given in the sub-caption.
Data is taken from the fiducial case ``1.0pc-PO-RAD-ION''.
}
\label{fig:PhaseDiagrams}
\end{figure*}

See Figs.~\ref{fig:visualization_disk} to \ref{fig:visualization_HII} for 3D iso-contour renderings of the associated objects.
The objects related to the dominant dynamics are shown in combination with their thermodynamics in Fig.~\ref{fig:PhaseDiagrams}, the ``dynamics phase diagram''.
The dynamics phase diagram can be first distinguished into an accretion and a feedback track, shown in blue-gray and orange-red-purple colors, respectively.
The accretion track gets further divided into the relative importance of rotation vs.~infall:
Regions, where the absolute value of the infall velocity from global collapse is higher than the rotational velocity ($|v_\mathrm{r}| > |v_\mathrm{\phi}|$ and $v_\mathrm{r} < 0$), are labeled as infall (dark blue contour).
Regions, where rotation velocity is higher than the absolute value of the infall velocity ($|v_\mathrm{r}| < |v_\mathrm{\phi}|$ and $v_\mathrm{r} < 0$), are associated with the torus (light blue contour).
Finally, regions close to gravito-centrifugal equilibrium ($v_\mathrm{\phi} = v_\mathrm{Kepler} \pm 10\%$) belong to the disk (gray contour).

The feedback track gets divided into a fast jet ($v_\mathrm{r} > 100~\kms$) and a slow outflow ($1\kms < v_\mathrm{r} < 100~\kms$) component as well as an HII region (determined as an ionization degree of $x > 0.5$).

The mass reservoir at $t = 0$ is set up with 2\% of rotational to gravitational energy, a negligibly small thermal pressure,
and no magnetic field contribution.
Hence, magnetic and thermal pressure forces are negligible, and the initial phase is marked by global gravitational collapse, i.e.~gravity is the dominant force and the system's evolution can be described by a free-fall solution albeit for a very short initial period in time.
The free-fall timescale is given by $t_\mathrm{ff} \approx 52~\kyr$ and $520~\kyr$ for the $0.1~\pc$ and $1.0~\pc$ mass reservoir, respectively.
During the initial phase of gravitational collapse, the chosen power-law of the initial density distribution of $\rho \propto r^{-2}$ results in a constant infall rate.
For the chosen configuration of the mass reservoirs, the resulting protostellar accretion rate is $\dot{M}_* = M_\mathrm{res} / t_\mathrm{ff} \approx 2 \times 10^{-3}~\Msolyr$ in both cases.

The second strongest force is the centrifugal force.
Due to angular momentum conservation, the infalling material approaches gravito-centrifugal equilibrium. 
Here, the term ``gravito'' includes the protostellar mass as well as the self-gravity of the disk and surrounding gas, i.e.~the total gravity due to the included mass at each radius.
In the following, we will refer to the associated rotation or angular velocity as being Keplerian.
As a consequence of this gravito-centrifugal equilibrium a circumstellar disk builds up and grows in time from the inside out (see Fig.~\ref{fig:visualization_disk} for a 3D iso-contour rendering example).
The first sign of this Keplerian rotation occurs in the simulations at $\approx 190~\yr$ at the inner rim $R_\mathrm{min} = 3~\au$ of the computational domain.
As shown in the dynamics phase diagrams of Fig.~\ref{fig:PhaseDiagrams}, the temperature and density slope leads to a reasonable correlation of higher density gas being hotter.
This correlation can be used in observational data analysis to correlate higher energy transitions of molecular line excitation with the associated orbit within the disk to check for rotational support (see e.g.~\citet{2015ApJ...813L..19J}, Fig.~4).

The disk itself is embedded in a torus environment, where rotation plays a significant role (in contrast to strongly gravity dominated larger scales), but the gas is not in gravito-centrifugal equilibrium and hence still infalling.
See Fig.~\ref{fig:visualization_torus} for a 3D iso-contour rendering example.

At $t = 4~\kyr$, 
\vONE{shortly after the formation of the circumstellar accretion disk, which in general is associated with the launching of a magneto-centrifugally launched jet and outflow,}
we inject the jet and protostellar outflow into the surrounding medium, see Fig.~\ref{fig:visualization_outflow} for a 3D iso-contour rendering example.
The disk has grown up to $\approx 63~\au$ at this point in time.
As a result of the injected momentum, the bipolar regions will be shaped into lower-density cavities with high outward velocity.
Hence, the protostellar outflow denotes the first feedback effect in the sense that it is able to push gas away from the accreting star and disk.
As a consequence, the phase diagram Fig.~\ref{fig:PhaseDiagrams} of the associated phase extends toward lower gas densities within the newly appearing regions of slow and fast outflowing material.
For a detailed analysis of the evolution and broadening of the outflow cavity and its dependence on the included physics and spatial scales, see Sect.~\ref{sect:results_Outflow} below.

The second feedback mechanism is the radiation force.
At $t \approx 9.2~\kyr$, the protostar becomes super-Eddington with respect to dust opacity $\Gamma_* = L_*/M_* \times \kappa_\mathrm{P}(T_*) / 4\pi G c > 1$, i.e.~the momentum input from the stellar luminosity into the dusty surrounding is higher than the gravitational attraction of the protostar.
At this point in time, the star has grown to $M_* \approx 10.8~\Msol$ with a luminosity of $L_* \approx 7500~\Lsol$.
Due to the strong bloating of the protostar during these early accretion epochs ($R_* \approx 140~\Rsol \approx 0.65~\au$), it has a resulting effective photospheric temperature of only $T_* \approx 4500~\K$.
From now on, the Eddington factor of the star will continuously increase, reaching a value of $100$ at $t \approx 30~\kyr$, i.e.~the radiation force is already 100 times stronger than the gravity of the star at this phase.

Once the stellar radius contracts toward the zero age main sequence, the stellar emission spectrum is shifted toward higher frequencies.
As a result, the strong EUV radiation from the star photoionizes the hydrogen in its surrounding, creating an HII region.
See Fig.~\ref{fig:visualization_HII} for a 3D iso-contour rendering example.
In HII regions of small size (so-called hyper-compact and ultra-compact HII regions), the soundspeed of the ionized gas remains below the escape velocity from the star, hence, the ionized gas is trapped.
For HII regions larger than the critical radius $R_\mathrm{crit,HII} = G M_* / 2 c_\mathrm{s,ion}^2$ \citep[][]{Keto:2003kk}, the thermal pressure of the hot, ionized gas exceeds the gravitational attraction of the star, hence, the ionization starts to be an important feedback mechanism.
In the simulation runs presented here, the HII region expansion starts at about $t \approx 25~\kyr$, when the $24~\Msol$ star has contracted down to $17.5~\Rsol$ with a luminosity and temperature of $1.1 \times 10^5~\Lsol$ and $25~\kK$, respectively.
For further details on the evolution and feedback effect of the expanding HII region, see Sect.~\ref{sect:results_HII}.

In the following subsections, we present and discuss in more detail the evolution of the star, the disk, the outflow cavity, and the HII region.
Most importantly, we quantify the importance of the different feedback mechanisms and their dependence on time and spatial scales.

\subsection{Evolution of the accretion disk}
\label{sect:results_Disk}
As addressed in the previous section, we use the terminology infall, torus, and disk to subdivide the accretion from cloud to stellar scales based on the relative importance of rotation compared to gravitational attraction.
On the largest scales, centrifugal forces are negligible compared to gravity, while on the smallest scales around the forming star, the medium approaches a gravito-centrifugal equilibrium.
We refer to the first one as the infall region, and to the latter one as the circumstellar disk.
In between these two extremes, the region is characterized by both motions -- infall and rotation.
In sync with the observational terminology, we refer to this region as the torus;
in our simulations we find that the increase of rotational support leads to flattened objects with densities $10^{-18}~\gccm \lesssim \rho_\mathrm{gas} \lesssim 10^{-15}~\gccm$, i.e.~in between pre-stellar cores and circumstellar disks.

\begin{figure*}[p]
\centering
{\bf t [kyr]} \\
\includegraphics[width=0.35\textwidth]{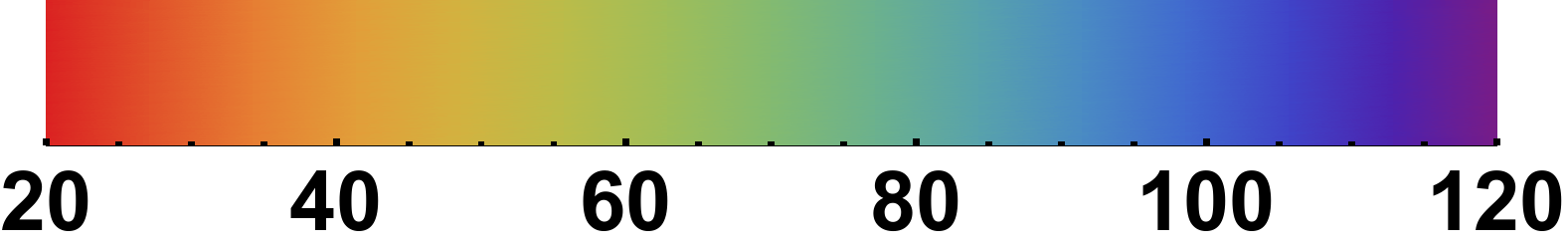}\\
\includegraphics[width=0.49\textwidth]{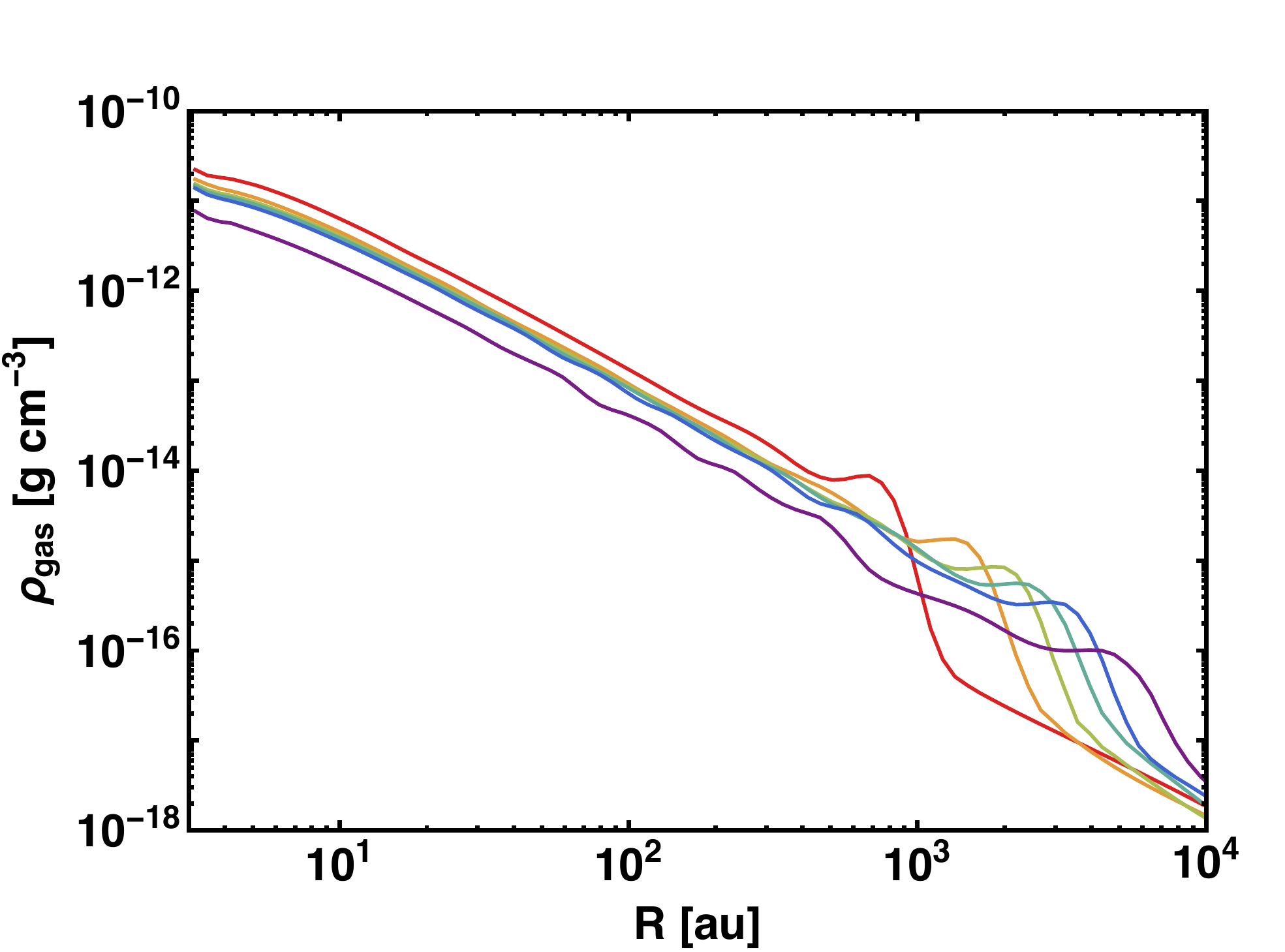}
\includegraphics[width=0.49\textwidth]{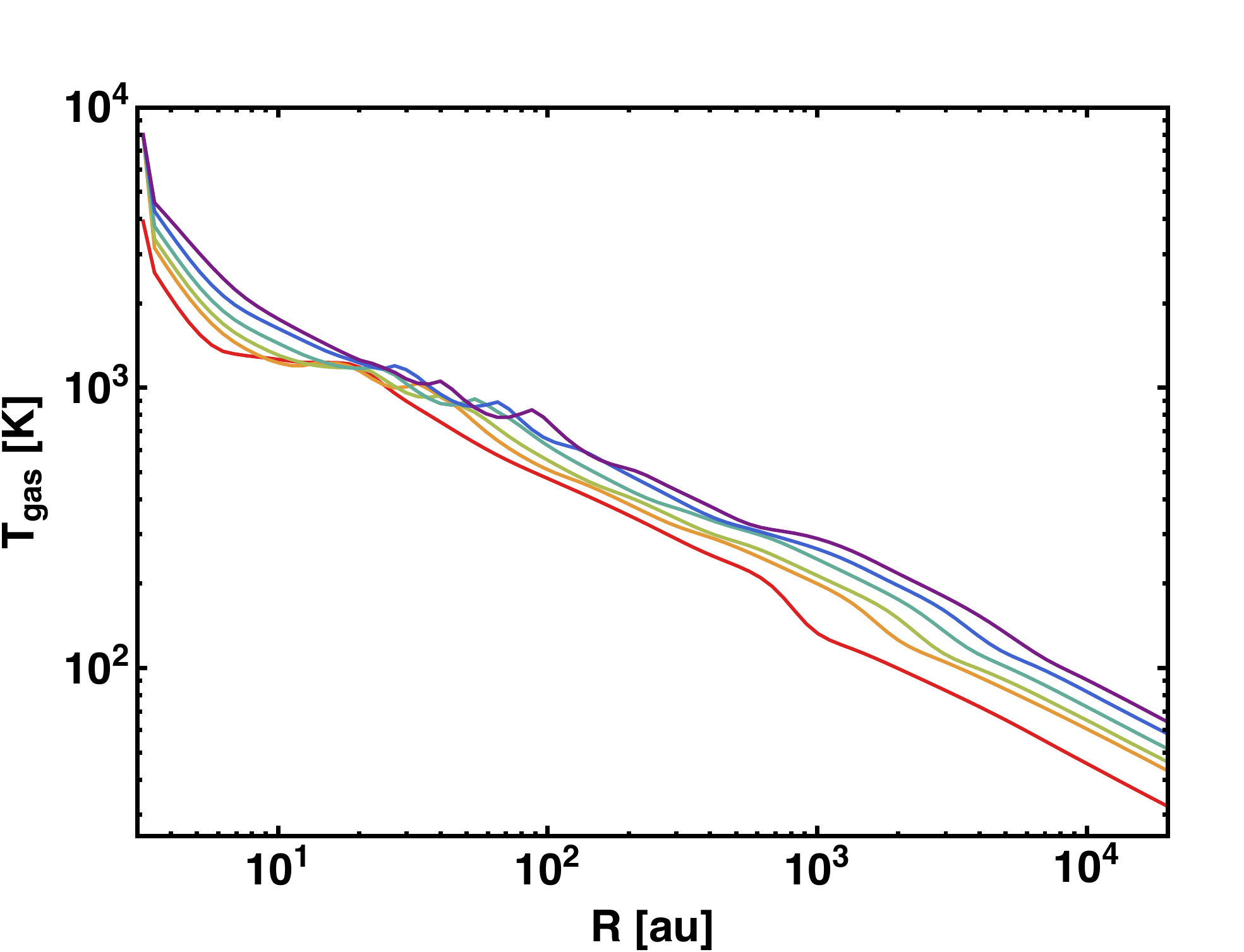}\\
\includegraphics[width=0.49\textwidth]{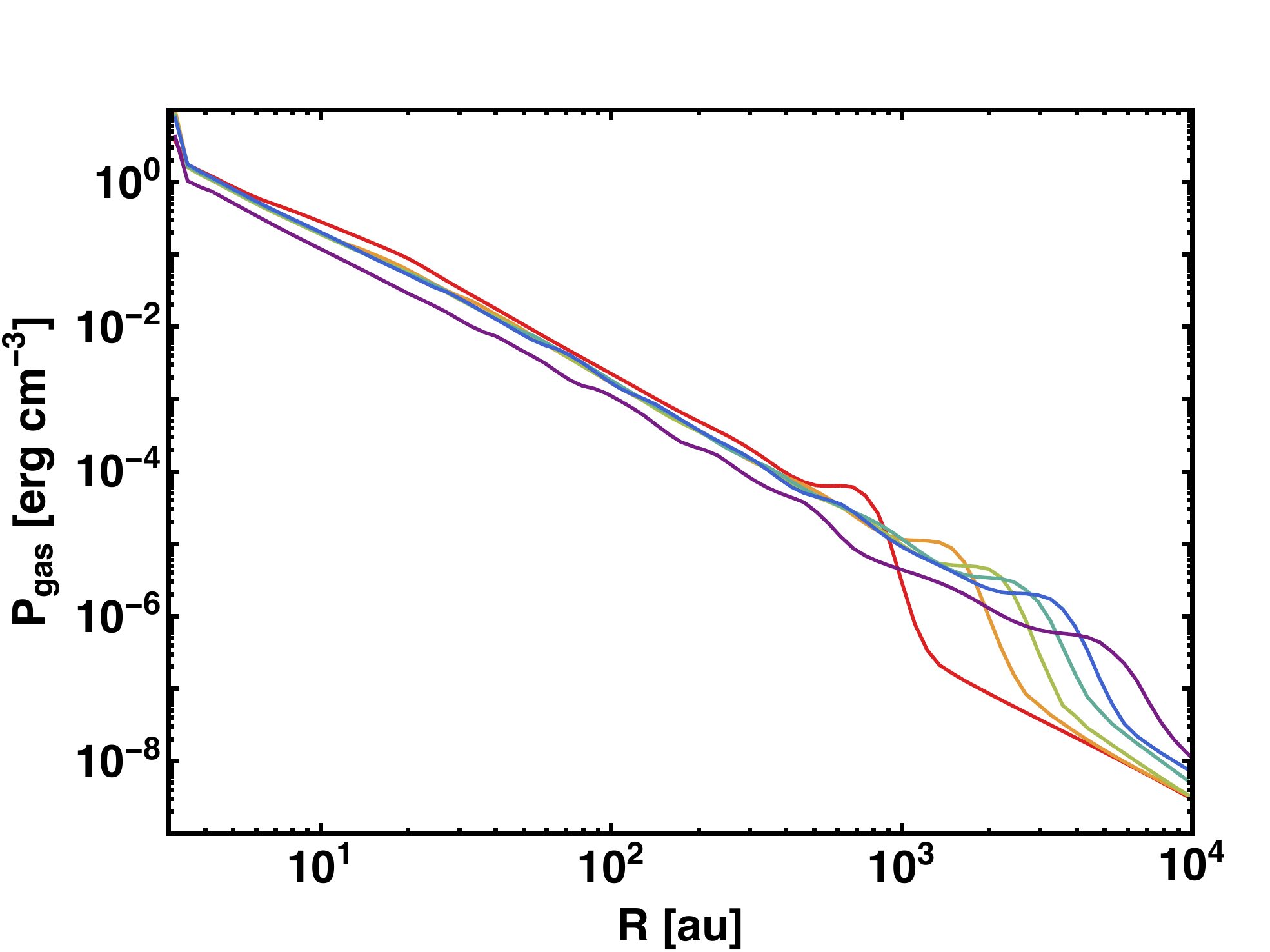}
\includegraphics[width=0.49\textwidth]{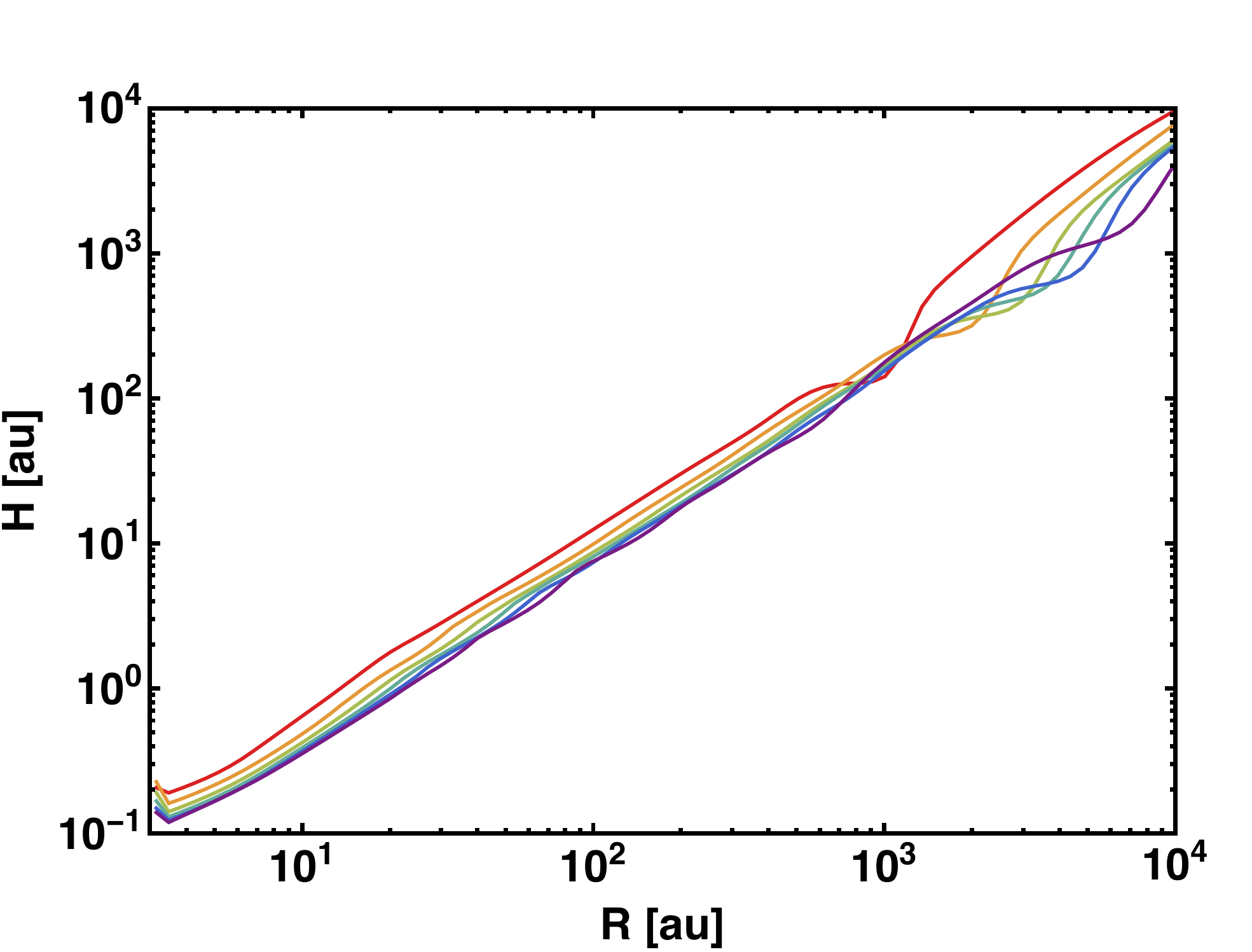}\\
\includegraphics[width=0.49\textwidth]{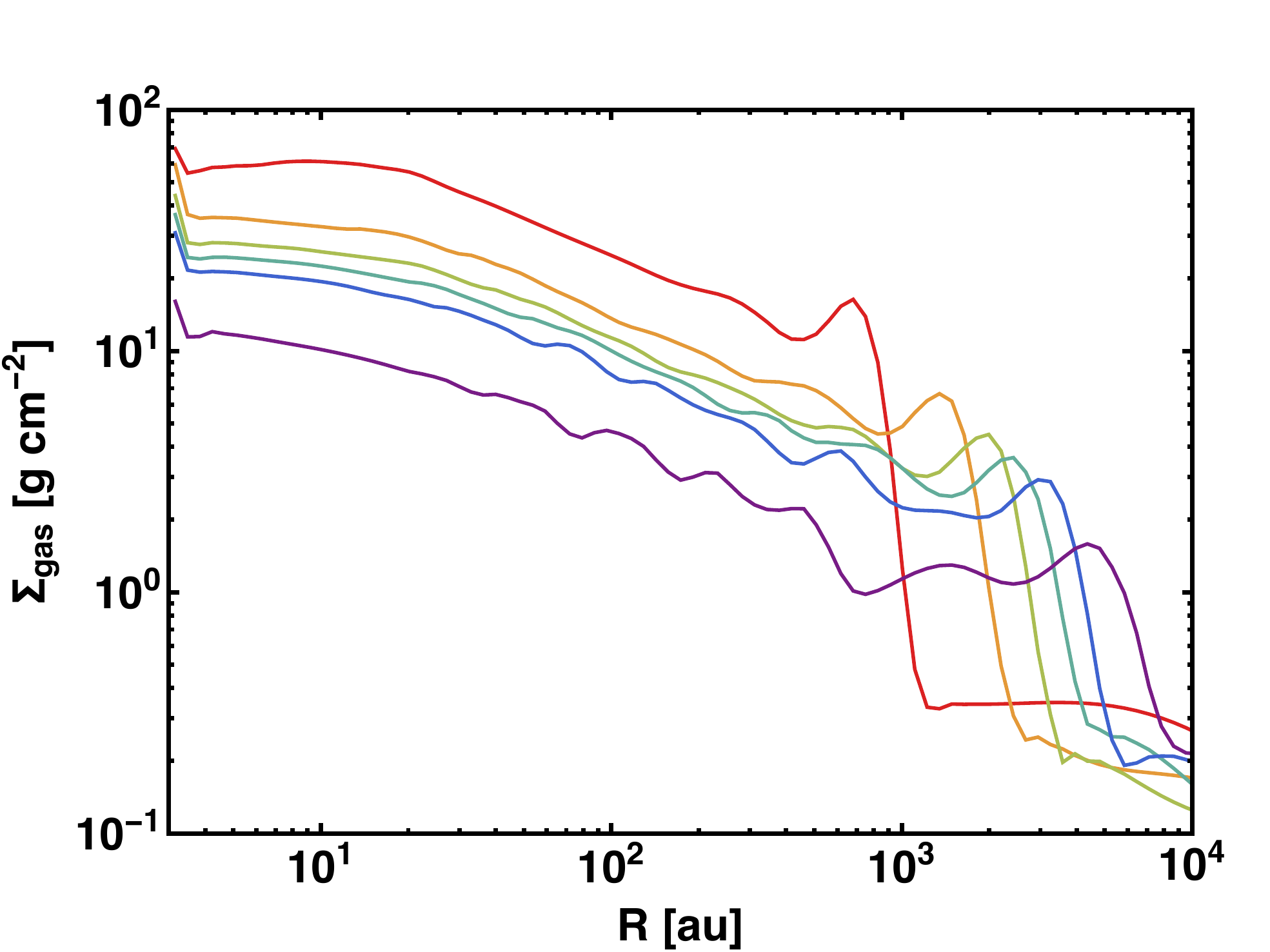}
\includegraphics[width=0.49\textwidth]{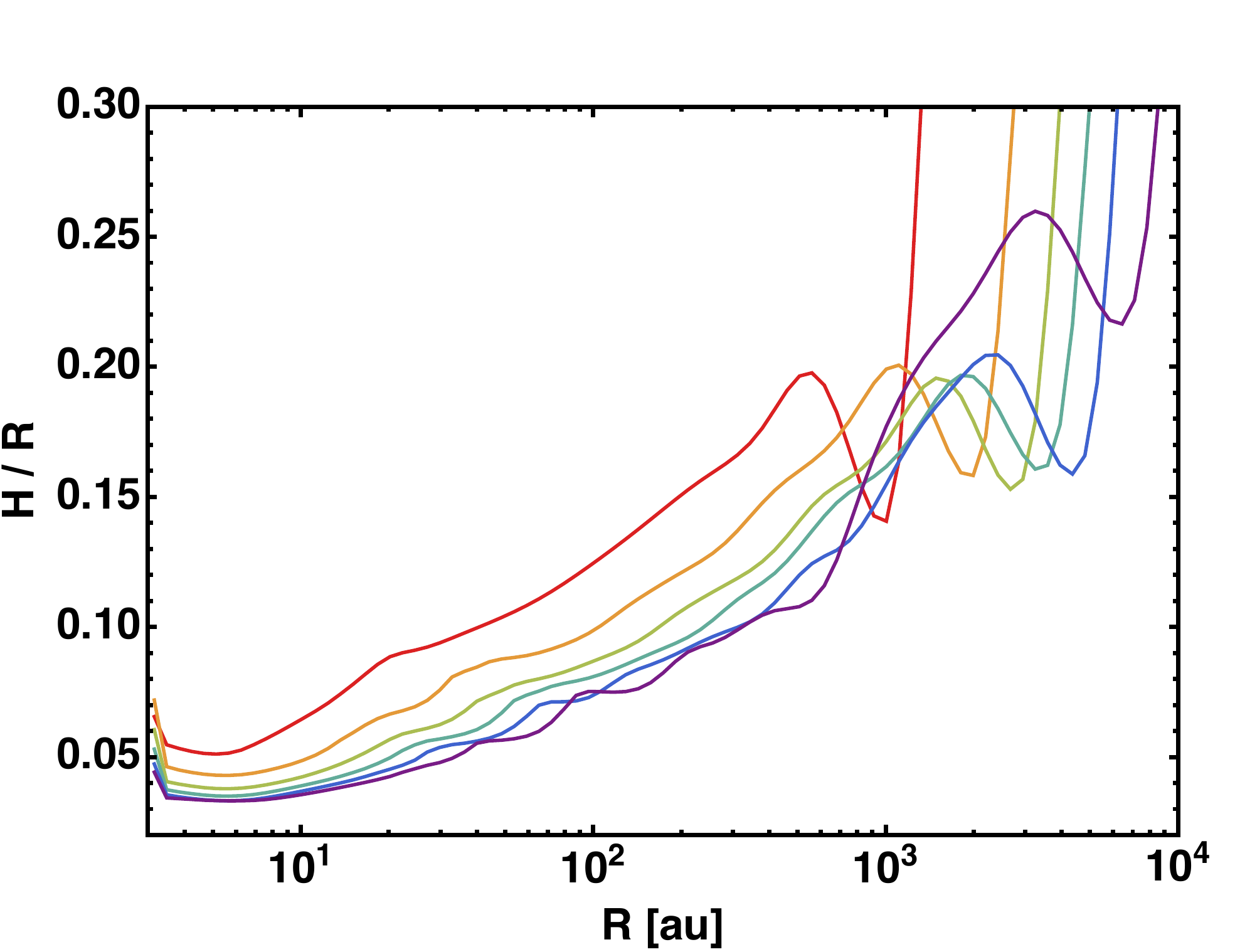}
\caption{
Basic disk diagnostic quantities in the disk's midplane as function of radius for several snapshots in time.
From left to right and top to bottom, the panels show
gas mass density,
gas temperature,
gas pressure,
gas pressure scale height,
gas surface density, and
the disk's aspect ratio.
The six snapshots in time correspond to
$20~\kyr$ (red),
$40~\kyr$ (orange),
$60~\kyr$ (light green),
$80~\kyr$ (dark green),
$100~\kyr$ (blue), and
$120~\kyr$ (purple) of evolution, as indicated in the color bar above the panels.
}
\label{fig:DiskDiagnostics}
\end{figure*}

\begin{figure}[htbp]
\centering
\includegraphics[width=0.49\textwidth]{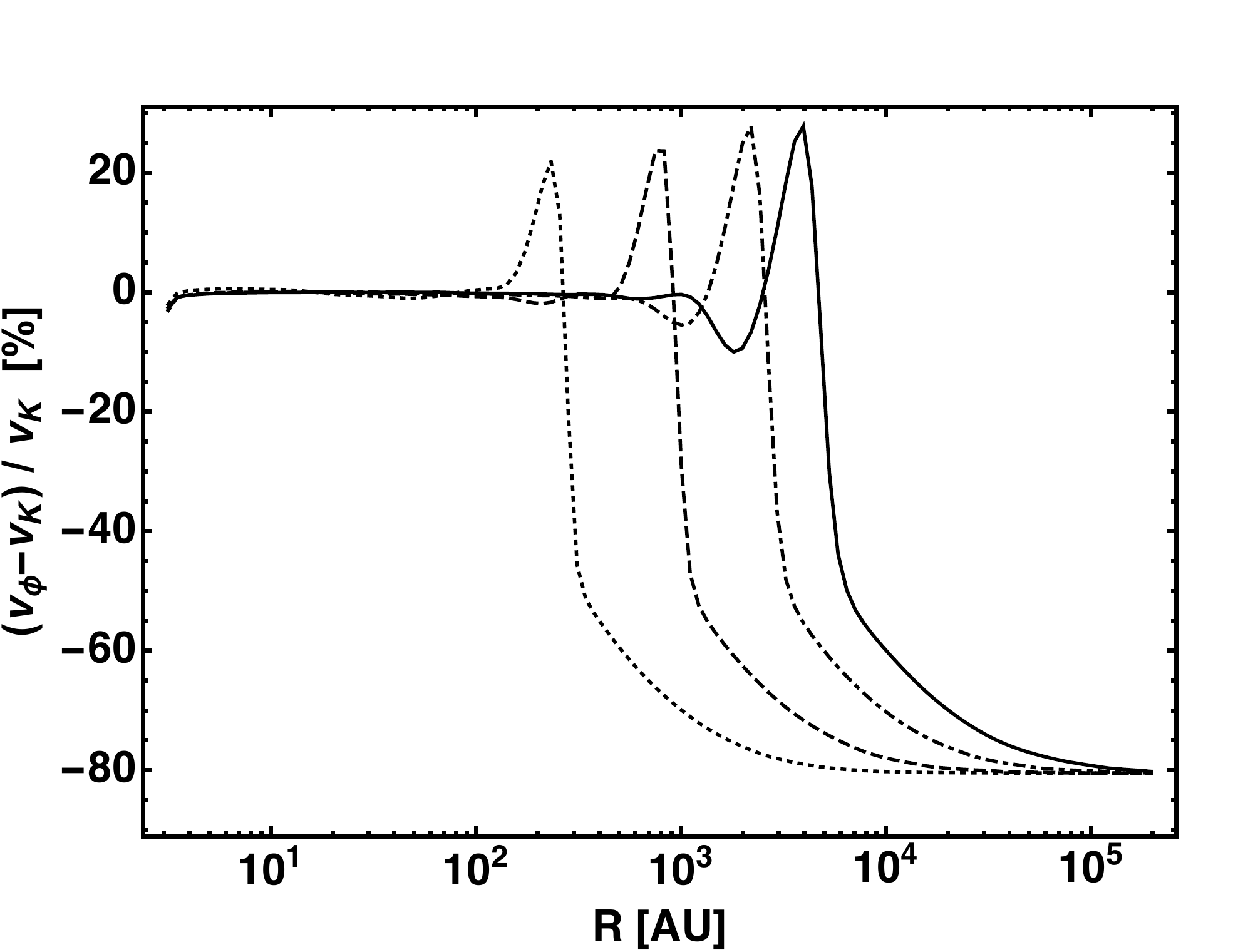}
\includegraphics[width=0.49\textwidth]{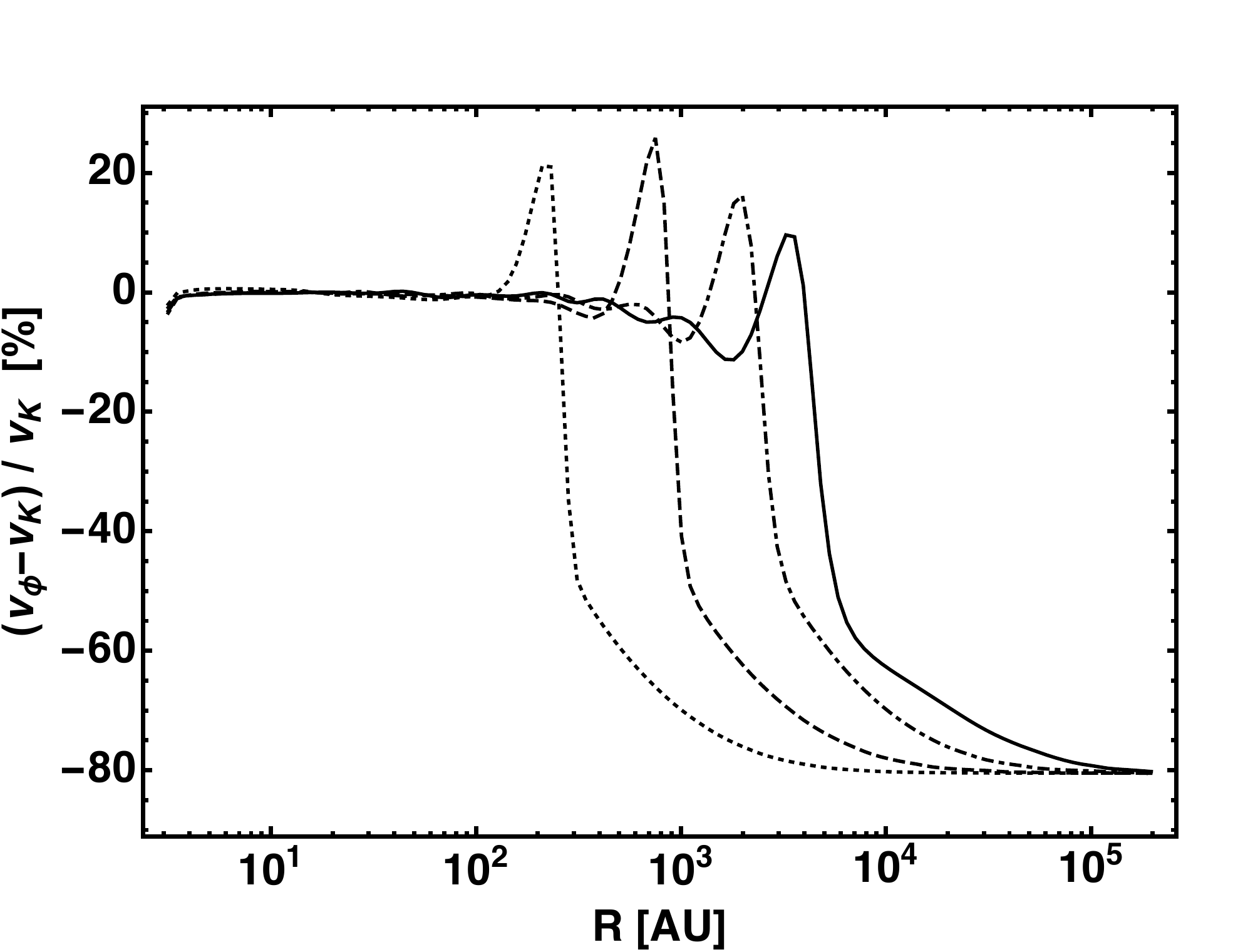}
\caption{
Visualization of disk formation: deviation from gravito-centrifugal equilibrium of the midplane gas as function of radius for four different snapshots in time.
Dotted, dashed, dot-dashed, and solid lines corresponds to $t = 5,15, 40, \mbox{ and } 80~\kyr$ of evolution, respectively.
The bottom panel refers to the simulation including all feedback components, radiation forces are neglected in the simulation data of the upper panel.
Outflows and photoionization is accounted for in both simulations shown, both panels present the data from the large-scale mass reservoir simulations.
}
\label{fig:diskformation}
\end{figure}

\vONE{
In Fig.~\ref{fig:DiskDiagnostics}, we show the basic disk diagnostic quantities, and how they evolve in time.
Most notably, the gas temperature and aspect ratio profiles reveal an inner ionized disk rim, an inner dust evaporated disk region ($10 ... 30~\au$), and a clear flaring of the irradiated circumstellar accretion disk toward larger radii.
}
The circumstellar accretion disk increases in radius with time as long as fresh material from larger and larger scales is infalling.
This inside-out growth of the disk holds as long as the new material from larger scales brings in angular momentum (the centrifugal radius of the initial gas reservoir increases toward larger spatial scales).
At the outer edge of the disk, the incoming gas does not simply arrive at its centrifugal radius in a tangential movement.
Due to the radial momentum from the infall, the approaching material feels the gravity of the star and the disk plus the ram pressure from the gravitational collapse.
As a consequence, higher angular velocity is required to bring the medium onto a stable orbit.
The net effect is that the outer disk edge is characterized by gas in a (gravito+ram pressure)-centrifugal equilibrium, and, hence, super-Keplerian velocities up to 20\% higher than expected for gravito-centrifugal equilibrium arise, as visualized in Fig.~\ref{fig:diskformation}.
This raises the question, is this super-Keplerian motion at the outer disk edge during the envelope-to-disk accretion phase strong enough to be observationally detectable?
For this, one should keep in mind that the gravito-centrifugal equilibrium as computed in Fig.~\ref{fig:diskformation} includes the contribution from the self-gravity of the disk, in other words an accurate determination of the stellar and disk mass is required to check for super-Keplerian speed at the outer disk edge.
Additionally, the forming disk is still embedded in the larger scale rotating and infalling torus, as described in the previous section, which has to be disentangled in for example the position-velocity diagram.
For an existing observational example of super-Keplerian motion at the outer disk edge around a young forming high-mass star, see \citet{Beuther:2008cc}, their Fig.~3 (but here, the gravity of the disk was neglected as well in determining the Keplerian gravito-centrifugal equilibrium velocity).

\begin{figure}[htbp]
\centering
\includegraphics[width=0.49\textwidth]{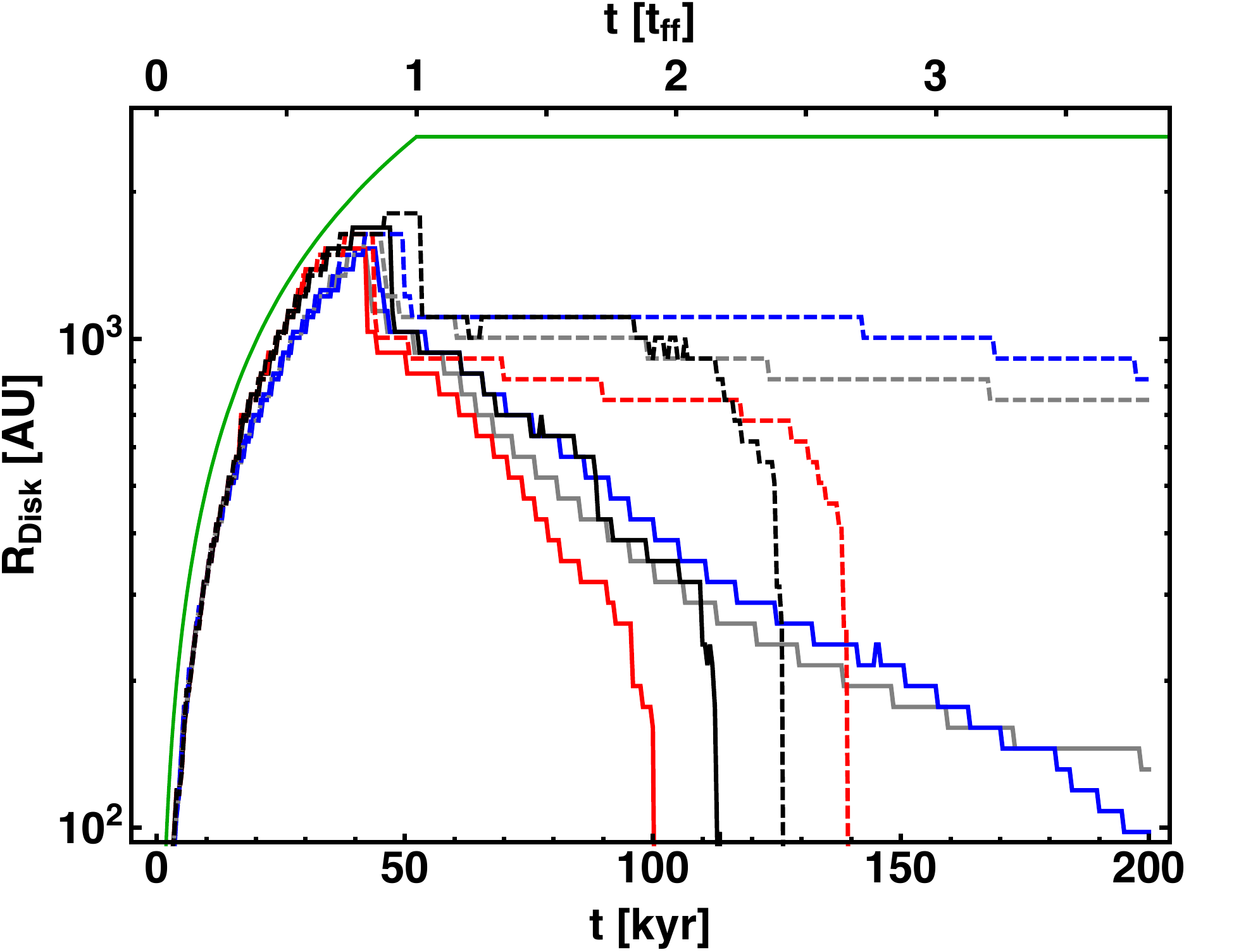}
\caption{
Disk radius as function of time.
Color and line styles are defined in Table~\ref{tab:sims}.
The additional green solid line denotes an analytical estimate of the maximum disk growth.
}
\label{fig:Rdisk_vs_time}
\end{figure}

In Fig.~\ref{fig:Rdisk_vs_time}, we present the evolution of the disk sizes with time.
Here, the disk size $R_\mathrm{disk}$ is determined as the maximum radius in the disk's midplane which is still in a $\pm 15\%$ range around gravito-centrifugal equilibrium.
The innermost $0.1~\pc$ of the computational domains are initially set up identically, hence, the growth phase of the accretion disk remains identical for roughly the free-fall time of this inner $0.1~\pc$ region containing $100~\Msol$, explicitly $t_\mathrm{ff} \approx 52~\kyr$.
After this initial free-fall time, the disk in the limited mass reservoir has reached its maximum radius by definition, because no material from larger scales can bring in higher angular momentum.
As a net effect of the ongoing (although decreasing in time) disk-to-star accretion and no feeding of the disk from large scales, the accretion disk decreases in mass and radius.

\begin{figure}[htbp]
\centering
\includegraphics[width=0.49\textwidth]{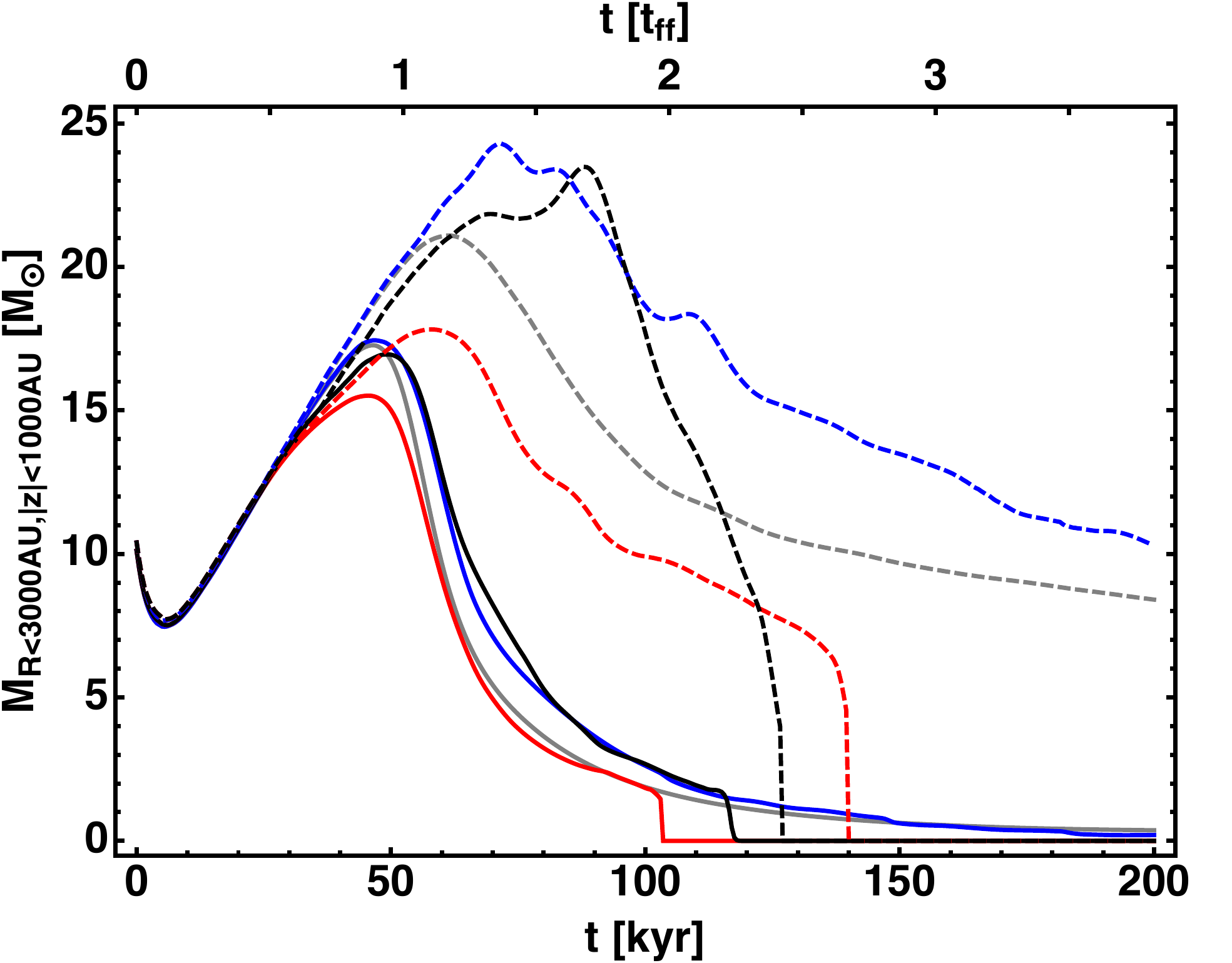}\\
\vspace{2mm}
\includegraphics[width=0.49\textwidth]{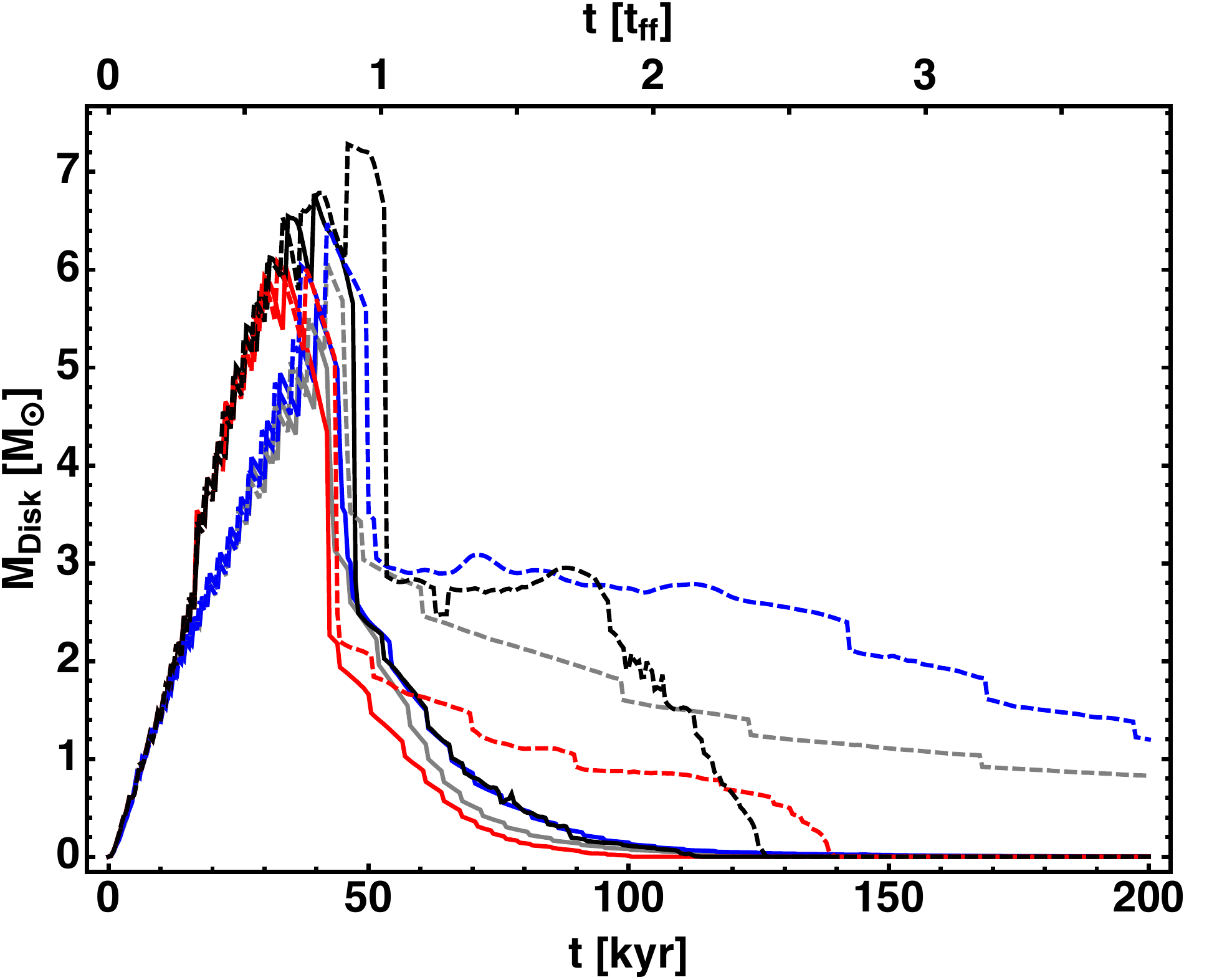}
\caption{
Disk mass (bottom panel) and the mass of the disk's neighboring environment (top panel) as function of time.
Color and line styles are defined in Table~\ref{tab:sims}.
}
\label{fig:Mdisk_vs_time}
\end{figure}

In the large scale, virtually unlimited mass reservoir simulation, the accretion disk would in principle be able to further increase in size, fed by the material from larger scales.
As depicted in Fig.~\ref{fig:Rdisk_vs_time} however, the evolution of the accretion disk at that point in time is already strongly impacted by the feedback from its high-mass host star.
During the course of stellar evolution, the stellar feedback onto its surroundings limits the accretion rate from larger scales onto the disk.
As a consequence, the feeding of the circumstellar disk from larger scales decreases.
The disk mass decreases and the optical depth of the circumstellar disk drops.
In Fig.~\ref{fig:Mdisk_vs_time}, we show the disk mass (bottom panel) as well as the mass of the disk surroundings (top panel) as function of time.
Here, the disk mass is determined as the gas mass in a $\pm 15\%$ range around gravito-centrifugal equilibrium.
The top panel shows the gas mass contained in a cylinder around the central star with a radius of $3000~\au$ and a total height of $2000~\au$, excluding the mass of the star.

\begin{figure}[htbp]
\centering
\includegraphics[width=0.49\textwidth]{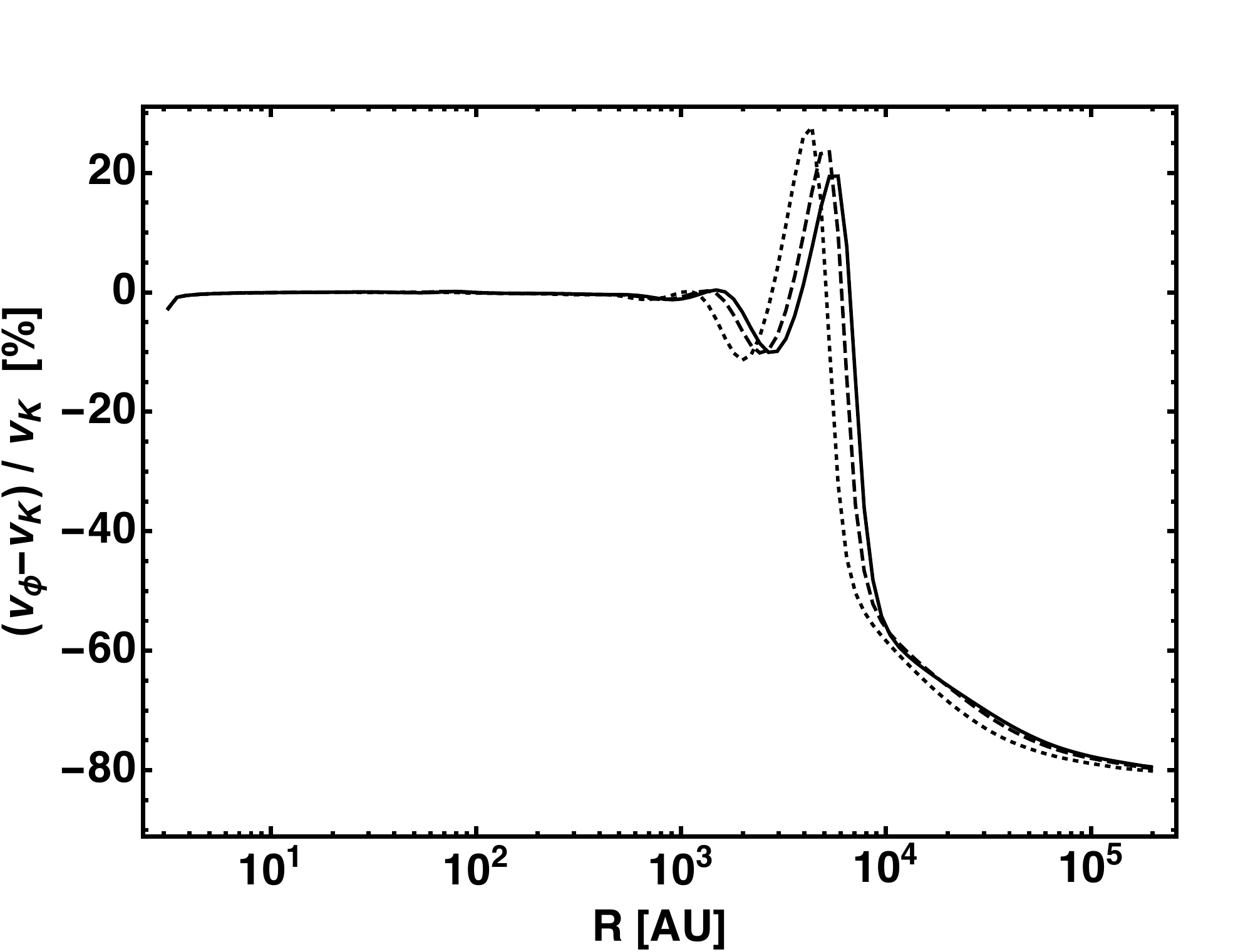}
\includegraphics[width=0.49\textwidth]{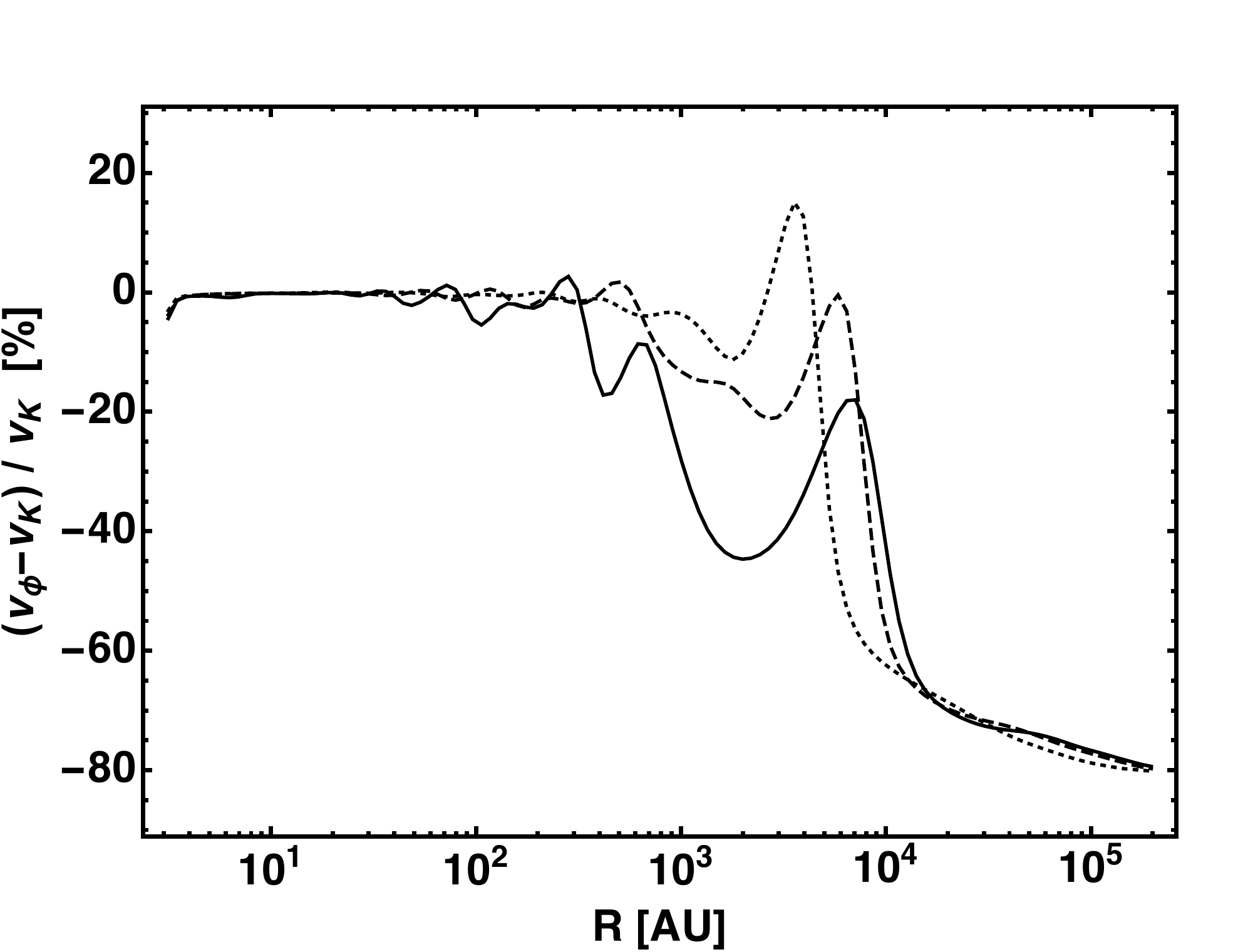}
\caption{
Visualization of disk destruction: same as Fig.~\ref{fig:diskformation}, but for later times of the disk evolution.
Dotted, dashed, and solid lines corresponds to $t = 90, 115, \mbox{ and } 125~\kyr$ of evolution, respectively.
The upper panel corresponds to a simulation without radiative forces, while the lower panel includes them.
Outflows and photoionization is accounted for in both simulations shown, both panels present the data from the large-scale mass reservoir simulations.
}
\label{fig:diskdestruction}
\end{figure}

\begin{figure}[htbp]
\centering
\includegraphics[width=0.49\textwidth]{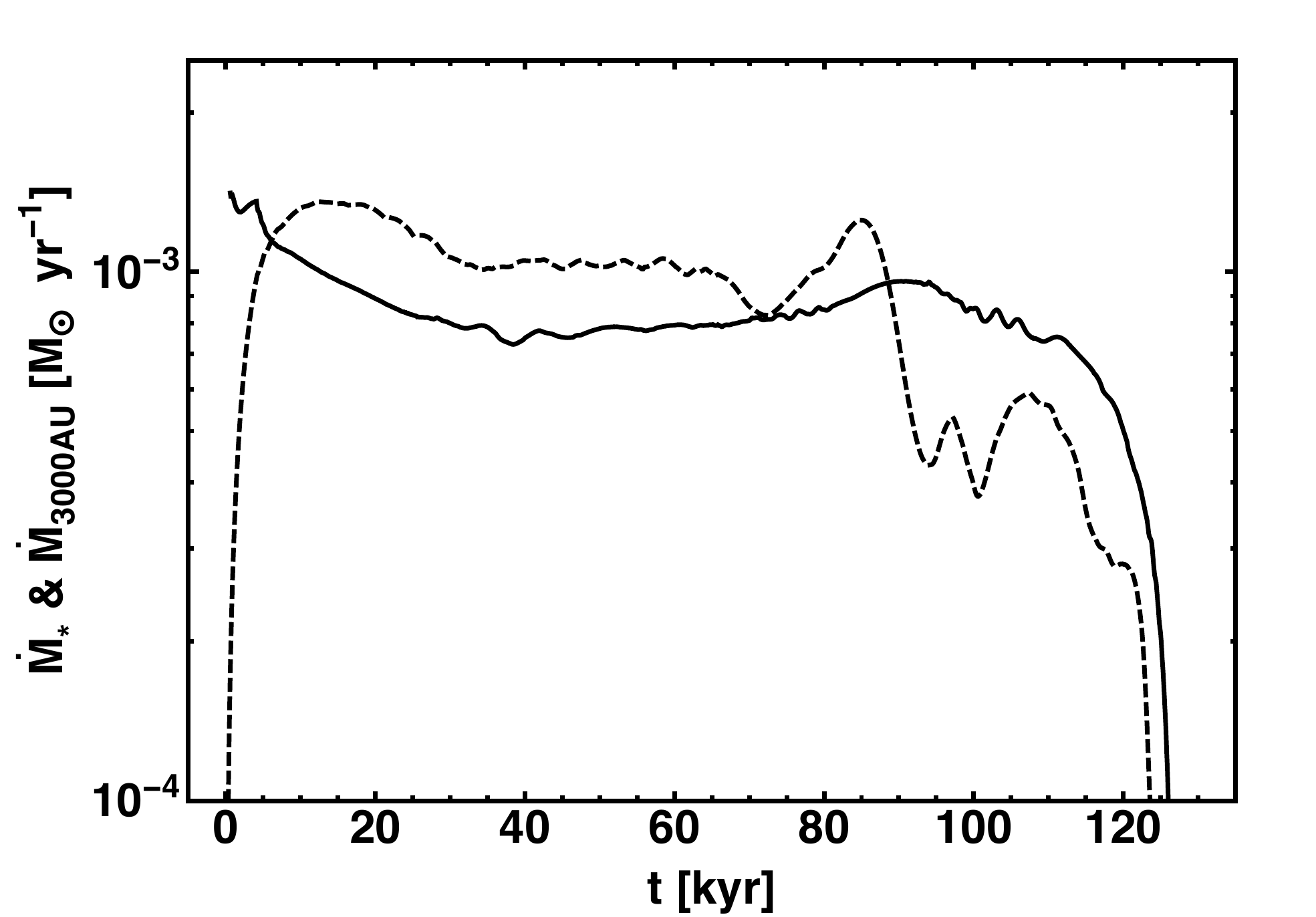}
\caption{
Disk-to-star $\dot{M}_*$ (solid line) and envelope-to-disk $\dot{M}_\mathrm{3000AU}$ (dashed line) accretion rate as function of time.
}
\label{fig:Mdotstardisk_vs_time}
\end{figure}

The high luminosity of the host star and the decrease in disk mass and optical depth leads to a stronger impact of radiative forces, which remove mass at the outer disk edge.
Fig.~\ref{fig:diskdestruction} shows the departure from gravito-centrifugal equilibrium as done in Fig.~\ref{fig:diskformation}, but here for the later times during the disk destruction phase.
The top and bottom panel of Fig.~\ref{fig:diskdestruction} compare directly the effect of radiation forces in the sense that the disk's rotation profile is shown at the same snapshots in time for the simulations with (lower panel) and without (upper panel) radiation forces taken into account.
Even in the simulation without radiation forces (Fig.~\ref{fig:diskdestruction}, top panel), the outflow and photoionization feedback limits the accretion from large scales toward the disk and hence its growth is limited to $\approx 1000~\au$ (compare Fig.~\ref{fig:Rdisk_vs_time}) and a few $\Msol$ (compare Fig.~\ref{fig:Mdisk_vs_time}) here.
Clearly visible in Fig.~\ref{fig:diskdestruction}, bottom panel, the centrifugally supported disk is eventually `destroyed' from the outside inward in the simulation including the effect of radiation forces.
The disk rotation profiles resemble a gravito-(centrifugal+radiation force) equilibrium.
This outside-in fashion of feedback stopping accretion is also true on larger scales, e.g.~when comparing the disk-to-star accretion rate with the envelope-to-disk accretion rate, see Fig.~\ref{fig:Mdotstardisk_vs_time}.
Here, the accretion rates are determined as the mass flux along the negative radial directions at $3~\au$ for the disk-to-star $\dot{M}_*$ and at $3000~\au$ for the envelope-to-disk $\dot{M}_\mathrm{3000AU}$ accretion rate.
Obviously, the stellar accretion ceases after the feedback has beaten down the envelope-to-disk accretion flow.
In simulations without radiation forces, the disk remains and the central host star gains more and more mass (see following section on stellar evolution).
The impact of feedback on the envelope-to-disk accretion is most dominant in the large-scale mass reservoir simulations, while in the small-scale mass reservoir simulation, the final disk and stellar growth is determined by the finite mass reservoir and the outflow feedback \citep[cp.~][]{2016ApJ...832...40K}; 
this can be seen e.g.~in Fig.~\ref{fig:Mdisk_vs_time}: while the simulations with different feedback components only show minor differences for the small-scale mass reservoir (solid lines), the different feedback components lead to distinctly different disk and envelope masses for the large-scale mass reservoir, which allows for sustained accretion from scales $> 0.1~\pc$.

\begin{figure*}[htbp]
\centering
\includegraphics[width=0.33\textwidth]{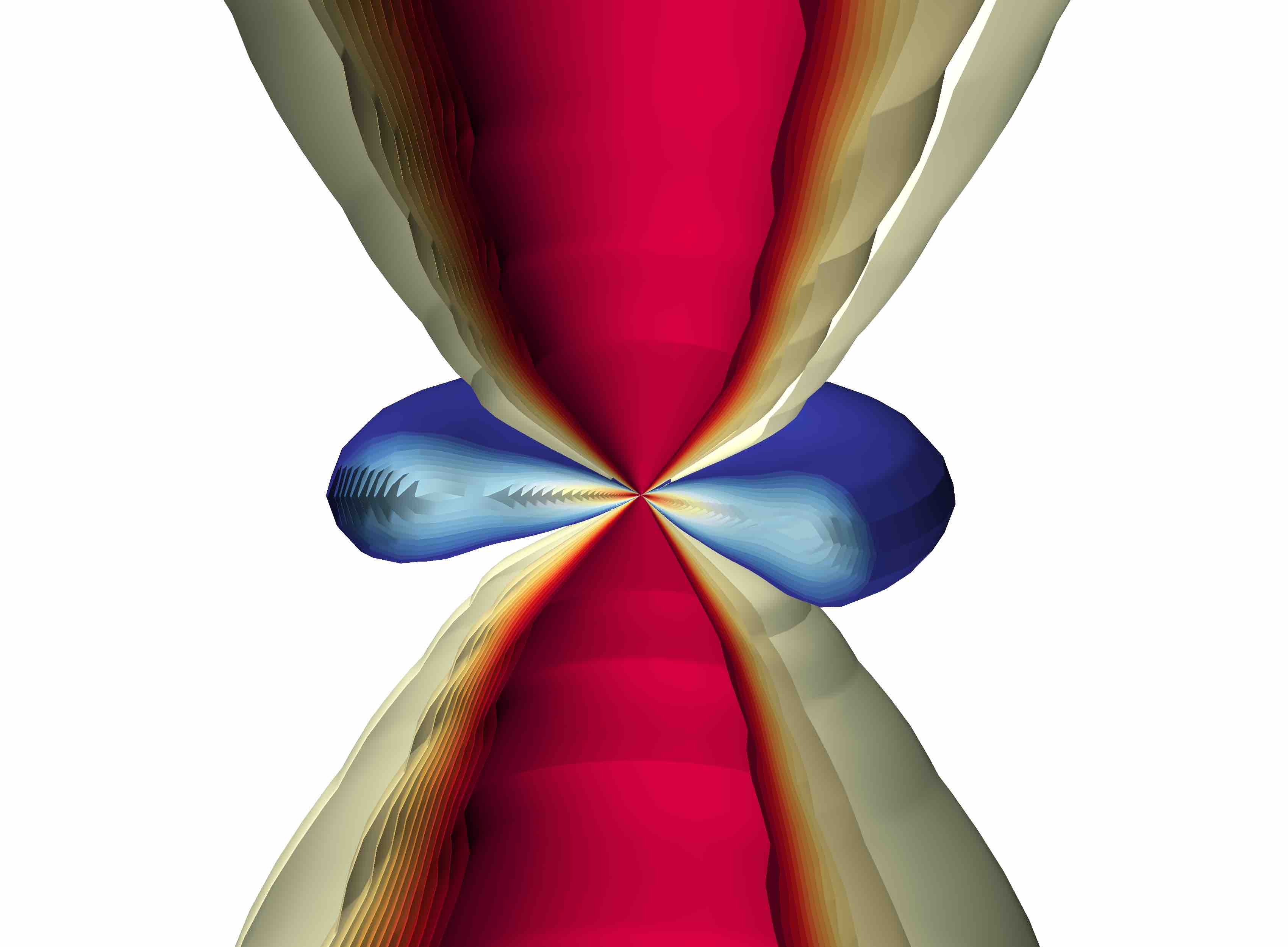}
\includegraphics[width=0.33\textwidth]{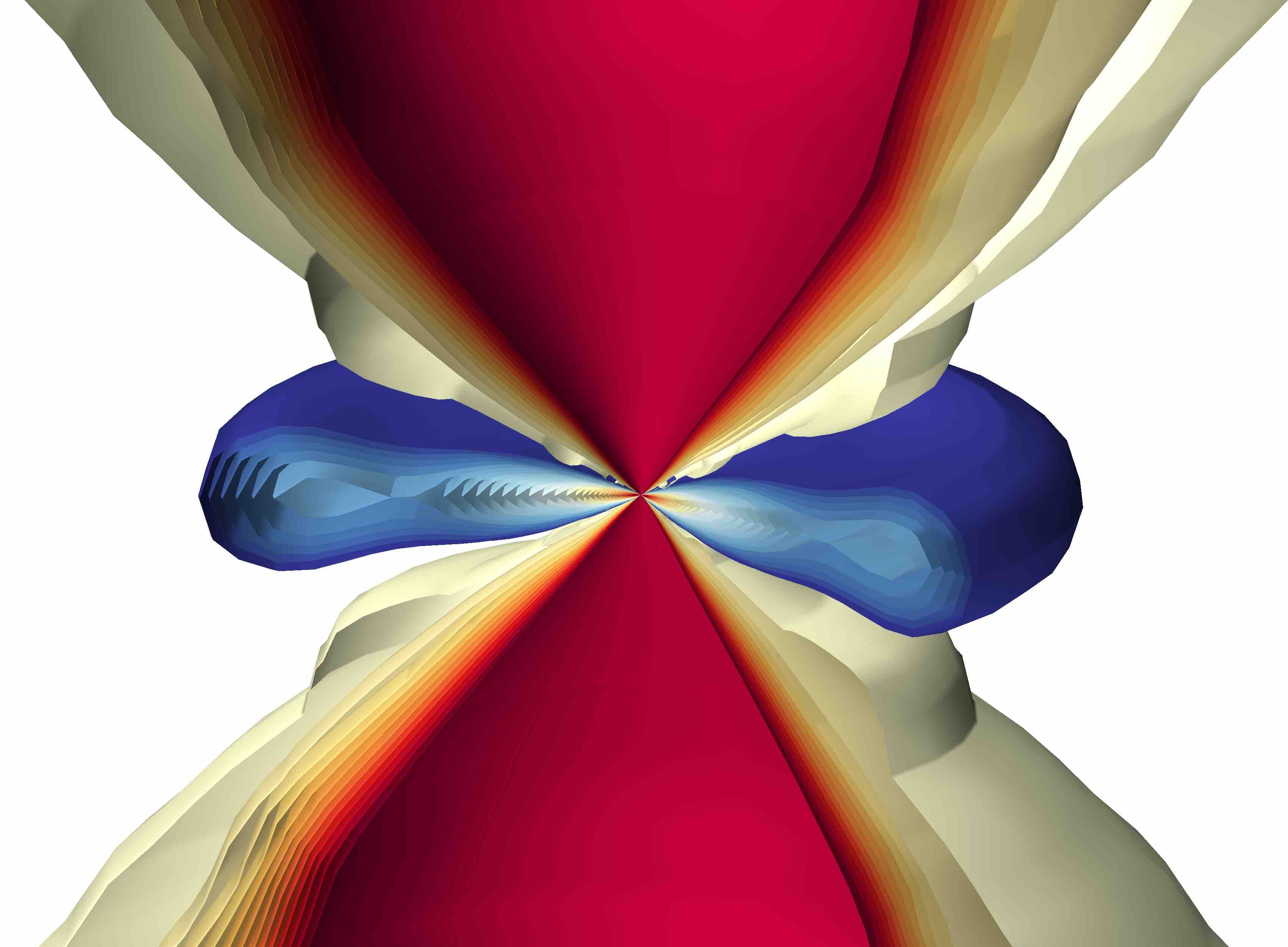}
\includegraphics[width=0.33\textwidth]{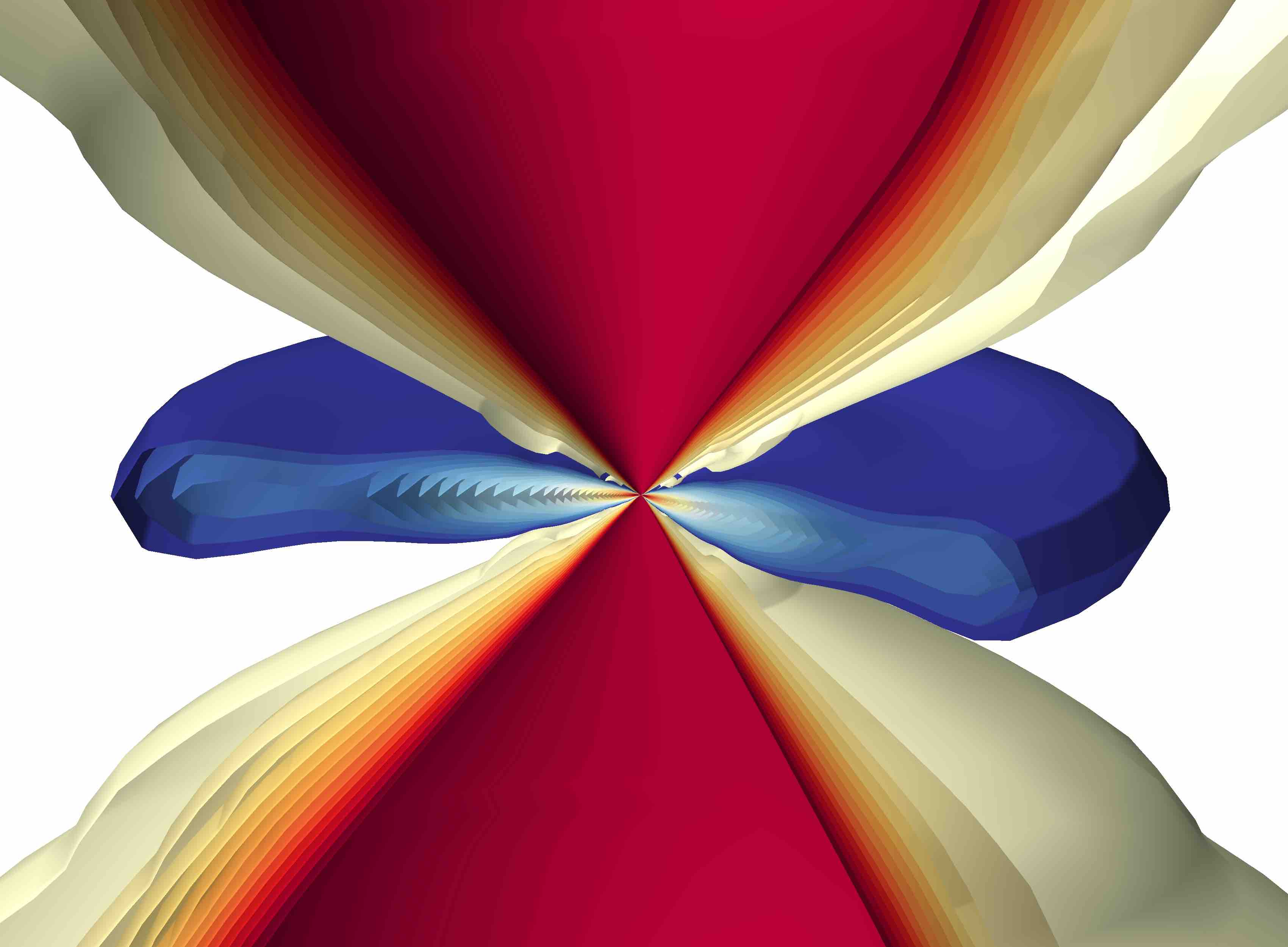}
\caption{
Scissor handle like push effect of the expanding HII region.
The panels show from left to right the accretion disk and protostellar outflow at $t = 20$, $30$, and $40~\kyr$ of evolution.
The disk density is color-coded as in Fig.~\ref{fig:visualization_disk}.
The outflow velocity is color-coded as in Fig.~\ref{fig:visualization_outflow}.
The images cover $\approx 5000~\au$ in width.
The outermost velocity iso-contour at each time resembles the extent of the HII region in the polar direction.
}
\label{fig:PushEffect}
\end{figure*}

\begin{figure*}[htbp]
\centering
\includegraphics[width=0.49\textwidth]{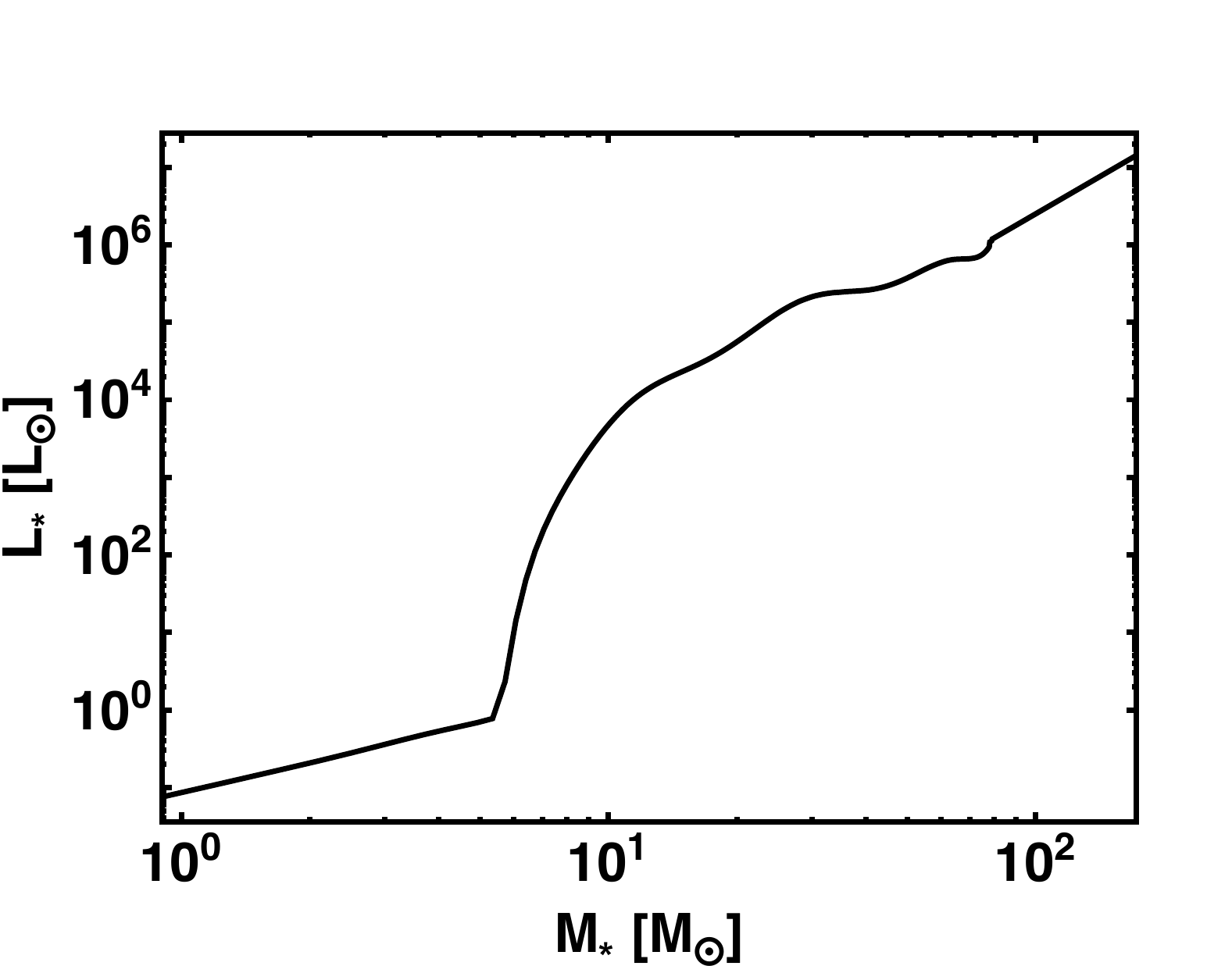}
\includegraphics[width=0.49\textwidth]{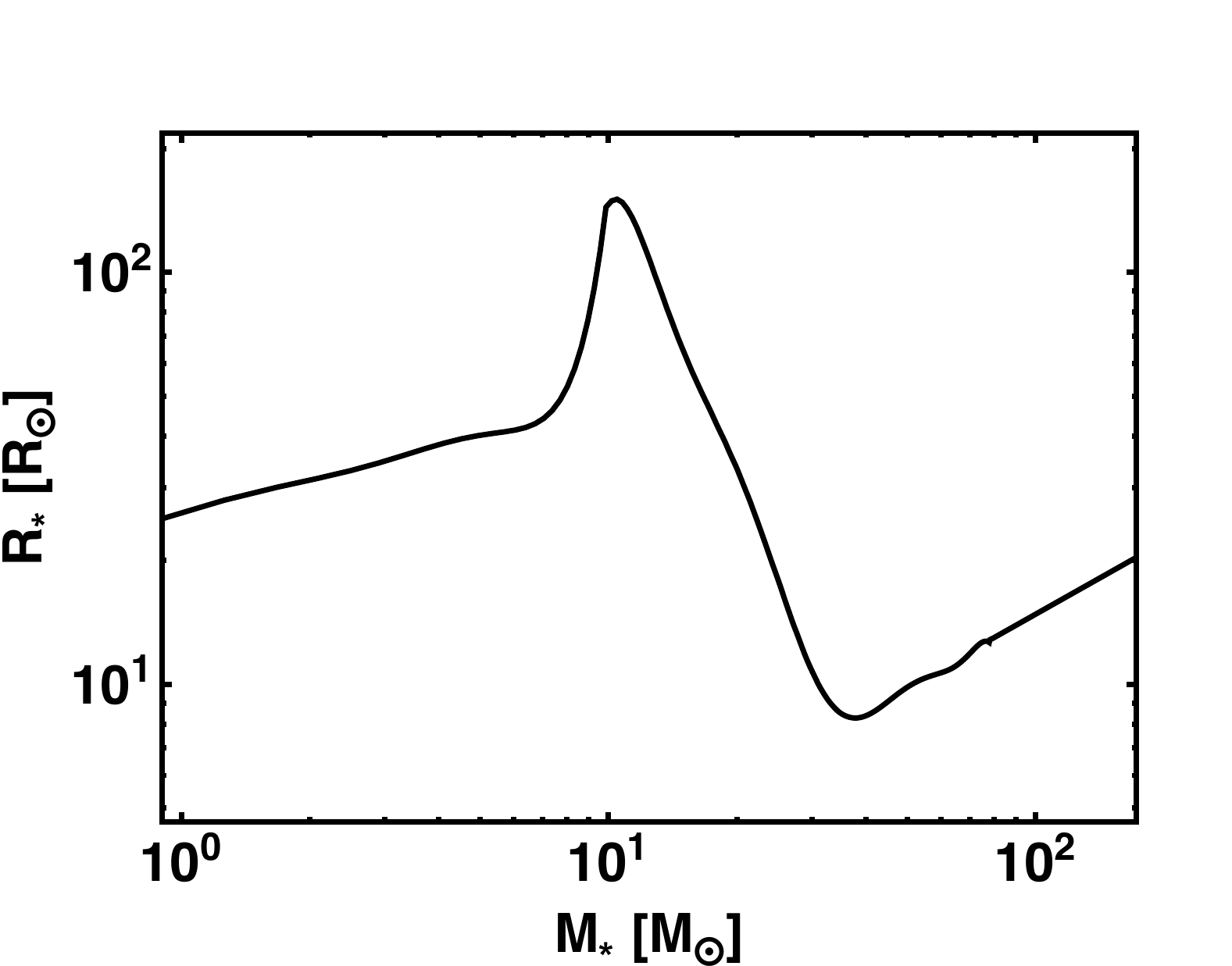}\\
\includegraphics[width=0.49\textwidth]{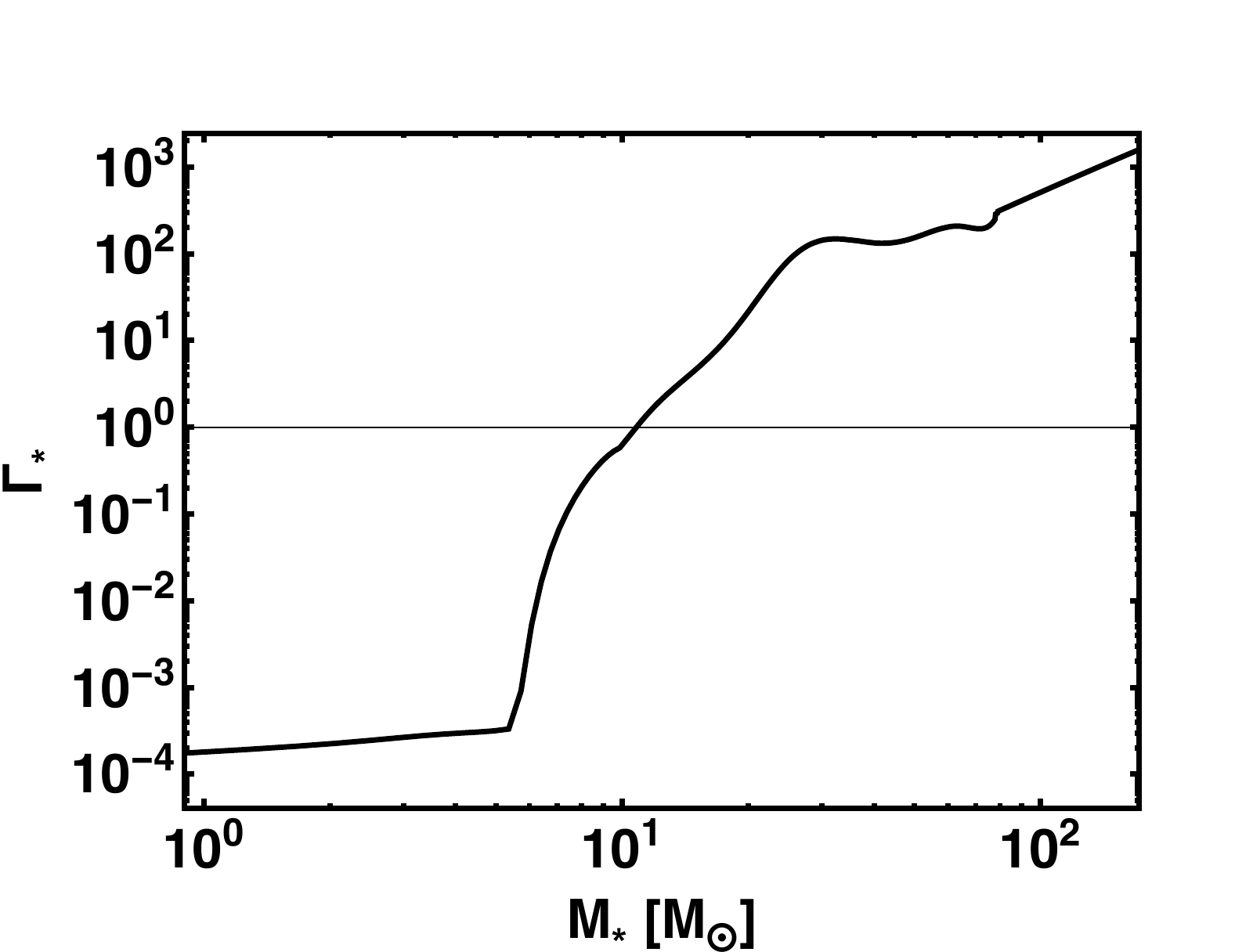}
\includegraphics[width=0.49\textwidth]{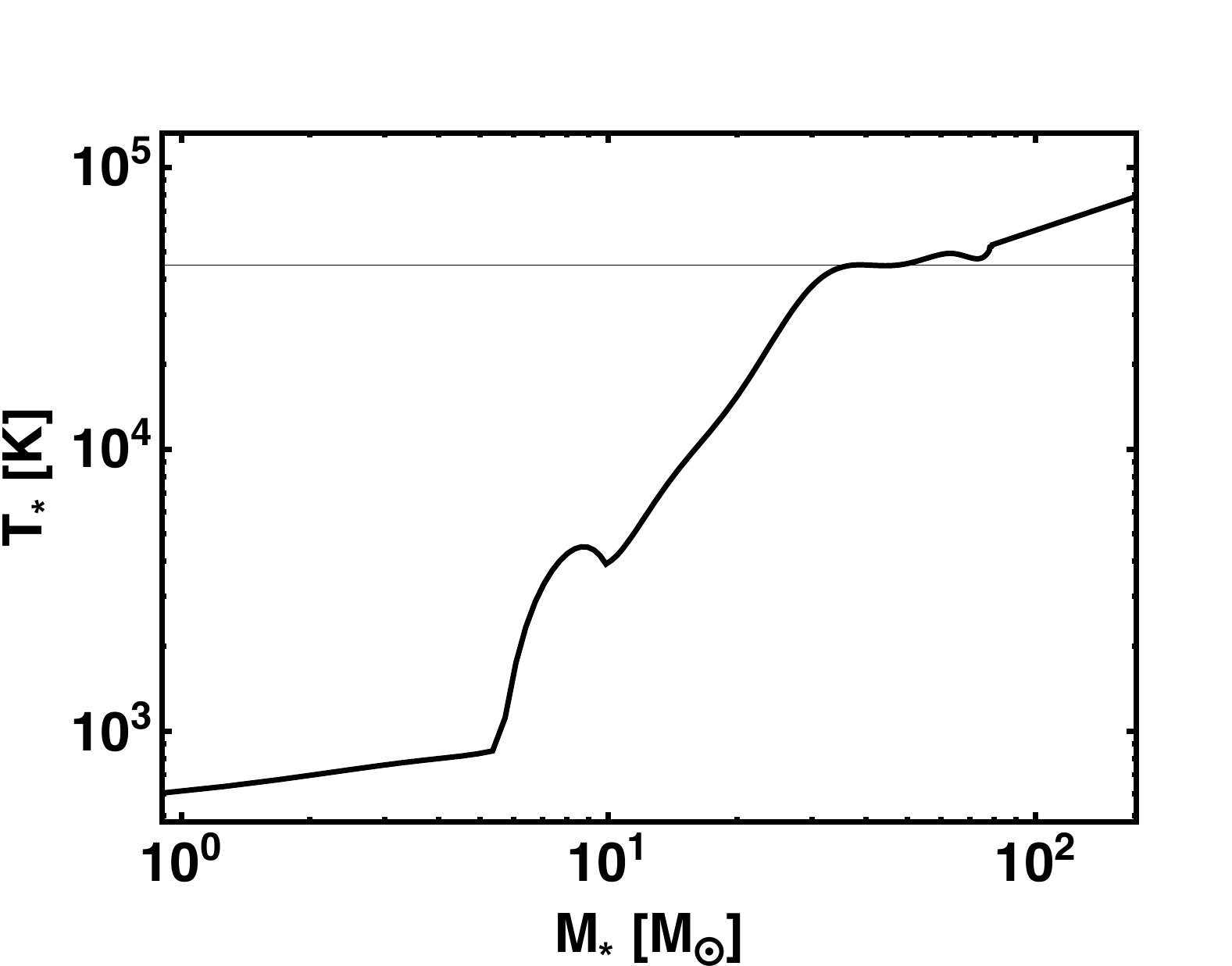}
\caption{
Stellar evolution. 
Luminosity (top left), radius (top right), Eddington factor with respect to dust absorption (bottom left), and photospheric temperature (bottom right) as function of stellar mass.
The visualized luminosity and temperature does not include the contribution from accretion, which varies among the different simulation runs.
}
\label{fig:StellarEvolution}
\end{figure*}

Comparing the evolution of disk mass and the surrounding mass for simulations with different feedback components (Fig.~\ref{fig:Mdisk_vs_time}) reveals another interesting effect: 
After the initial buildup phase of the disk, simulations with photoionization feedback always yield higher mass concentration in the disk and its close surroundings compared to the same configuration without photoionization (black vs.~red and blue vs.~gray lines). 
This behavior is independent of the inclusion of radiation forces (black vs.~red lines with radiation forces and blue vs.~gray lines without radiation forces), and independent of the size of the initial mass reservoir (solid lines for $0.1~\pc$ and dashed lines for $1~\pc$ mass reservoirs).
The reason for this increase in gas mass density is that the HII region generated by photoionization from the disk's host star quickly fills the bi-polar low-density cavity regions (a detailed analysis of the evolution of the HII regions can be found in Sect.~\ref{sect:results_HII}).
As a result, the thermal pressure feedback from the HII region not only leads to a radial expansion (as further discussed in the feedback Sect.~\ref{sect:results_HII}), but also pushes the envelope gas along the polar angle directions from top and bottom toward the disk, like scissor handles.
See Fig.~\ref{fig:PushEffect} for a visualization of the interaction of the HII region and the accretion disk.
This push effect generated by the HII regions has further consequences for the evolution of the disk and the star: The redistribution of gas mass from high to low latitudes implies an increase in the anisotropy of optical depth  in the full solid angle of the forming star.
Hence, the thermal pressure from photoionization leads to additional gas in the shadowed regions of and behind the optically thick disk. 
This mass remains shielded from the direct radiation feedback of the star.
So finally, all simulations with photoionization feedback result in epochs of enhanced disk-to-star accretion rates compared to the corresponding simulations without photoionization (see following section for a more detailed discussion of the stellar evolution and their final masses).

Other than this ``scissor handle'' effect, the impact of photoionization on the highly optically thick massive disks is limited.
Only the disk's inner rim and its atmospheric layers get photoionized.
The disk's midplane, for instance, remains fully unaffected (Fig.~\ref{fig:diskdestruction}, top panel).
The effect of disk mass loss due to photo-evaporation seems to play a minor role for these disks, since, when compared to simulations without photoionization feedback, the scissors handle like push effect described above \vONE{yields a stronger impact than the photo-evaporated mass loss.}

\subsection{Stellar evolution}
\label{sect:results_Star}
As depicted in Fig.~\ref{fig:Mdotstardisk_vs_time}, stellar accretion and therefore stellar evolution is tightly coupled to the evolution of the accretion disk.
Speaking only somewhat hyperbolically, the disk is not evolving around the star, but the star is evolving inside its accretion disk, leading to our choice to discuss the evolution of the accretion disk in the previous section before now turning to the evolution of the central host star.
Effects on larger scale and \vONE{associated morphologies} will be discussed in the following sections.

In contrast to our earlier studies in \citet{2013ApJ...772...61K}, here we do not solve for the interior stellar structure simultaneously to the gravitational collapse.
Instead, we use the stellar evolutionary tracks from \citet{Hosokawa:2009eu} to determine the stellar properties from the current stellar mass and its accretion rate (see \citet{2010ApJ...722.1556K} for technical details).
For completeness, the evolution of the main stellar properties is also visualized here in Fig.~\ref{fig:StellarEvolution}.

We discuss three points in time which denote special milestones in the stellar evolution:
At about $M_* \approx 5~\Msol$, the internal luminosity wave arrives at the stellar surface, and as a consequence, the stellar luminosity rapidly increases.
Earlier times are dominated by accretion luminosity.

After that, the importance of radiation feedback increases as depicted in Fig.~\ref{fig:StellarEvolution}, bottom left panel, in terms of the Eddington factor with respect to dust opacities $\Gamma_* = L_*/M_* \times \kappa_\mathrm{P}(T_*) / 4\pi G c$.
Here, the Eddington factor denotes the ratio of the radiative force exerted on directly irradiated dust (i.e.~for a medium optical depth less than unity) compared to the gravitational attraction of the star.
Due to the fact that both forces scale with the inverse of the radius squared, their ratio $\Gamma_*$ is scale-independent.
The star becomes super-Eddington ($\Gamma_* > 1$), and thus radiative repulsion is able to compensate the stellar gravitational attraction, at about $M_* \approx 10~\Msol$.
From now on, radiation forces denote (by far) the dominant force in directly irradiated regions until at later times and larger scales, thermal pressure from photoionization dominates.

Despite the strong increase in luminosity in between $M_* \approx 5~\Msol$ and $M_* \approx 10~\Msol$, which gives rise to the radiation forces, photoionization is still negligible during this phase of stellar evolution, because the star is still on the track toward its maximum bloating radius of $R_*^\mathrm{max} \approx 160~\Rsol$.
Consequently, the photospheric temperature is far too low for there to be a substantial amount of photoionizing EUV radiation.
We estimate a turnover point as the point in time, where a corresponding black body would emit half of its energy in the EUV regime ($h\nu > 13.6~\eV$), which is about $T_* \approx 45000 \mbox{ K}$.
This critical value is added as a horizontal line to Fig.~\ref{fig:StellarEvolution}, bottom right panel.
This estimate is only used here in the analysis of the stellar evolution to get an idea of the onset of photoionization feedback; the simulation code always allows for the emission of the EUV fraction of the stellar spectrum using a Kurucz atmosphere model to properly determine the stellar spectrum instead of a black body assumption (see \citet{MakemakeSedna} for details).
Examining where the stellar temperature crosses this threshold shows that the onset of strong photoionization feedback can be estimated as $M_* \approx 30~\Msol$, i.e.~directly at the end of Kelvin-Helmholtz contraction toward the zero age main sequence.
After this point photoionization will create an HII region, which expands into the surrounding medium due to its high thermal pressure.
We will further discuss the morphology of these HII regions and their feedback efficiency in Sect.~\ref{sect:results_HII}.

\subsection{Star formation and small-scale feedback}
\label{sect:results_StarFormation}
\begin{figure}[ht!]
\centering
\includegraphics[width=0.49\textwidth]{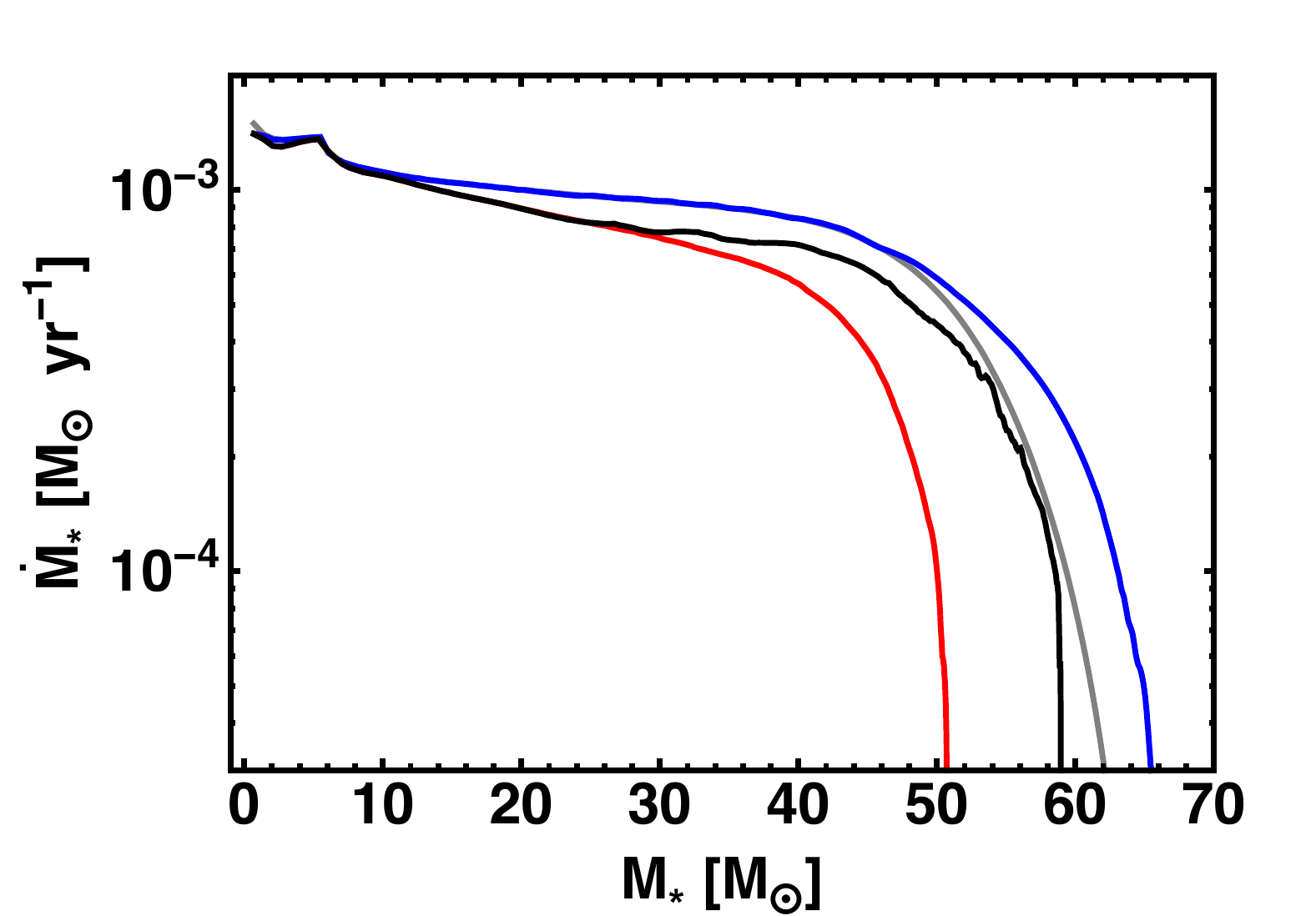}
\includegraphics[width=0.49\textwidth]{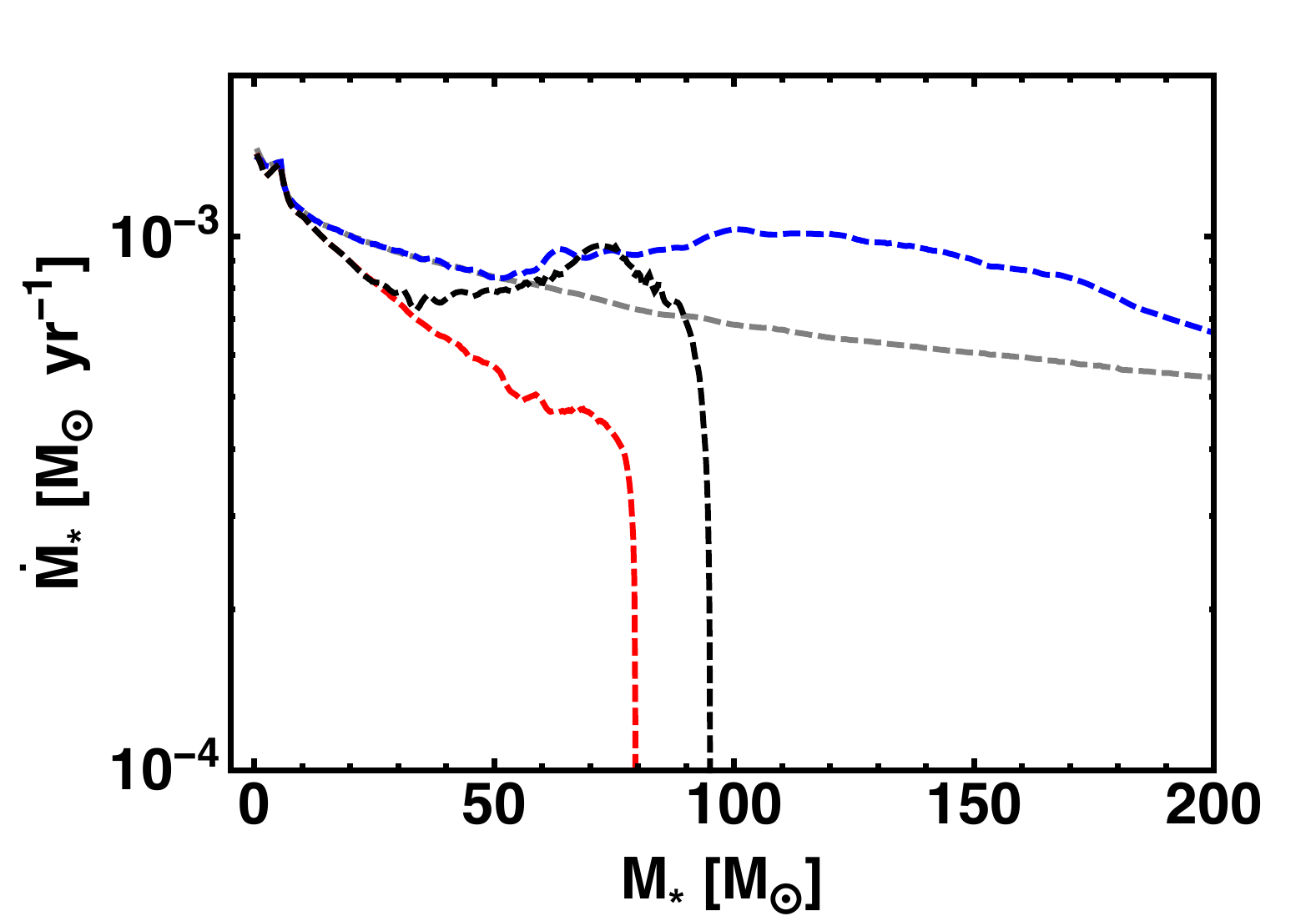}
\caption{
Stellar accretion rate as function of the actual stellar mass for models with a
finite mass reservoir of $100~\Msol$ (upper panel) and
a virtually infinite, extended mass reservoir (lower panel).
Color and line styles are given in Table~\ref{tab:sims} as follows:
Line style refers to the type of initial mass reservoir: 
solid lines for $100~\Msol$ in a $0.1~\pc$ sphere and 
dashed lines for $1000~\Msol$ in a $1~\pc$ sphere.
Colors denote the inclusion of feedback components:
Gray lines for accounting only for protostellar outflows,
blue lines for protostellar outflows plus photoionization.
red lines for protostellar outflows plus radiation forces, and
black lines for protostellar outflows plus radiation forces plus photoionization.
}
\label{fig:Mdot_vs_Mstar}
\end{figure}

These basic stellar evolutionary tracks describe the stellar properties only as functions of the actual stellar mass.
During the course of the simulations, the stellar evolution also depends on the accretion history, which can be severely altered by means of feedback.
The resulting accretion histories are shown for the different mass reservoirs and different feedback components accounted for in the individual simulation runs in Fig.~\ref{fig:Mdot_vs_Mstar}.

In models using a finite mass reservoir (upper panel), representing e.g.~isolated pre-stellar cores of $100~\Msol$, the central star grows up to $M_* \approx 51 - 66~\Msol$, depending on the feedback components taken into account.
As expected from the above estimates of evolutionary milestones, the differences between the tracks are insubstantial in the regime $M_* < 10~\Msol$.
After that, the simulations with radiation forces (red and black lines) yield a lower accretion rate than simulations that neglect these forces (gray and blue lines).
Next, at $M_* \approx 30~\Msol$, the two simulations with radiation forces further split based on the inclusion (black line) or omission (red line) of photoionization feedback.
While it may at first seem counterintuitive, the scissor handle like push effect of the HII region feedback on spatial scales of the accretion disk (described in the previous section on the disk evolution), causes the accretion rate to remain higher in simulations that account for photoionization feedback than in simulations that neglect ionization.
The push effect yields additional mass flow from the envelope toward the disk, the disk mass increases and the disk-to-star accretion rate rises accordingly.
As a remarkable result of this effect, the final stellar mass in the small-scale mass reservoir simulation with all feedback components is $M_*^\mathrm{final} \approx 59~\Msol$ compared to only $M_*^\mathrm{final} \approx 51~\Msol$ in the simulation that neglects photoionization.
The resulting duration of disk-to-star accretion is also extended in the simulation with ionization ($t_\mathrm{acc}^\mathrm{tot} \approx 115~\kyr$) compared to the simulation without ionization ($t_\mathrm{acc}^\mathrm{tot} \approx 100~\kyr$).
This relation changes when taking into account accretion/infall from larger spatial scales as included in the model of a virtually infinite mass reservoir (see discussion below).

In the finite mass reservoir scenario, the central star can by definition not grow beyond $100~\Msol$, and only the feedback effect of protostellar outflows (upper panel, gray solid line) already yields a reduction to $M_*^\mathrm{final} \approx 64~\Msol$ due to redirection of the disk-to-star accretion flow and entrainment of the surrounding gas (see \citet{2015ApJ...800...86K, 2016ApJ...832...40K} for detailed analyses of the outflow feedback efficiency).
Adding radiation forces and photoionization alters the accretion history, and hence fine-tunes the final stellar mass, but the final mass of the star is not controlled by these feedback mechanisms.

This conclusion changes completely, if we take into account the infall or gravitational collapse of the larger-scale environment (bottom panel of Fig.~\ref{fig:Mdot_vs_Mstar}).
Here, the reservoir contains $1000~\Msol$ in a sphere of $1~\pc$ and, hence, the star would be allowed to accrete material up to the observed upper mass limit of stars and beyond, if no feedback effects would be considered.
If we now allow for feedback from protostellar outflows and jets only, but neglect radiation forces and photoionization (gray dashed line), the star still does not stop accretion until we stop the simulation (at $M_* > 300~\Msol$).
The same conclusion holds if protostellar outflows and photoionization feedback are included, but radiation forces are excluded (blue dashed line).
Does this imply that outflows and photoionization do not effect star formation?
No, as we will show later, outflows and photoionization directly or indirectly inject momentum into the surroundings, which alters the large-scale collapse.
And despite the fact that they do not shut off accretion onto the source star, outflows and photoionization will certainly act as feedback mechanisms on secondary stars and the cluster environment.
See, e.g., \citet{Peters:2014bm} and \citet{2016JPhCS.719a2002F} for recent studies on outflows and \citet{2017MNRAS.467.1067D} and  \citet{2017MNRAS.466.3293P} as well as \citet{2016MNRAS.461.2953H, 2017MNRAS.470.3346H} for recent studies on photoionization feedback on stellar cluster scales.
For recent reviews, please see \citet{Frank:2014kz} and \citet{2015NewAR..68....1D} for outflows and photoionization feedback, respectively.

In contrast, if radiation forces are added to the simulations, the shut-off of stellar accretion is an intrinsic part of the system's evolution (red and black dashed lines for simulations without and with additional photoionization feedback, respectively).
The resulting final mass of the star is $M_*^\mathrm{final} \approx 79~\Msol$ without photoionization and $M_*^\mathrm{final} \approx 95~\Msol$ with photoionization included.
As in the finite mass reservoir runs, via the push effect photoionization yields an increase in disk mass and stellar accretion.
But here in the infinite mass reservoir scenario, the photoionization feedback on larger scales limits the infall toward small scales, and the resulting duration of stellar accretion is $t_\mathrm{acc}^\mathrm{tot} \approx 140~\kyr$ without taking photoionization feedback into account and $t_\mathrm{acc}^\mathrm{tot} \approx 125~\kyr$, if photoionization is considered.
We discuss the large-scale feedback effect of photoionization and the expansion of HII regions in more detail in Sect.~\ref{sect:results_HII}.
In summary, photoionization and HII region expansion yields a shorter stellar disk accretion phase, but at a higher mass accretion rate.
See Table~\ref{tab:sims} for an overview of $M_*^\mathrm{final}$ and $t_\mathrm{acc}^\mathrm{tot}$ for all simulation results.

In a high-mass star formation scenario in which the large scale cloud or filament fragmentation leads to isolated moderately massive (e.g.~$\le 100~\Msol$) pre-stellar cores, 
%\citep[such as the turbulent core model, ][]{2003ApJ...585..850M}, 
the final mass of the forming high-mass star is set, first, by the large-scale fragmentation (determining the mass of the finite reservoir) and, secondly, by the feedback effects, which further reduce the available mass within the finite reservoir.
On the other hand, in a high-mass star formation scenario in which high-mass stars form from an open mass reservoir through significant accretion from larger spatial scales,
%\citep[such as competitive accretion,][]{Bonnell:2001fk}, 
the final mass of a forming super-Eddington star is intrinsically controlled by its own radiation force.
Taken together, radiation force feedback naturally explains the rareness of massive stars with masses of the order of $\approx 100~\Msol$;
even in the virtually infinite mass reservoir simulation, no higher mass star forms.
The formation of even higher mass stars up to the observed upper mass limit either requires environments of initially higher density, yielding higher accretion rates, or star formation regions with lower metallicity, reducing the radiation feedback, which could be important to understand the high-mass cluster R136 in the Tarantula Nebula aka 30 Doradus or NGC 2070.
See also \citet{2014ApJ...781...60H, 2015MNRAS.448..568H}, \citet{2016ApJ...824..119H}, and \citet{Hirano:2017hq} for similar simulations as ours for the case of primordial star formation.
Higher density or higher mass reservoirs as considered herein are certainly rare (maybe as rare as those high-mass stars).
See e.g.~\citet{2012A&A...547A..49R}, their Fig.~9. bottom panel for a core mass distribution of the EPoS sample.

\begin{figure*}[htbp]
\centering
\includegraphics[width=0.24\textwidth]{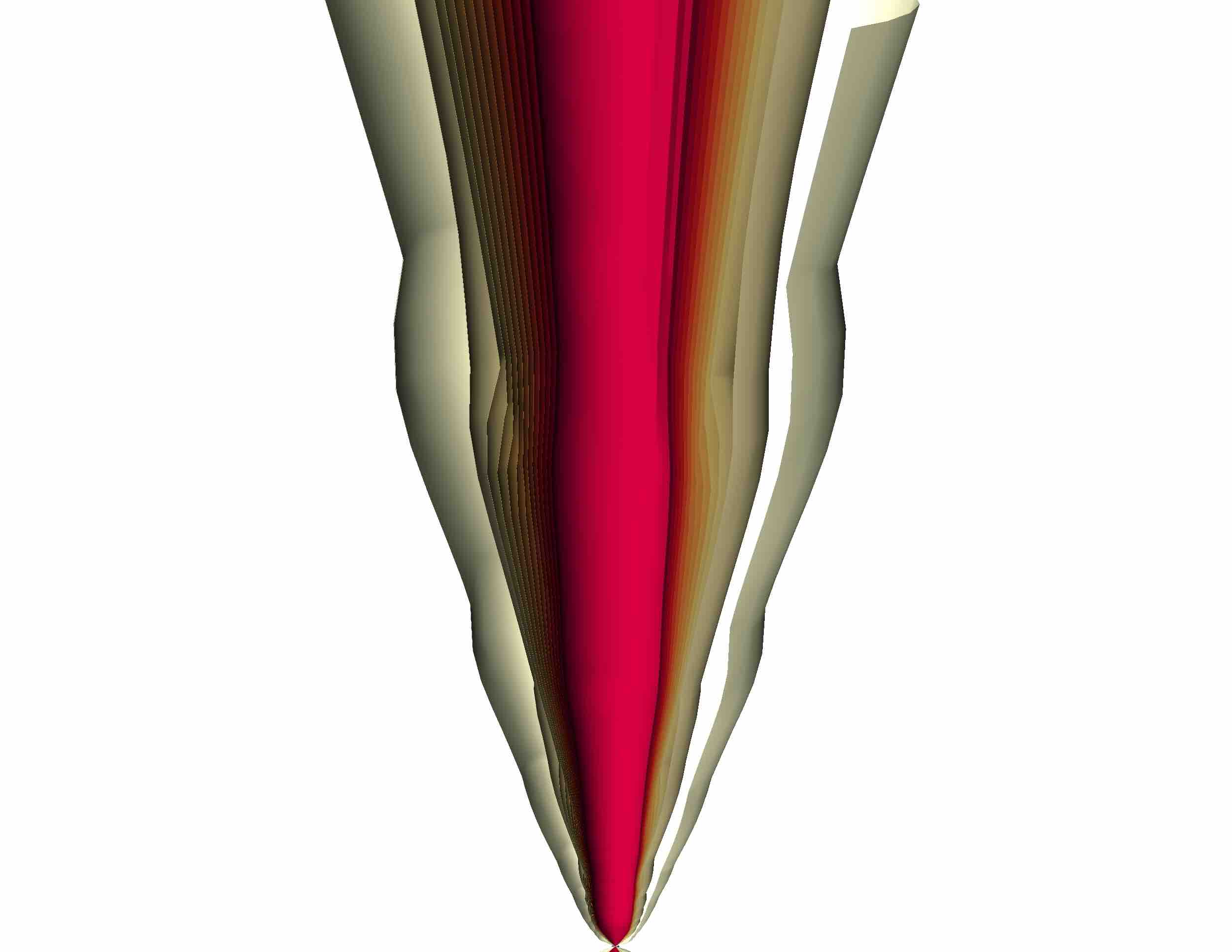}
\includegraphics[width=0.24\textwidth]{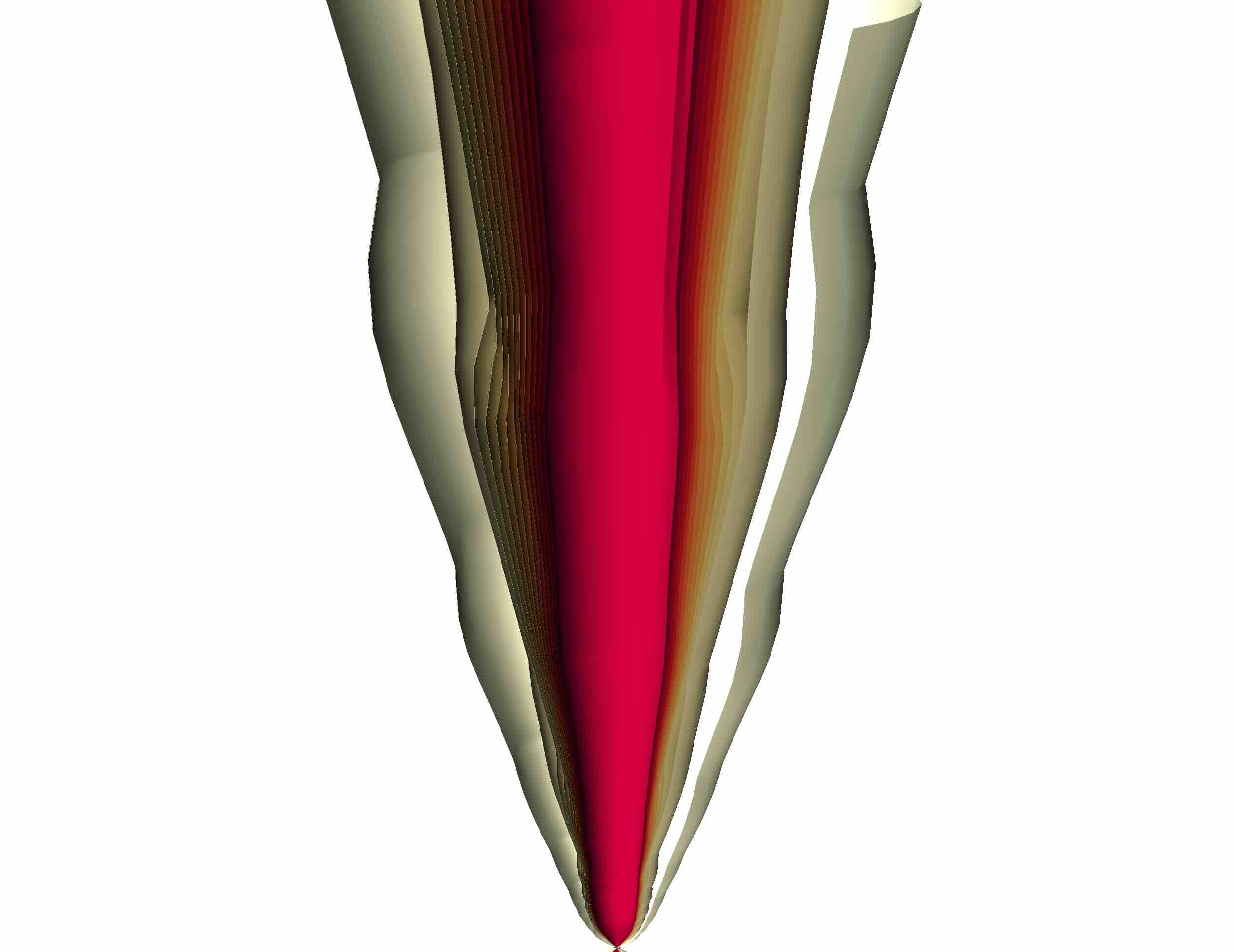}
\includegraphics[width=0.24\textwidth]{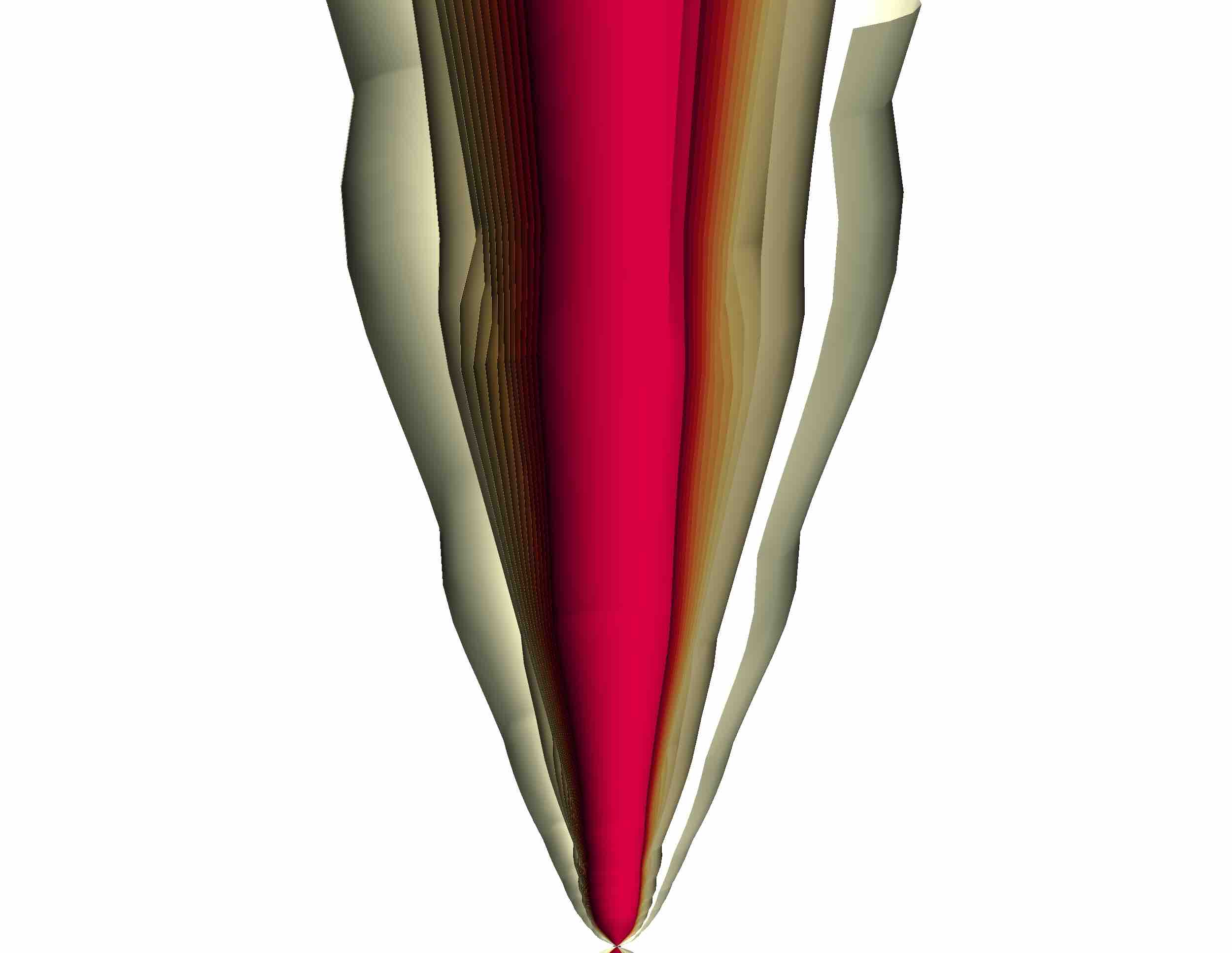}
\includegraphics[width=0.24\textwidth]{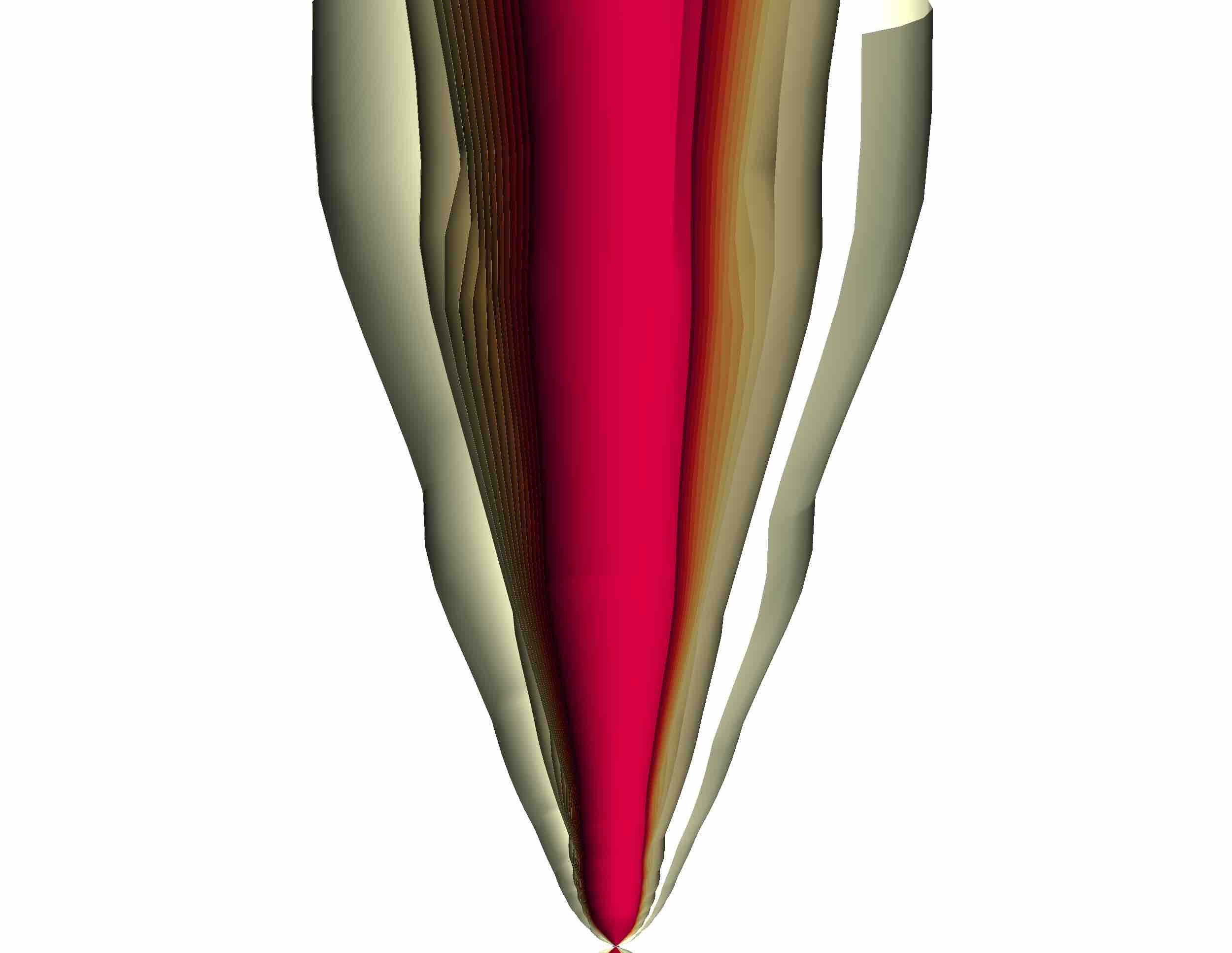}
\includegraphics[width=0.24\textwidth]{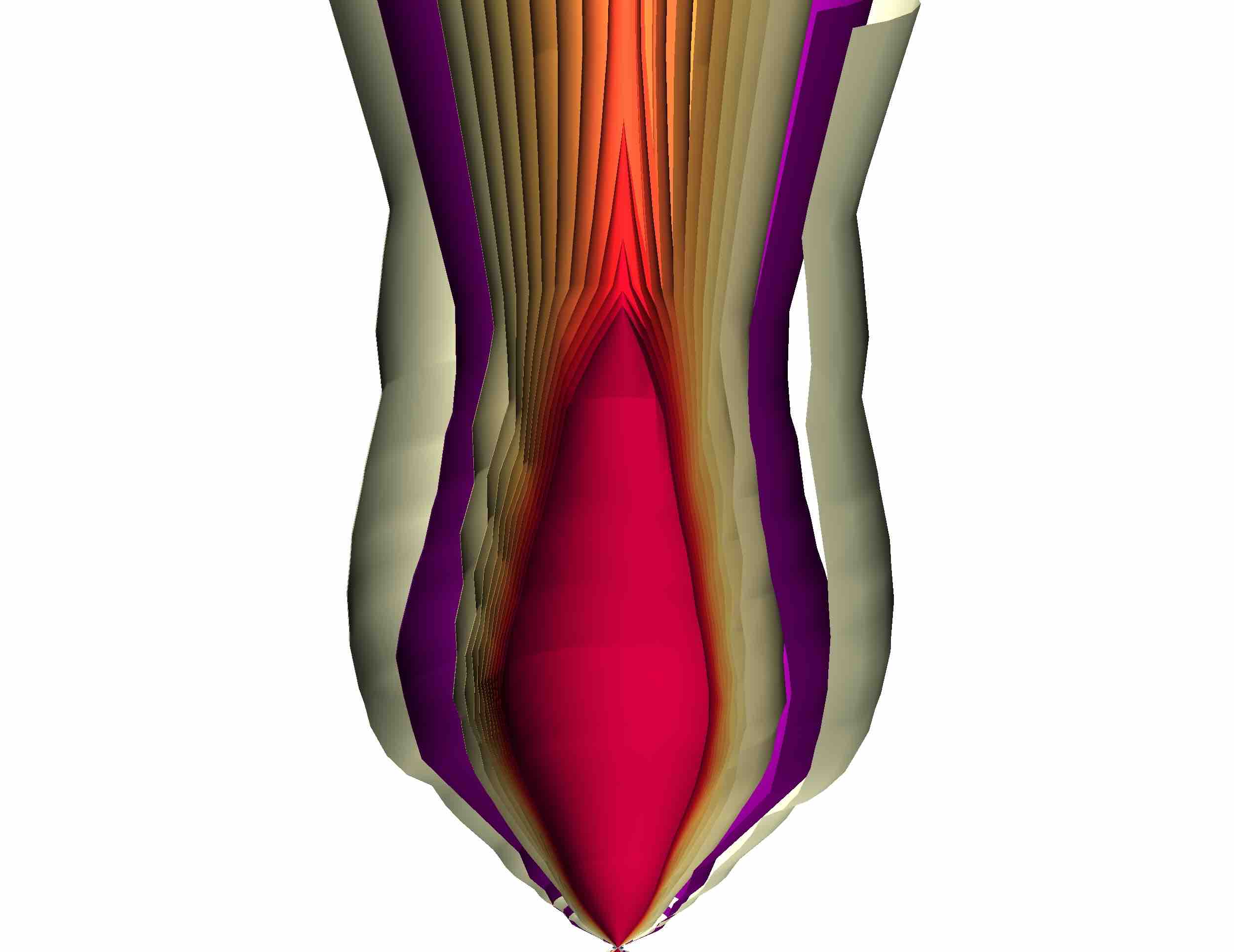}
\includegraphics[width=0.24\textwidth]{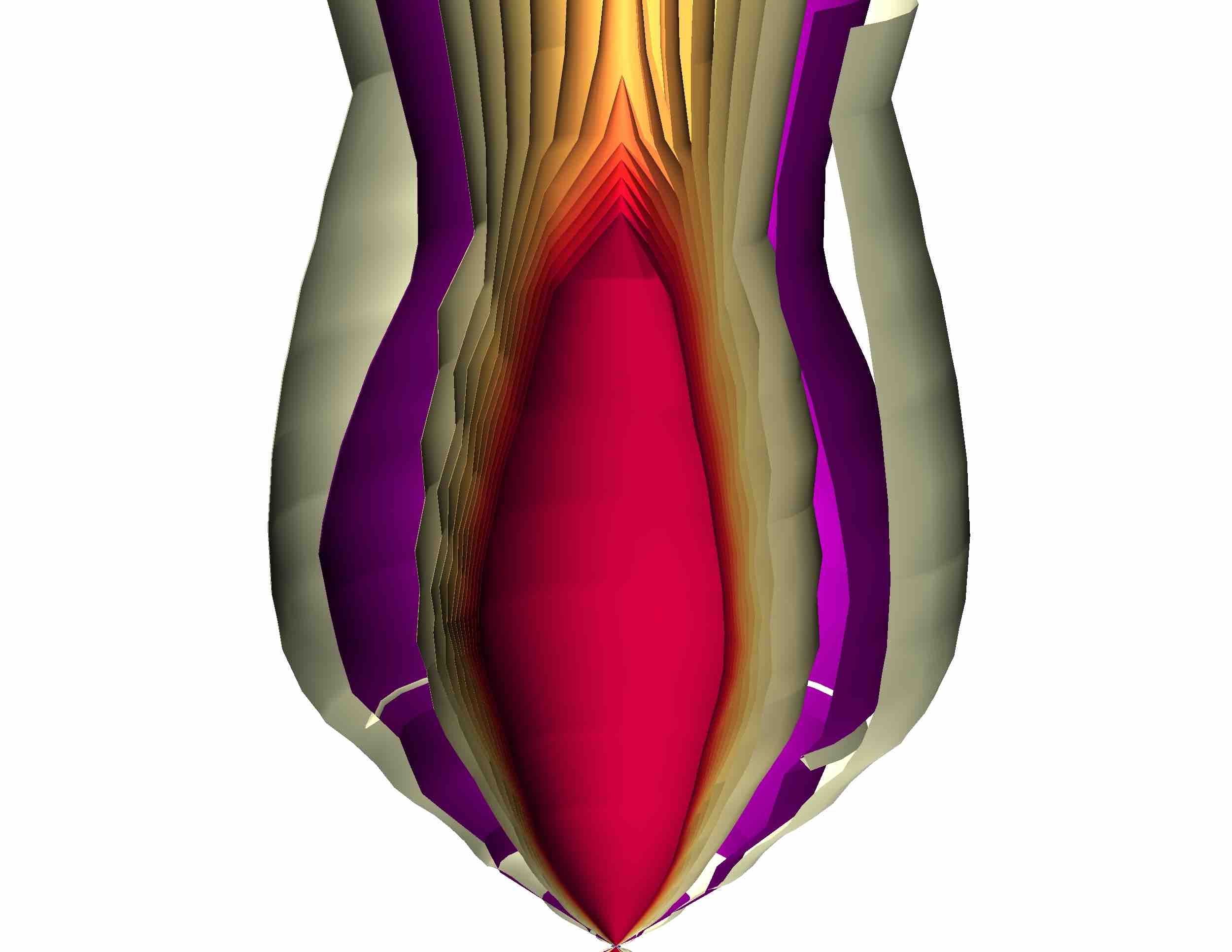}
\includegraphics[width=0.24\textwidth]{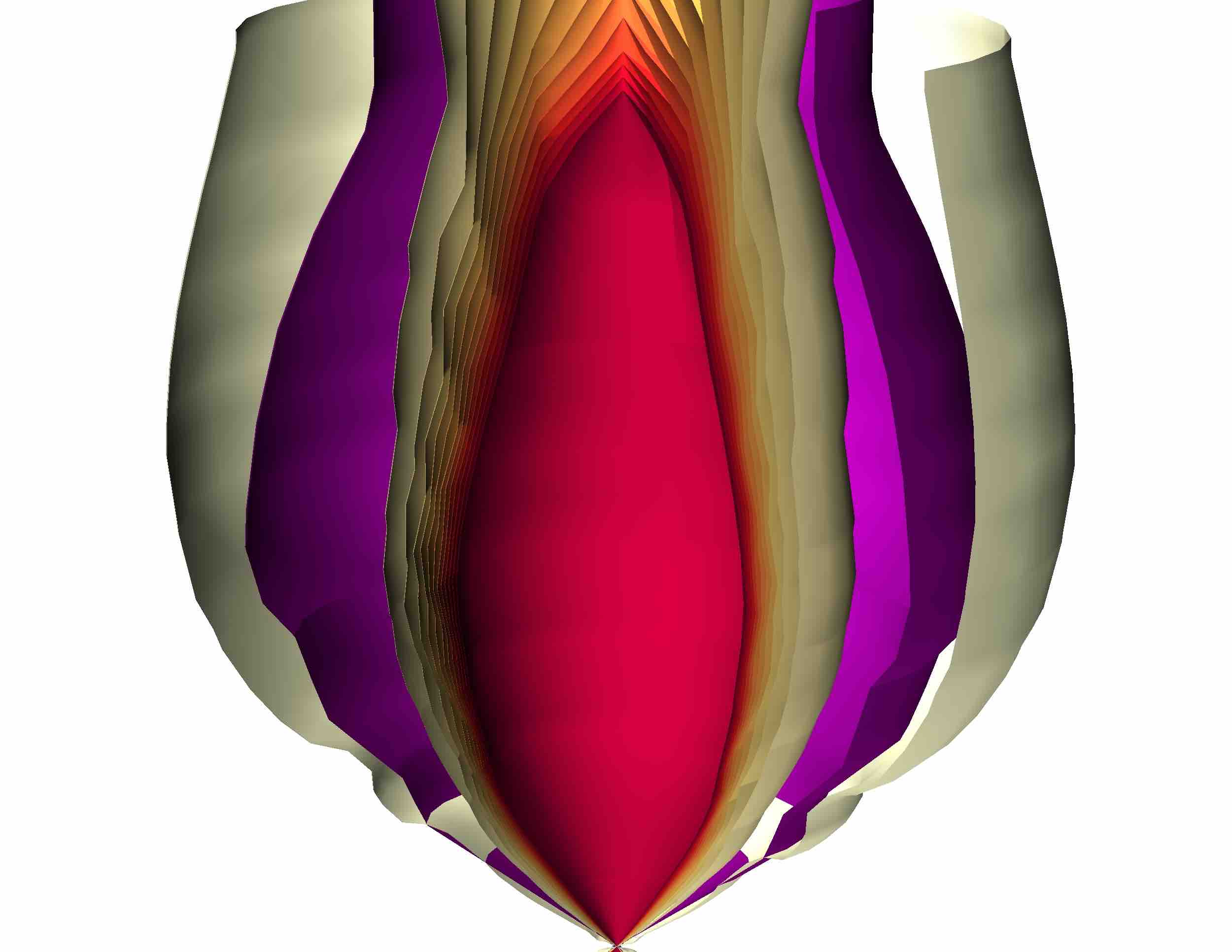}
\includegraphics[width=0.24\textwidth]{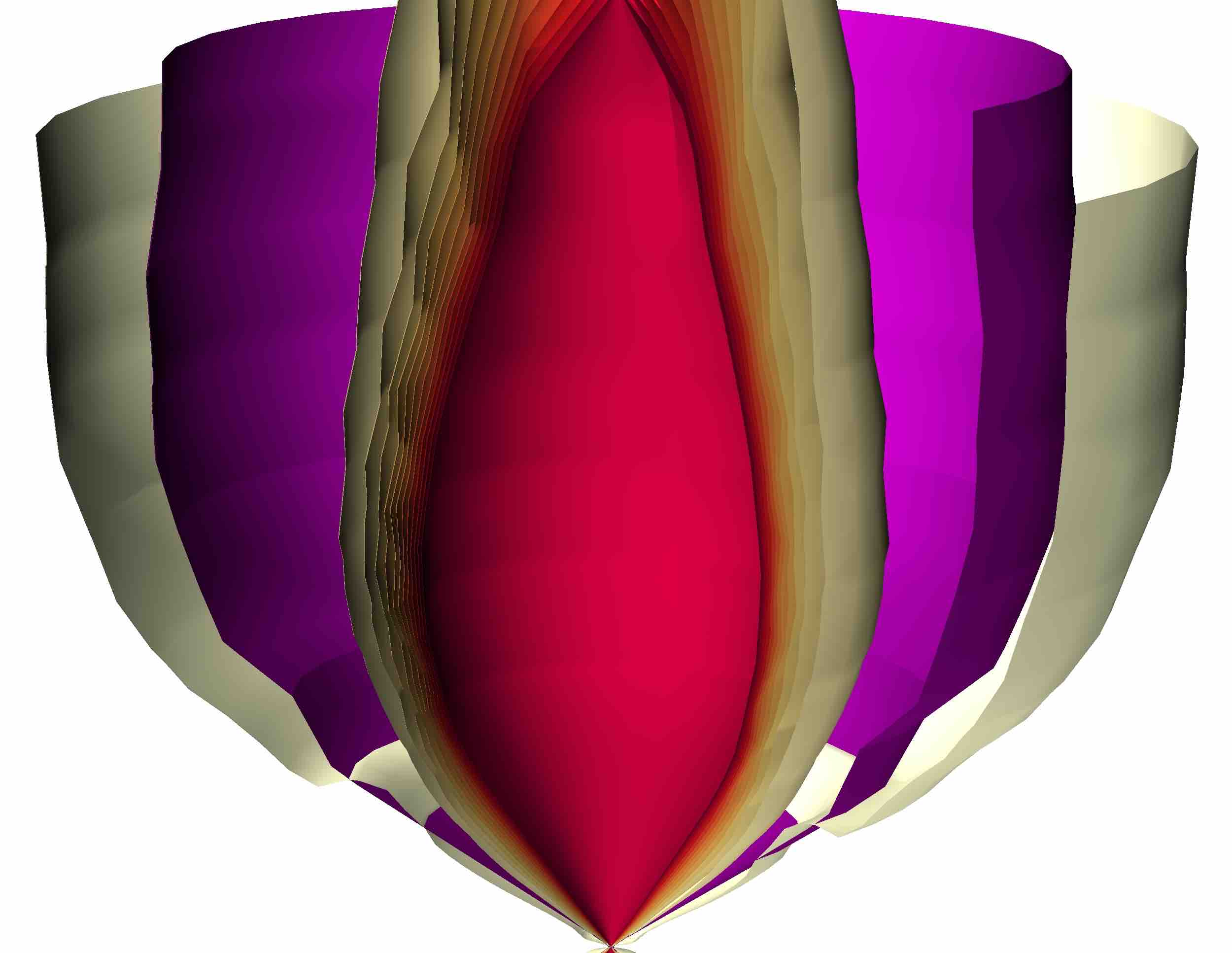}
\caption{
Outflow broadening at late evolutionary times on large scales without (top panels) and with (bottom panels) photoionization feedback.
All panels show the protostellar outflow and HII region; the accretion disk covers only a few pixels here.
The outflow velocity is color-coded as in Fig.~\ref{fig:visualization_outflow}.
The extent of the ionized HII region is shown as a purple contour.
The disk density is color-coded as in Fig.~\ref{fig:visualization_disk}.
The panels show from left to right the systems at $t = 90$, $100$, $110$, and $120~\kyr$ of evolution.
The images cover $\approx 1~\pc$ in height.
Data is taken from the simulation runs ``1.0pc-PO-RAD'' and ``1.0pc-PO-RAD-ION'', respectively.
}
\label{fig:OutflowBroadeningHII}
\end{figure*}

\begin{figure*}[htbp]
\centering
\includegraphics[width=0.24\textwidth]{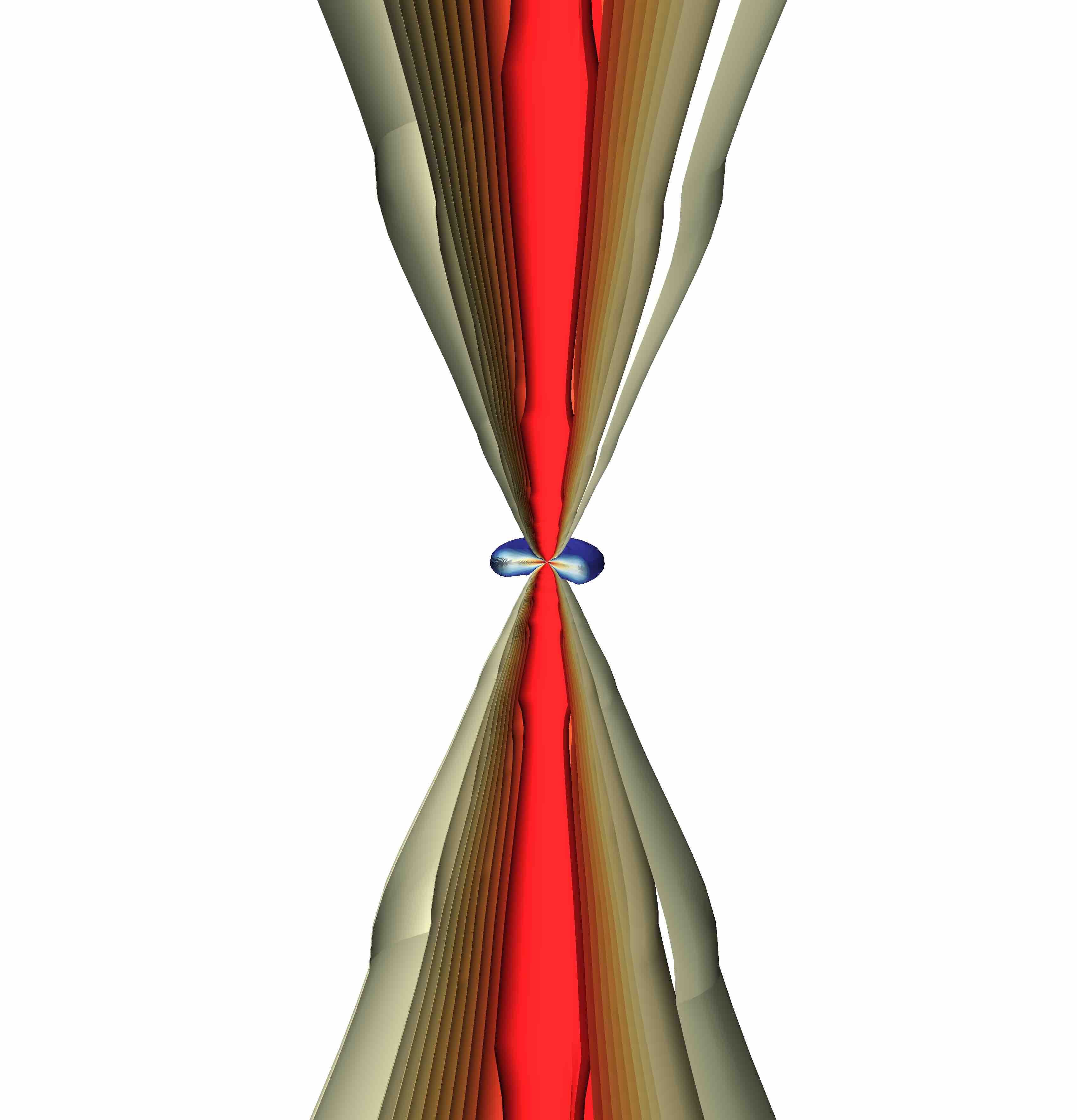}
\includegraphics[width=0.24\textwidth]{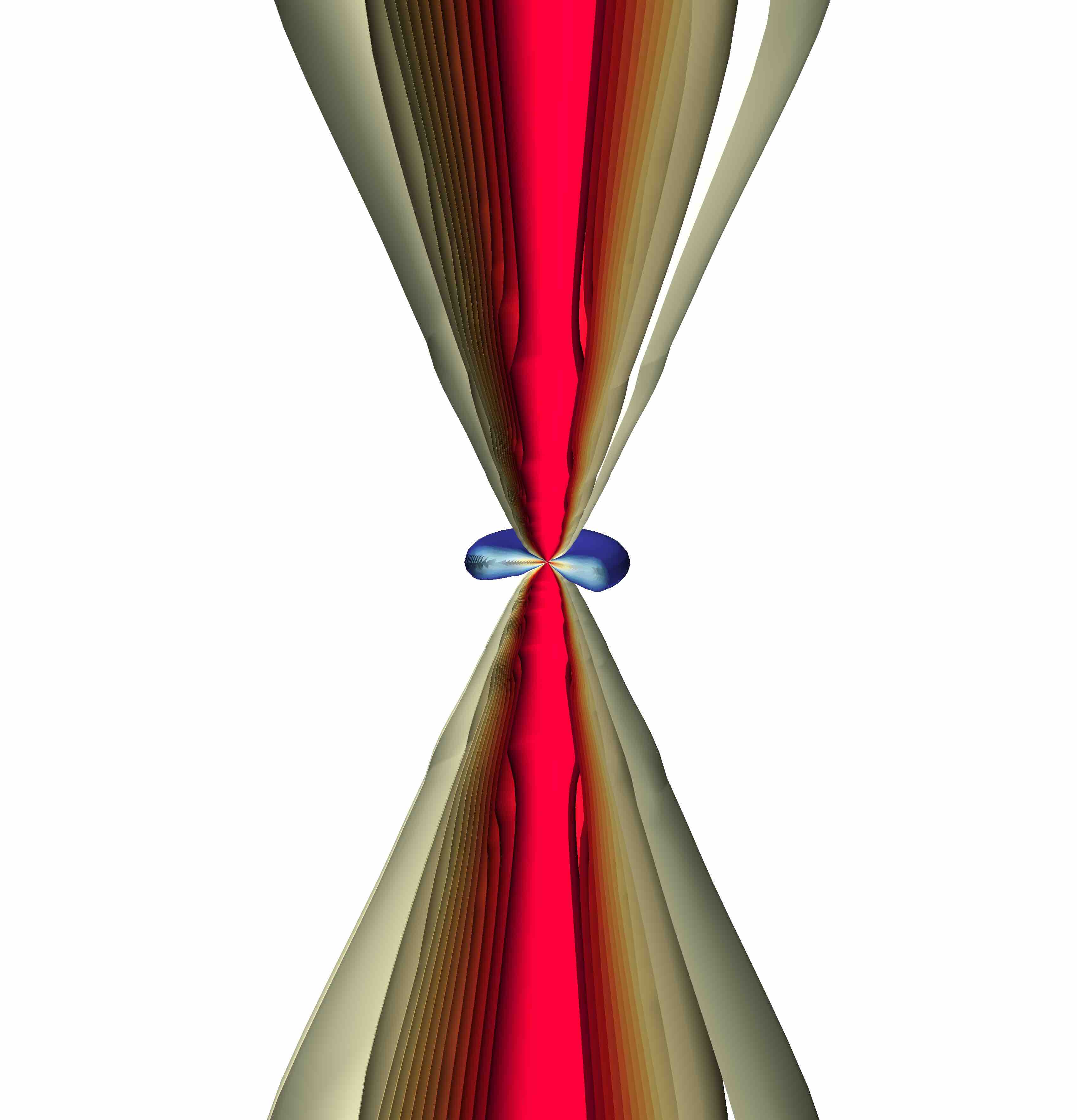}
\includegraphics[width=0.24\textwidth]{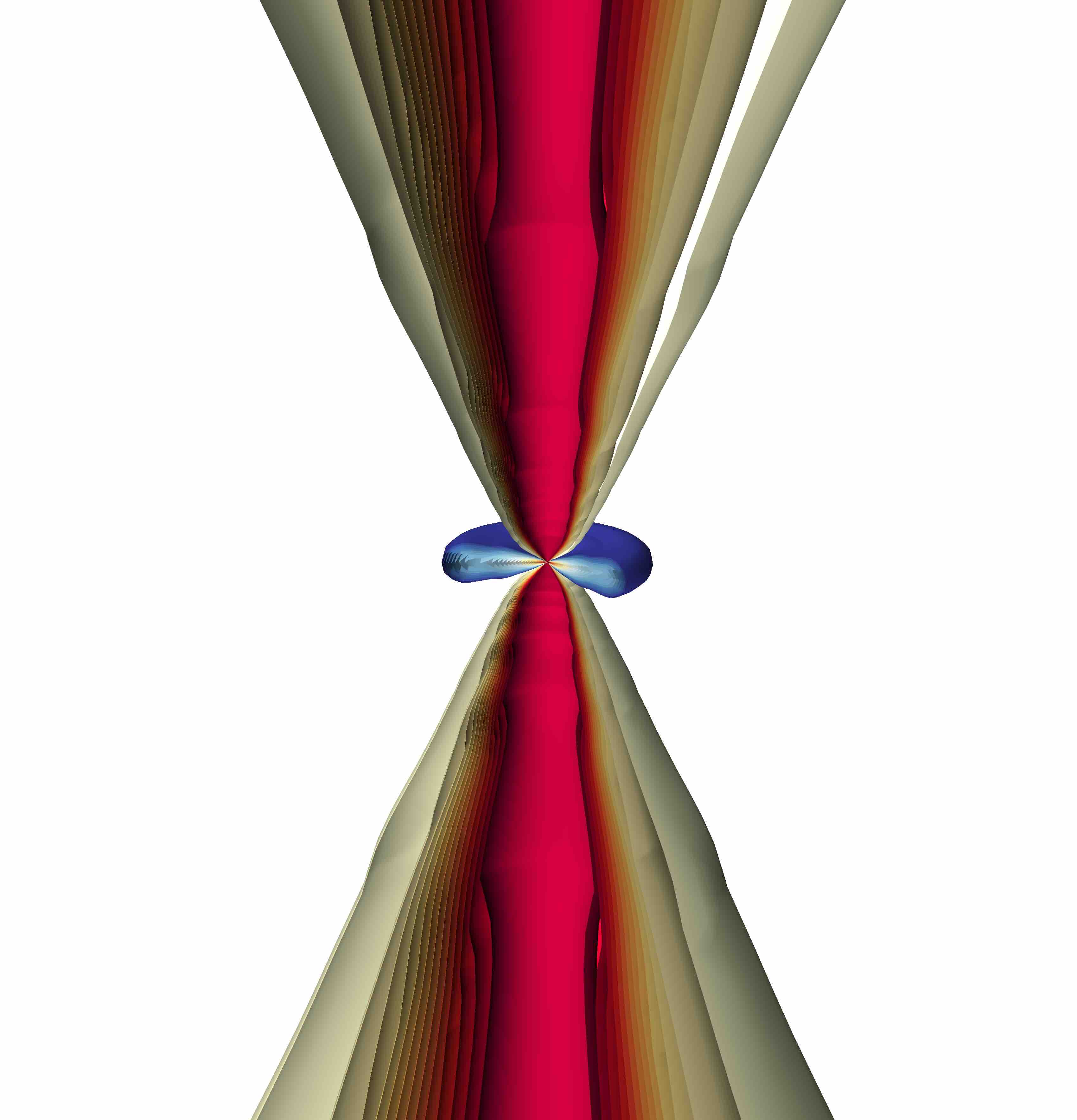}
\includegraphics[width=0.24\textwidth]{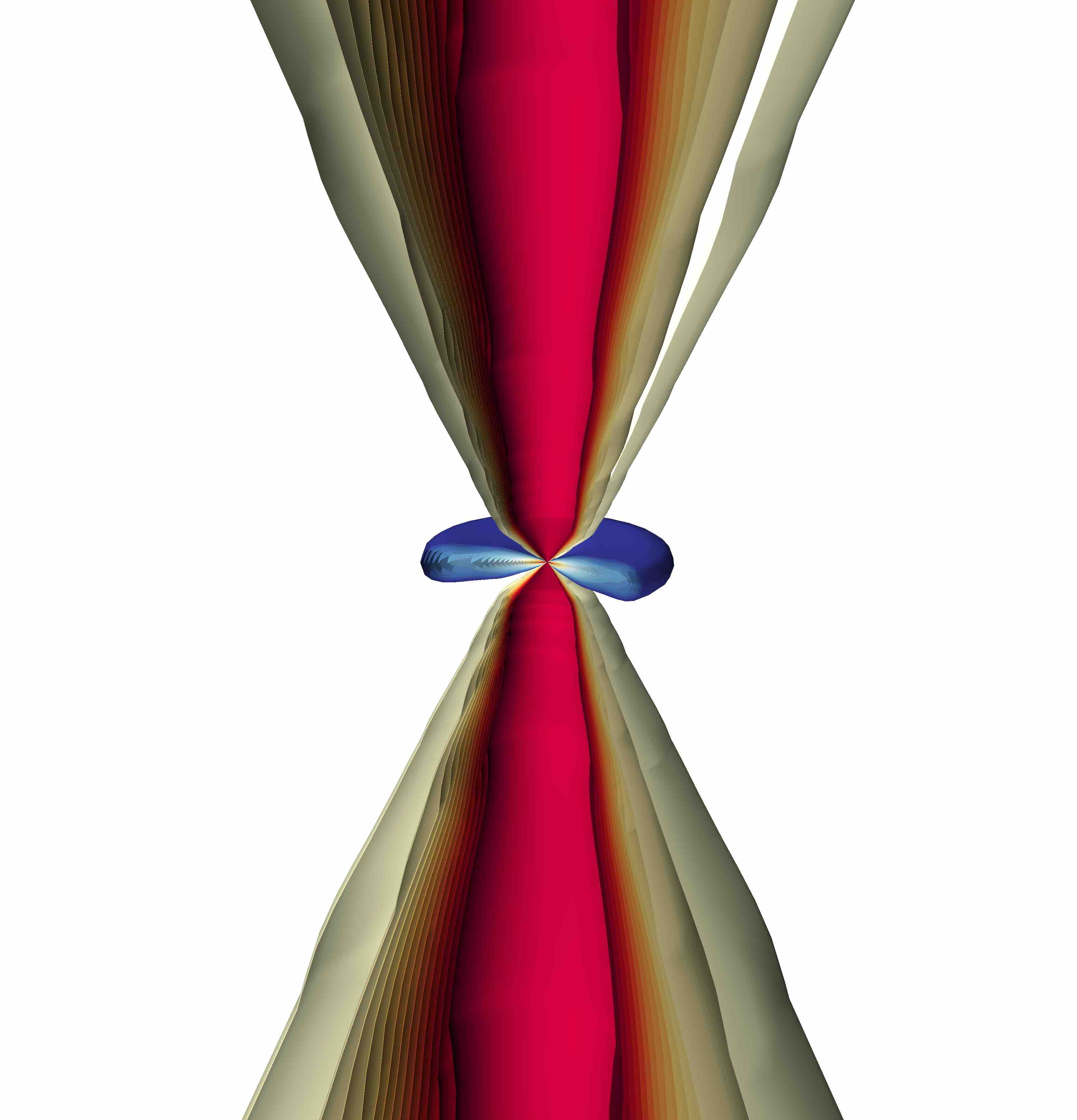}
\includegraphics[width=0.24\textwidth]{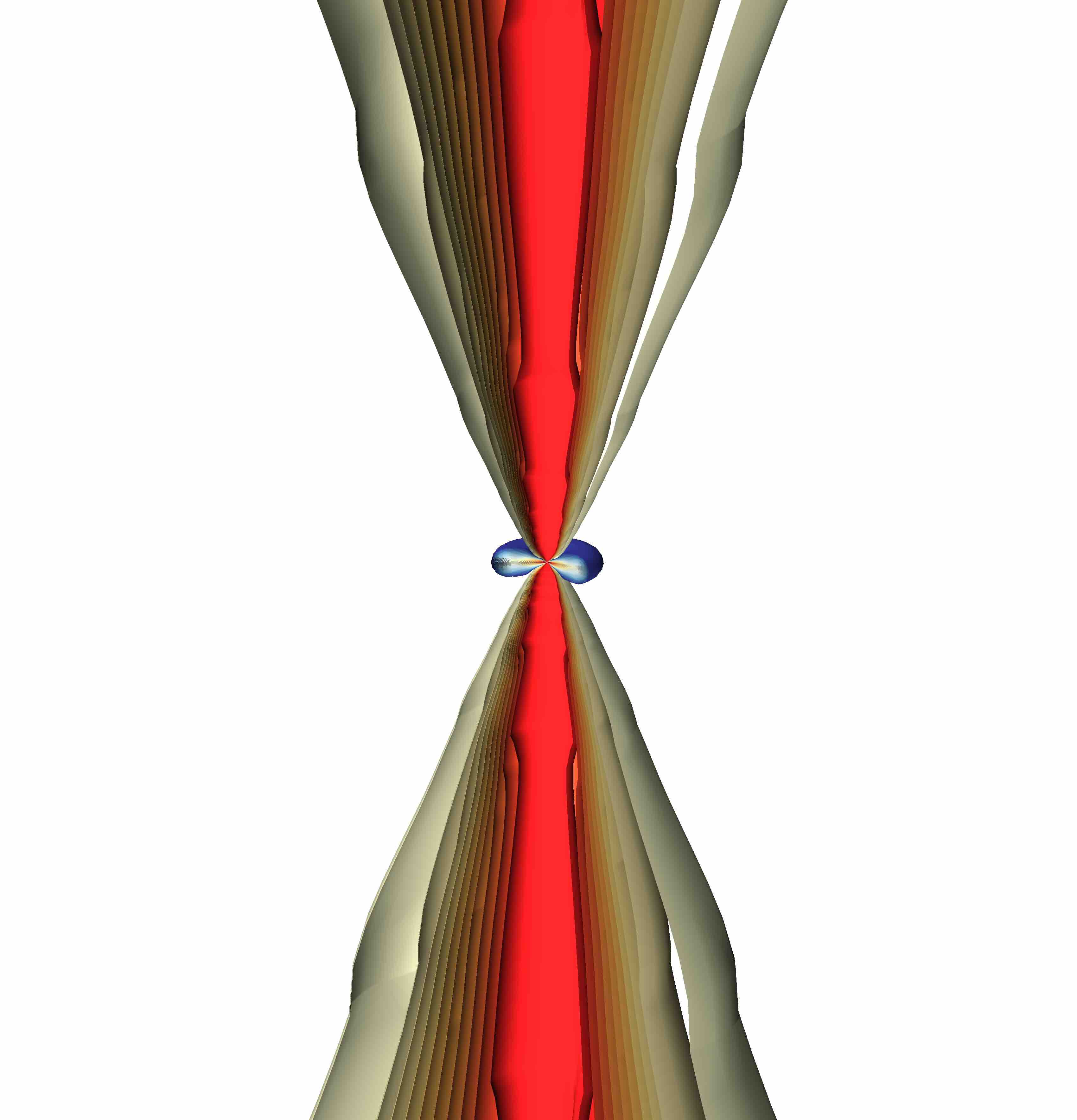}
\includegraphics[width=0.24\textwidth]{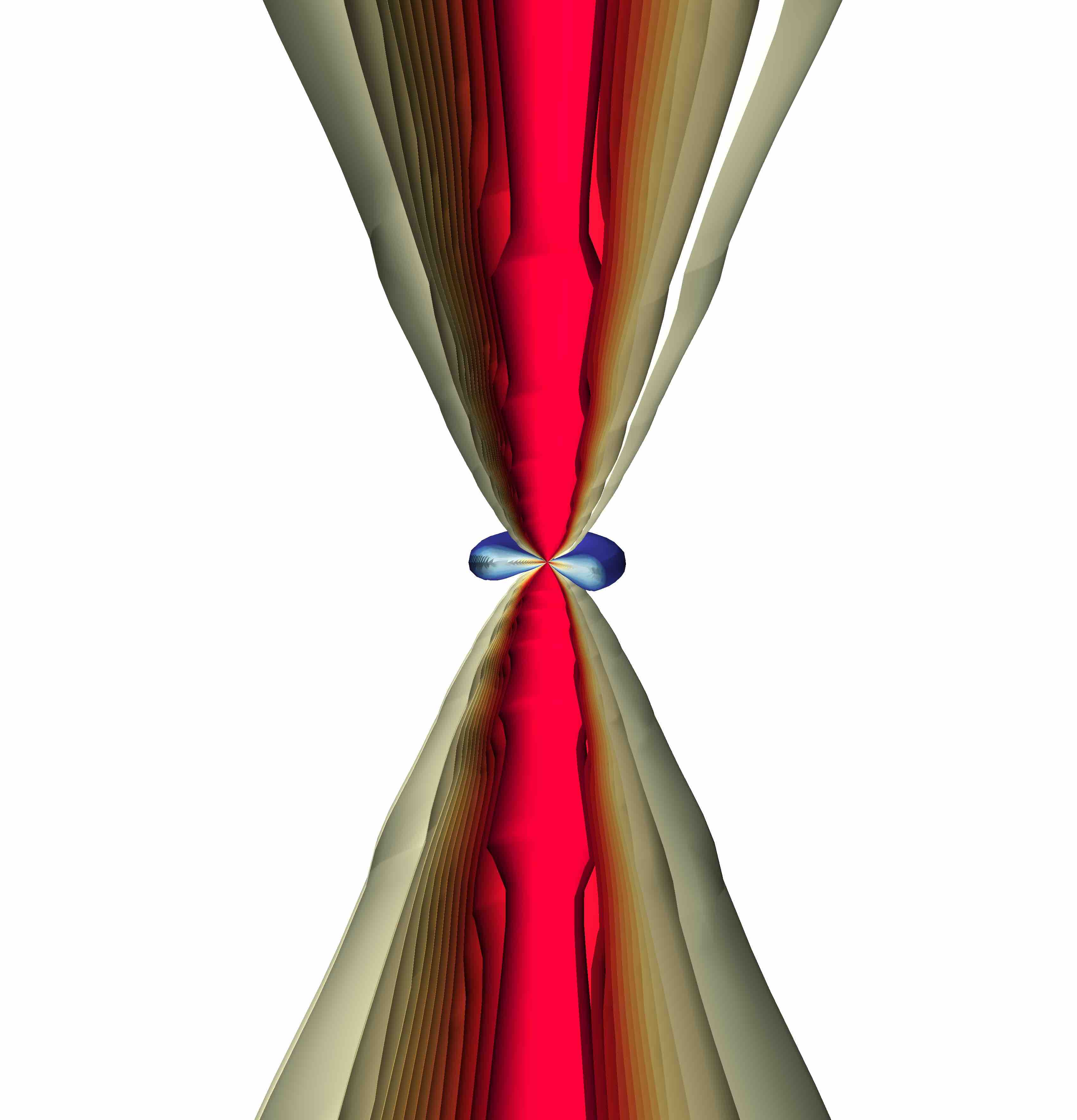}
\includegraphics[width=0.24\textwidth]{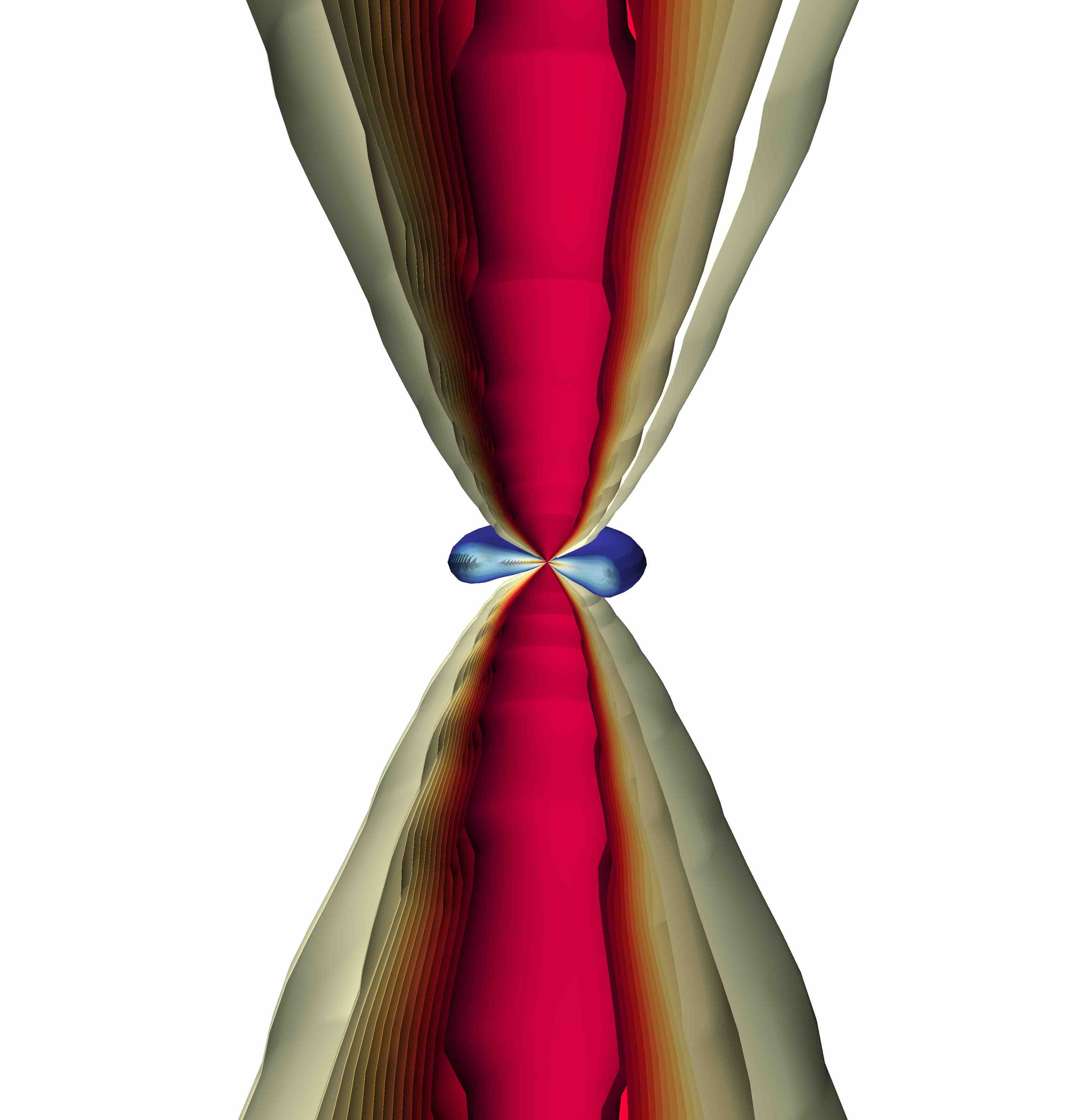}
\includegraphics[width=0.24\textwidth]{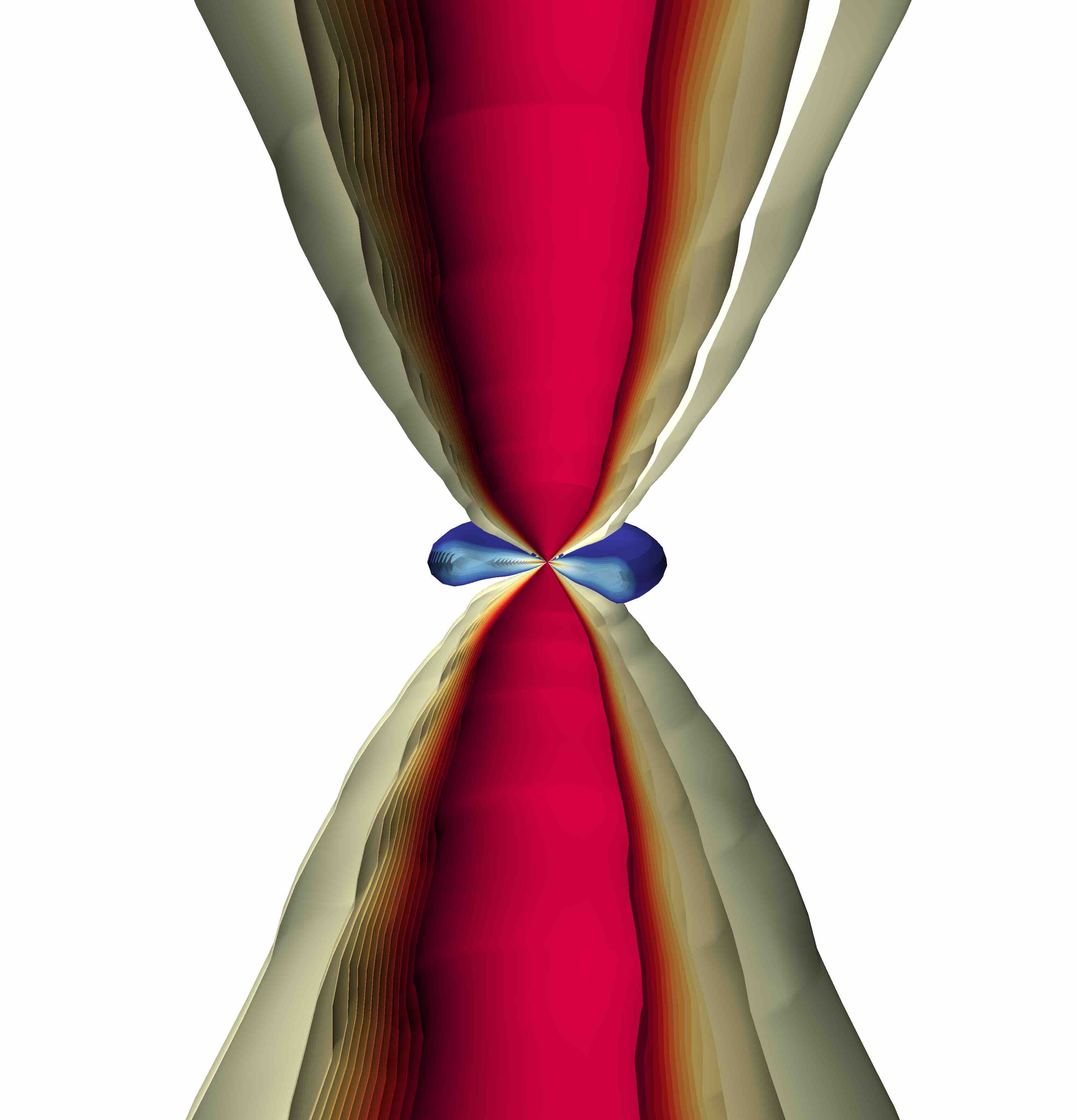}
\caption{
Outflow broadening at early evolutionary times on small scales without (top panels) and with (bottom panels) radiation force feedback.
The panels show from left to right the accretion disk and protostellar outflow at $t = 10$, $15$, $20$, and $25~\kyr$ of evolution.
The disk density is color-coded as in Fig.~\ref{fig:visualization_disk}.
The outflow velocity is color-coded as in Fig.~\ref{fig:visualization_outflow}.
The images cover $\approx 0.05~\pc$ in height.
Data is taken from the simulation runs ''1.0pc-PO-ION'' and ``1.0pc-PO-RAD-ION'', respectively.
}
\label{fig:OutflowBroadeningRadiationForce}
\end{figure*}

We conclude this section on the formation and evolution of the high-mass star by introducing the concept of the maximum accretion radius or region of influence.
We use the phrase ``virtually infinite'' mass reservoir for the $R_\mathrm{res} = 1~\pc$ sized simulation, because this reservoir is much larger than the region which has participated in the formation of the central star.
We can estimate the maximum radius, from which gas could have been traveled toward the forming high-mass star as follows:
From the initial conditions, we can compute the free-fall time as function of radius 
\begin{equation}
t_\mathrm{ff}(r) = \sqrt{ \frac{\pi^2 ~ r^3}{8 ~ G ~ M(r)} }
\end{equation}
with the included gas mass per radius $M(r) = 1000~\Msol \times r / \pc$ at the beginning of the simulations.
If we now identify the free-fall time in the equation above with the duration of stellar accretion, we can invert the equation to solve for the maximum accretion radius $R_\mathrm{acc}^\mathrm{max}$.
From radius $R_\mathrm{acc}^\mathrm{max}$, only gas in free-fall motion could have reached the central star during its accretion epoch, which sets an upper limit on the region of potential influence.
The larger scale environment $r > R_\mathrm{acc}^\mathrm{max}$ can be guaranteed to have not participated in the formation of the high-mass star.
Unsurprisingly, in the case of the finite small-scale mass reservoir simulations, the maximum accretion radius is larger than the initial pre-stellar core, hence, the full reservoir participated in the star formation process.
In the case of the virtually infinite large-scale mass reservoir, simulations without radiation forces have not stopped stellar accretion at the end of the simulation, and the region of influence has already grown to $1~\pc$ at the point in time, when these simulations where stopped.
The simulations which account for radiation forces have accretion epochs of $t_\mathrm{acc}^\mathrm{tot} \approx 140~\kyr$ and $t_\mathrm{acc}^\mathrm{tot} \approx 125~\kyr$ for the case without and with photoionization feedback, respectively.
These accretion times correspond to maximum accretion radii of $R_\mathrm{acc}^\mathrm{max} \approx 0.27~\pc$ and $R_\mathrm{acc}^\mathrm{max} \approx 0.24~\pc$.
Even if the gravitational collapse of larger regions are accounted for, the additional gas mass reservoir at $r > R_\mathrm{acc}^\mathrm{max}$ has no impact on the forming star.
I.e.~the finally $79~\Msol$ and $95~\Msol$ high-mass stars have formed from actual mass reservoirs of at most $270~\Msol$ and $240~\Msol$, respectively (the first value for simulations neglecting photoionization, the latter accounting for photoionization feedback on top of the radiation force feedback).
Hence, to form a super-Eddington star of a specific mass, an initial mass reservoir of about two to three times higher mass is required.

Do these results on the mass reservoir favor a competitive accretion or a monolithic collapse scenario for the formation of very massive stars?
We do not think that this question can be securely answered at this point.
To answer such a general question conclusively, many more simulations for different initial density profiles are required to cover the broad parameter space.
E.g.~an initially shallower density profile will lead to a slower collapse and star formation process, which in turn leads to a larger maximum accretion radius, and initially steeper density profiles would speed up the star formation process thereby reducing the maximum accretion radius accordingly.
\vONE{
Additionally, we have to keep in mind that gas from large scales is brought to the forming star and its host core by more complex morphologies than considered here, such as filaments within a turbulent molecular cloud complex, see for comparison the simulation studies of \citet{Smith:2011ip}.
}
%\subsection{From Infall to Outflow}
%\begin{figure*}[ht!]
%\centering
%\includegraphics[width=0.49\textwidth]{./images/MassFluxes/Mdotin30000AU_vs_time_1000}
%\includegraphics[width=0.49\textwidth]{./images/MassFluxes/Mdotout30000AU_vs_time_1000}\\
%%\includegraphics[width=0.49\textwidth]{./images/MassFluxes/Mdotin10000AU_vs_time_1000}
%%\includegraphics[width=0.49\textwidth]{./images/MassFluxes/Mdotout10000AU_vs_time_1000}\\
%\includegraphics[width=0.49\textwidth]{./images/MassFluxes/Mdotin3000AU_vs_time_1000}
%\includegraphics[width=0.49\textwidth]{./images/MassFluxes/Mdotout3000AU_vs_time_1000}\\
%%\includegraphics[width=0.49\textwidth]{./images/MassFluxes/Mdotin1000AU_vs_time_1000}
%%\includegraphics[width=0.49\textwidth]{./images/MassFluxes/Mdotout1000AU_vs_time_1000}\\
%\includegraphics[width=0.49\textwidth]{./images/MassFluxes/Mdotin300AU_vs_time_1000}
%\includegraphics[width=0.49\textwidth]{./images/MassFluxes/Mdotout300AU_vs_time_1000}\\
%%\includegraphics[width=0.49\textwidth]{./images/MassFluxes/Mdotin100AU_vs_time_1000}
%%\includegraphics[width=0.49\textwidth]{./images/MassFluxes/Mdotout100AU_vs_time_1000}\\
%\caption{
%}
%\label{fig:MassFluxes}
%\end{figure*}

\subsection{HII regions and large-scale feedback}
\label{sect:results_HII}

After the end of the bloating phase of the protostar at about $M_* \approx 30~\Msol$, the star contracts toward the zero age main sequence.
The ultimate decrease in radius from $\approx 160~\Rsol$ to $8-9~\Rsol$ together with the increase in luminosity during this time yields a change in temperature of about one order of magnitude, see Fig.~\ref{fig:StellarEvolution}.
The stellar spectrum of the photospheric temperature of $T_* \approx 45000~\K$ results in a strong release of EUV radiation, which photoionizes the hydrogen of the stellar surrounding.
To release the electron from the neutral hydrogen, the absorbed EUV photon requires an energy $h\nu \ge 13.6~\eV$.
Any excess energy of an ionizing photon is first transformed into kinetic energy of the electron, which afterwards thermalizes via collisions.
This process denotes a dominant heating source of the gas in the stellar surrounding.
The resulting increase in thermal pressure yields an expansion of the hot, ionized gas.

The inclusion of simulations with and without photoionization feedback as well as with and without radiation forces in this paper allows us to investigate the impact of photoionization in direct comparisons of the evolution of the system with and without photoionization feedback.
Additionally, we can study the impact of radiation forces on the evolution of the HII region by directly comparing its evolution in simulations with and without radiation force feedback.

The rate of EUV photons required to photoionize a neutral medium of hydrogen number density $n_H$ scales quadratically with $n_H$.
Hence, the low-density cavity quickly becomes ionized, while only the innermost rim of the accretion disk and its atmospheric layers get ionized later in the course of the evolution of the system.
In turn, this means that the morphology of the initiated and expanding HII region highly depends on the pre-existing morphology of the stellar surroundings, especially any anisotropy in optical depth.
For the observational implications of such outflow-confined HII regions please see the post-processing radiative transfer models of \cite{2016ApJ...818...52T, 2017ApJ...849..133T}.

In general, including photoionization feedback leads to a broadening of the bipolar cavities on large scales (see Fig.~\ref{fig:OutflowBroadeningHII} for a visualization) as well as to a decrease in density of these regions.
The thermal pressure yields a compression of the gaseous medium at the HII region boundaries.
This compression is responsible for the increase in disk mass, and hence stellar accretion rate, as discussed in the previous sections.

Further including the effect of radiation forces leads to an additional broadening on small scales (see Fig.~\ref{fig:OutflowBroadeningRadiationForce} for a visualization).
Without radiation forces, no radial acceleration of the medium outside of the HII region is visible on small scales, while including radiation forces quickly removes this gas.
The combination of both feedback effects yields a much larger opening angle of the cavity on small scales.
With radiation forces becoming less important on larger scales, this effect vanishes with increasing radius.
A quantitative investigation of the broadening of the bipolar cavities is presented and discussed in the following Sect.~\ref{sect:results_Outflow}.

Does the ionization prevent or reduce the accretion flow on disk or core scales?
On large scales, the HII regions expand into the low-density bipolar regions cleared by the previous protostellar outflow.
The radial extent of the HII regions in these low-density regions very quickly after their formation grows to be much larger than their critical radius (cite Keto 2003):
\begin{equation}
R_\mathrm{crit}^\mathrm{HII} = \frac{G ~ M_*}{2 ~ c_s^2}
\end{equation}
\begin{figure}[htbp]
\centering
\includegraphics[width=0.49\textwidth]{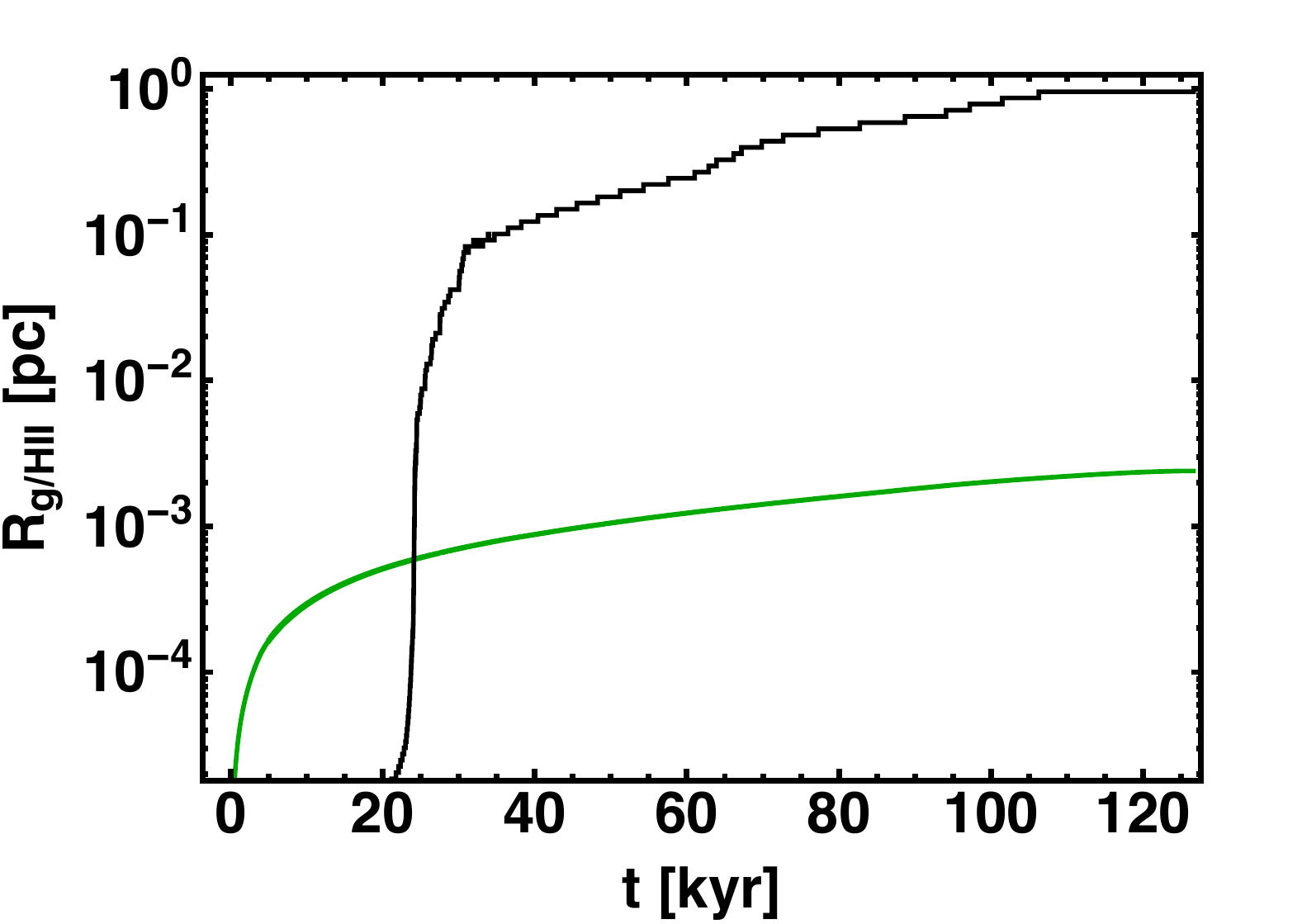}
\caption{
Critical radius $R_\mathrm{crit}^{HII}$ (green) of HII region expansion and actual size of the expanding HII region $R_\mathrm{HII}$ (black), explicitly the extent of the HII region along the polar axis.
Simulation data is taken from the large-scale mass reservoir simulation series including all feedback components
}
\label{fig:RgHII_vs_time}
\end{figure}
In Fig.~\ref{fig:RgHII_vs_time} we show the radial expansion of the HII region in contrast to the evolution of the critical radius.
On large scales, the expansion velocity of the HII region of $\approx 11~\kms$ is -- as a consequence of protostellar outflow and radiation force momentum feedback -- slightly higher than the sound speed of the ionized gas of $\approx 9~\kms$.
As expected in the case of HII regions larger than the critical radius, we do not detect any sign of gas accretion thru the large-scale HII region.
Instead, the HII region reduces the infall momentum of the gravitational collapse, see Fig.~\ref{fig:m-tot_vs_time} and discussion below.
This would change of course in the case of gravitationally trapped hyper-compact or ultra-compact HII regions, see \citet{Keto:2003kk, Keto:2007jy}.
And in fact, at the ionized inner disk rim, the distance to the star is less than the critical radius, hence, we measure a sustained disk-to-star accretion rate thru this ionization layer.
As depicted in Figs.~\ref{fig:Mdisk_vs_time} and \ref{fig:diskdestruction} and analyzed in the previous Sect.~\ref{sect:results_StarFormation} on small-scale feedback, on spatial scales of the accretion disk the effects of photoionization sum up to a positive feedback for the forming star.

\begin{figure}[htbp]
\centering
\includegraphics[width=0.49\textwidth]{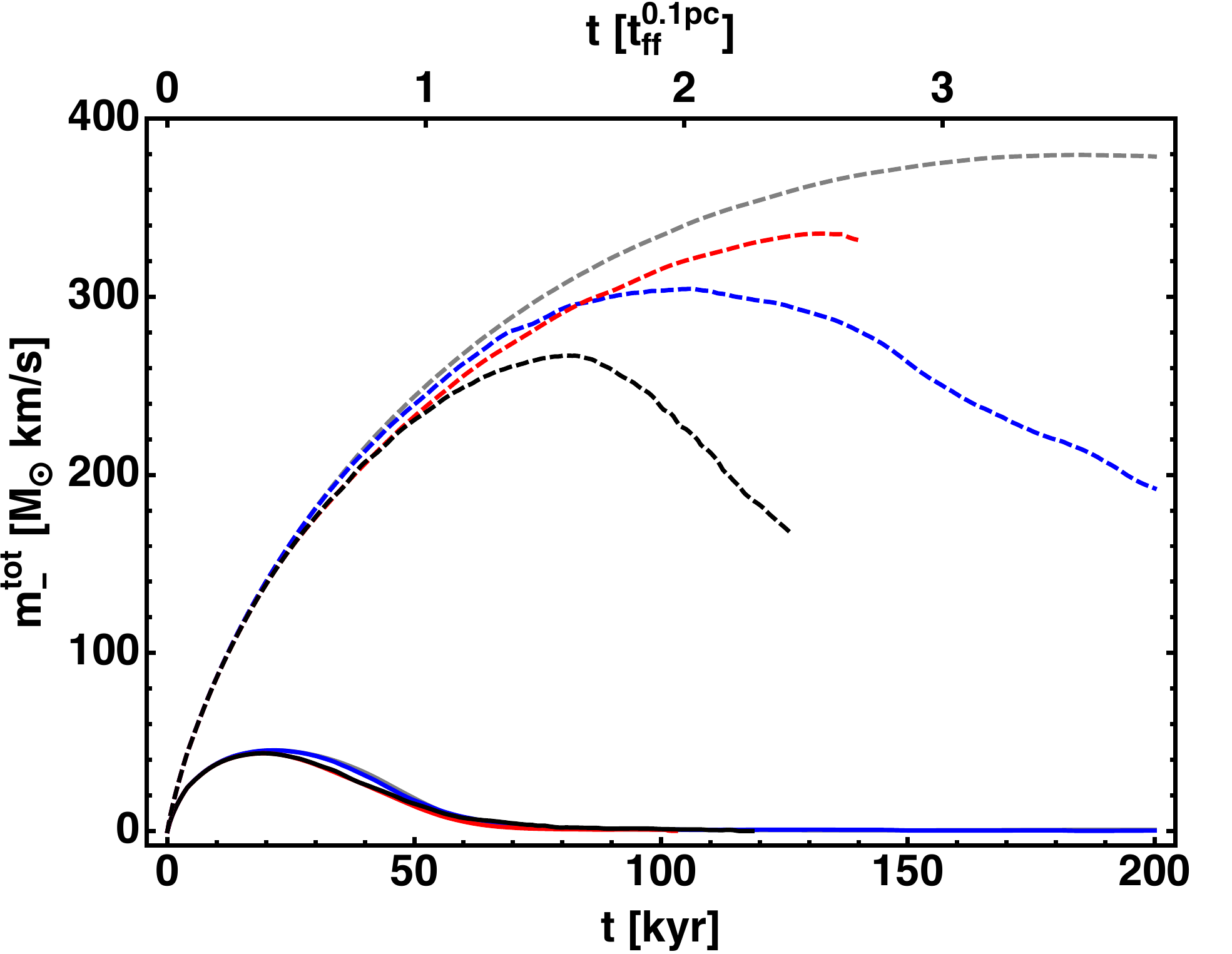}
\caption{
Total momentum of the radially infalling gas as function of time.
Color and line styles are defined in Table~\ref{tab:sims}.
}
\label{fig:m-tot_vs_time}
\end{figure}
On scales larger than $0.1~\pc$, the expansion of the HII region is the dominant feedback effect which slows down the global gravitational collapse.
Fig.~\ref{fig:m-tot_vs_time} shows the evolution of the radially inward total momentum $m_\mathrm{-}^{tot}$ in the different simulations.
To compute $m_\mathrm{-}^{tot}$, the infall momentum is integrated over the full computational domain, radially positive momenta do not contribute. 
In the case of the small-scale reservoir simulations, the gravitational collapse leads to an increase of the total inward momentum for roughly half of a free-fall time, decreasing afterwards due to centrifugal forces and outflow, radiation, and photoionization feedback.
This final decrease is slightly faster for simulations including radiation forces, but the differences are minor compared to the large-scale mass reservoir.

In the case of large-scale mass reservoirs, the global gravitational collapse leads to a monotic increase in infall momentum in all cases neglecting photoionization feedback; although the amount of total infall momentum seems to have reached its maximum in evolution at the end of the simulations.
Including the photoionization feedback yields a decrease in infall momentum after $t \ge 80~\kyr$ with radiation forces included and $t \ge 100~\kyr$ without radiation forces.
At the final stage of both large-scale simulations including ionization feedback the total infall momentum is reduced by roughly 50\% compared to the pure outflow feedback case.
The difference between the small-scale and the large-scale reservoir simulations means that photoionization generates this momentum feedback on the scales larger than $0.1~\pc$, the outer reservoir radius of the small-scale simulations.

These results are in agreement with the expected reduction of the efficiency of star cluster formation due to photoionization feedback \citep{2017MNRAS.470.3346H}.

%\begin{figure}[ht!]
%\centering
%\includegraphics[width=0.49\textwidth]{./images/Momenta/m+tot_vs_time}
%\caption{
%}
%\label{fig:m+tot_vs_time}
%\end{figure}
%
%\begin{figure}[ht!]recor
%\centering
%\includegraphics[width=0.49\textwidth]{./images/Momenta/mtot_vs_time_100Msol}
%\caption{
%}
%\label{fig:mtot_vs_time_100Msol}
%\end{figure}

\subsection{Outflow broadening as a multi-scale and multi-physics phenomenon}
\label{sect:results_Outflow}
The opening angle of the bipolar low-density cavity is affected by a variety of different physical processes, namely 
the launching and collimation of magneto-centrifugally driven jets and outflows, 
the ram pressure from the infalling material from large-scales,
the radiation forces, and
the photoionization.
Except for the first one, each of these processes is explicitly and self-consistently treated in our simulations, and its influence has been investigated in the simulation series by performing simulations in which each individual component is by purpose neglected.
The launching and collimation of the magneto-centrifugally driven jets and outflows is included in the simulation series via means of a subgrid module, i.e.~we inject the expected momentum feedback in an previously defined way.
For the effects of varying the free parameters of the associated subgrid module, we refer the interested reader to previous publications \citep{2015ApJ...800...86K, 2016ApJ...832...40K}.

The polar angle distribution of the injected momentum at the base of the protostellar outflow is an input parameter, hence, we note that cases and epochs where the opening angle is controlled by the protostellar outflow physics alone, should be taken with care.
In the results shown, this applies to the first $15~\kyr$ of evolution for all simulations and to the outflow broadening on smallest spatial scales in simulations excluding radiation forces.
In other epochs, where the protostellar outflow denotes by far a subdominant contribution to the cavity width, the simulation series offers a rich data set to study the effects and resulting trends in dependence on ram pressure, radiation forces, and photoionization feedback.

\begin{figure}[htbp]
\centering
\includegraphics[width=0.49\textwidth]{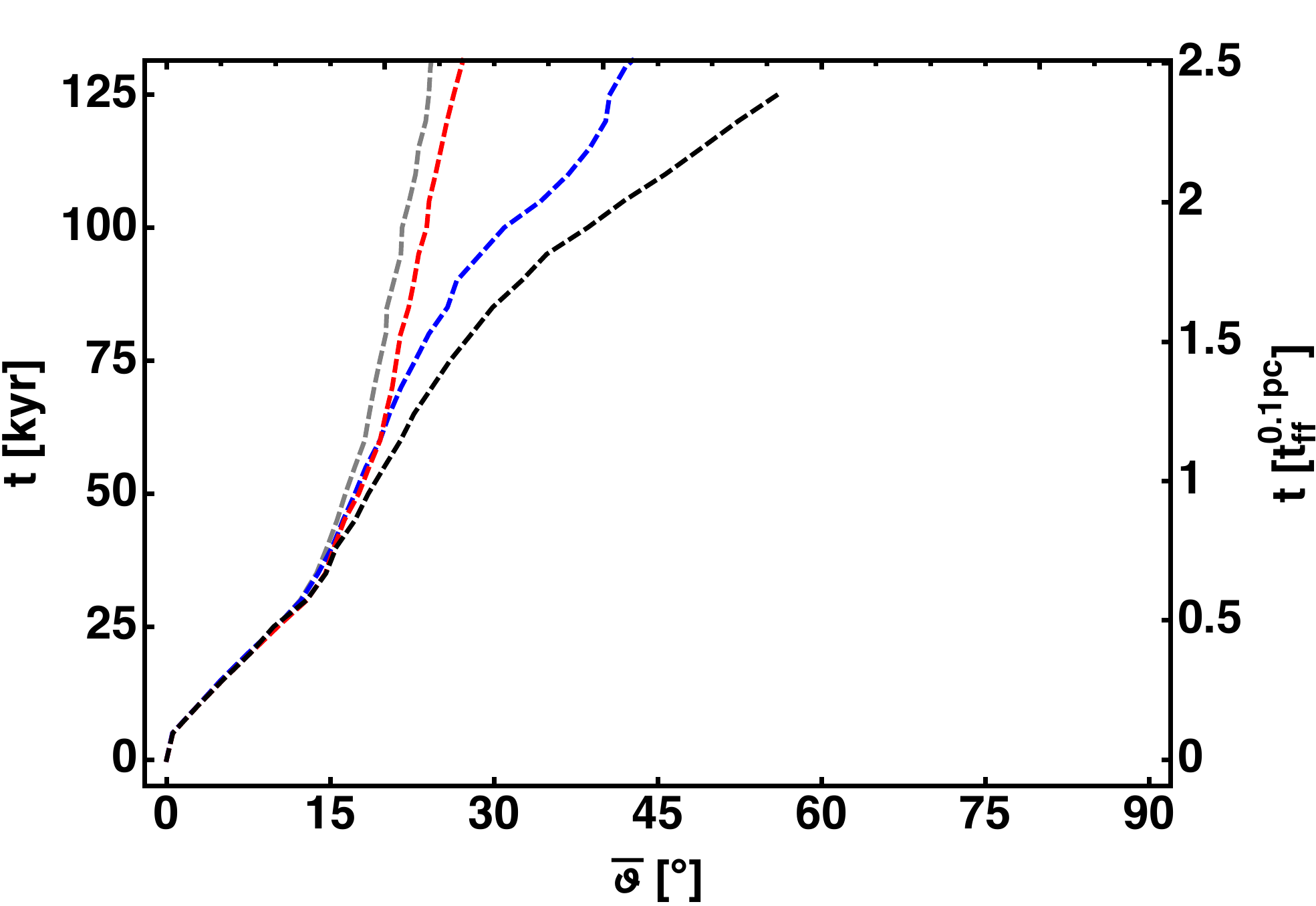}
\includegraphics[width=0.49\textwidth]{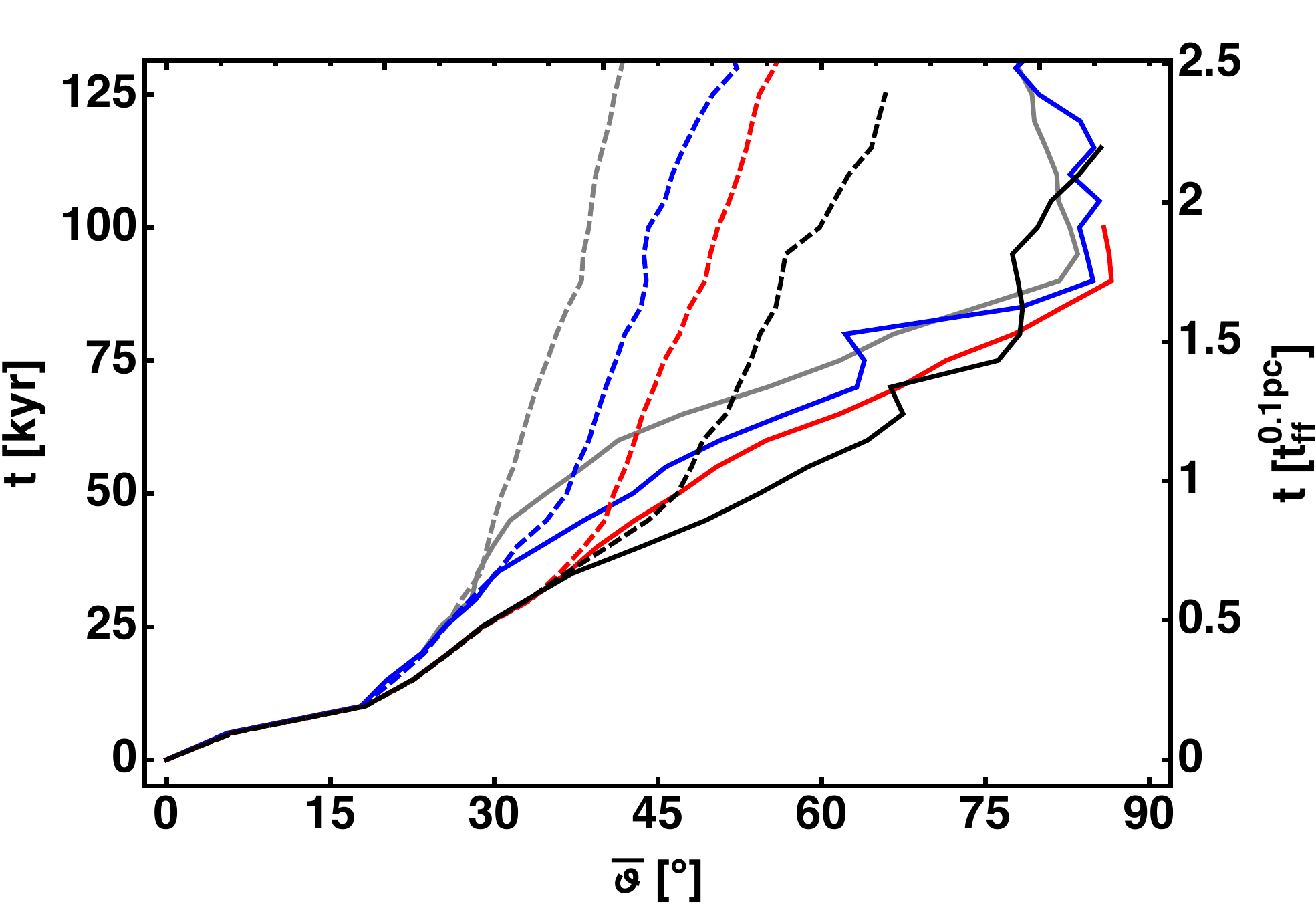}
\includegraphics[width=0.49\textwidth]{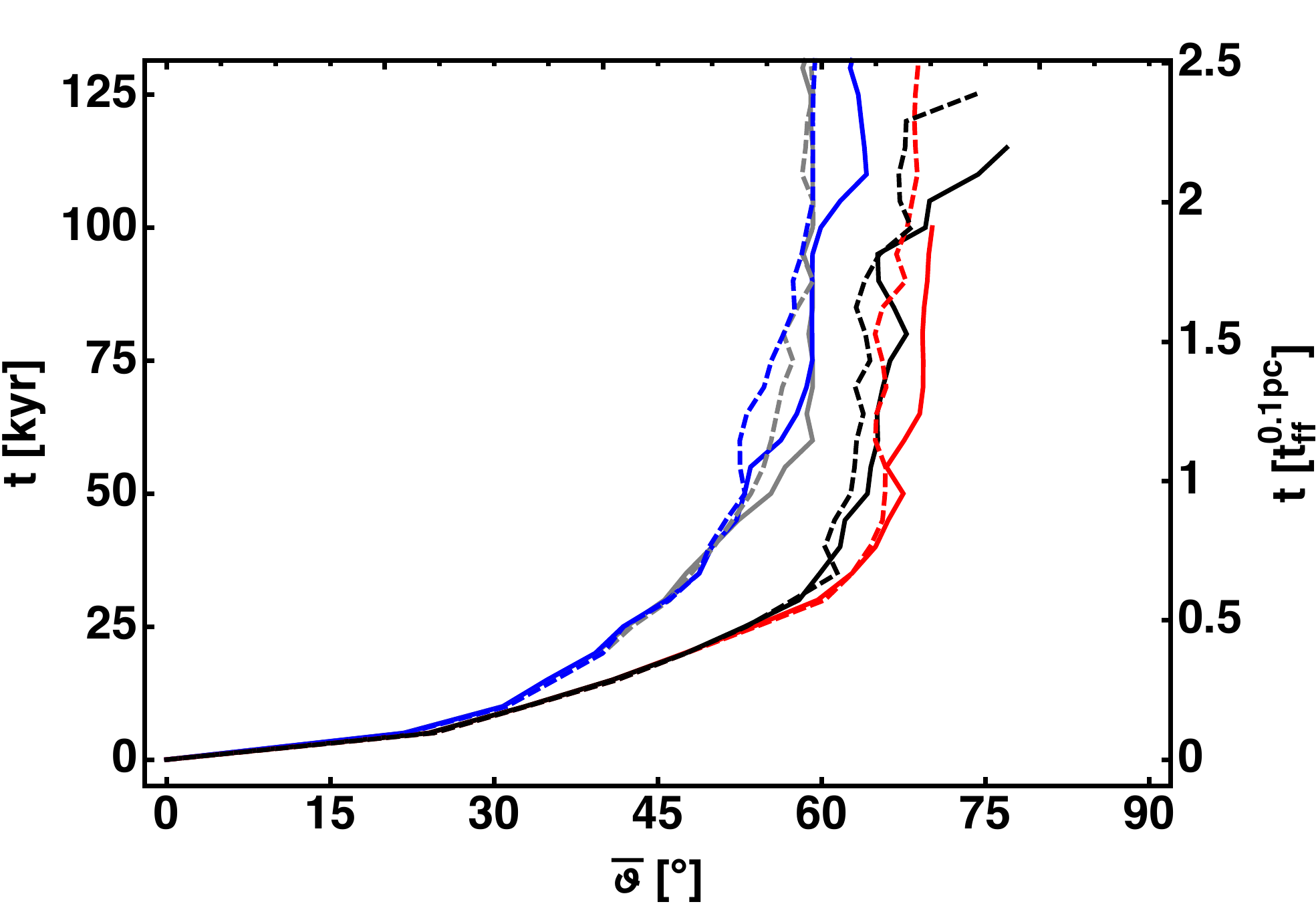}
\caption{
Outflow broadening:
Mean semi-opening angle $\bar{\theta}$ of the outflow cavity as function of time (vertical axis), averaged over three different spatial scales.
From top to bottom, the panels show the mean semi-opening angle $\bar{\theta}$ of the outflowing gas on $1.0~\pc$, $0.1~\pc$, and $0.01~\pc$ ($\sim$ 2000~au) scales.
Color and line styles are defined in Table~\ref{tab:sims}.
}
\label{fig:OutflowBroadening}
\end{figure}

First of all, just from the visualization of the bipolar region as e.g.~done in Fig.~\ref{fig:OutflowBroadeningHII}, we know that the opening angle $\theta$ of the low-density cavity is in fact not a single value, which is evolving in time, but a function of radius as well ($\theta =  \theta(r, t)$).
The underlying reason for this is the relative importance of the different broadening effects on the different spatial scales.
Therefore, we distinguish the following analysis into three spatial scales: $0.01~\pc \approx 2000~\au$, $0.1~\pc$, and $1~\pc$.
These scales can be roughly interpreted as the spatial scales of the disk, the pre-stellar core, and the large-scale infall.
Based on this classification, we then compute at each point in time three mean opening angles
\begin{equation}
\bar{\theta}(t) = \frac{1}{R_\mathrm{scale} - R_\mathrm{min}} \int_{R_\mathrm{min}}^{R_\mathrm{scale}} \theta(r, t) ~ dr
\end{equation}
with the inner rim of the computational domain at $R_\mathrm{min} = 3~\au$ and $R_\mathrm{scale} =$ $0.01$, $0.1$, or $1~\pc$, respectively.
$\theta(r,t)$ defines at each point in time and each radius the width of the cavity in angle measured from the bipolar axis; within this region, the radial velocity is positive, the adjacent region further away from the outflow axis has negative radial velocity, i.e.~still infalling.
The resulting broadening as function of time is shown in Fig.~\ref{fig:OutflowBroadening}.

\begin{figure}[htbp]
\centering
\includegraphics[width=0.49\textwidth]{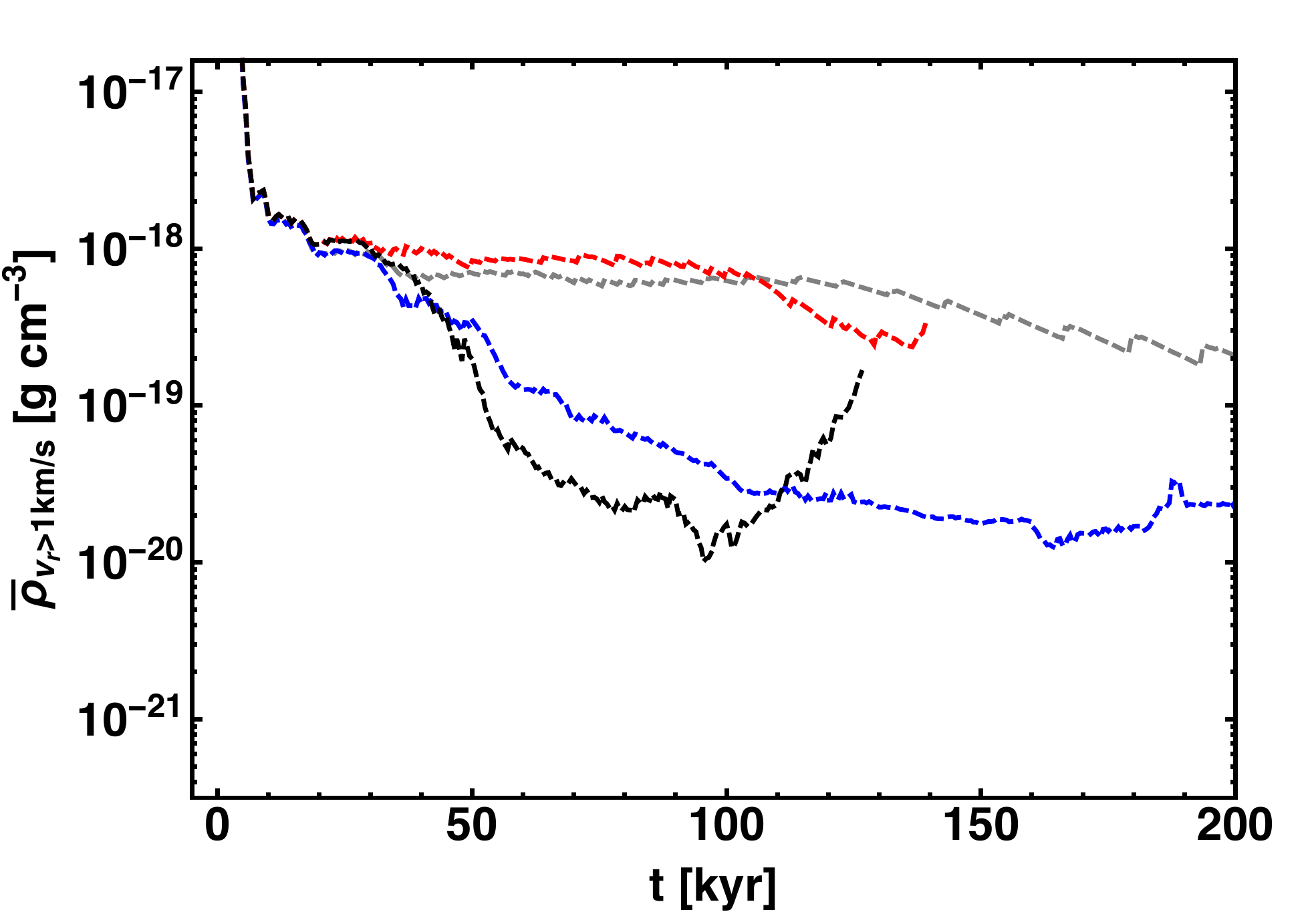}
\includegraphics[width=0.49\textwidth]{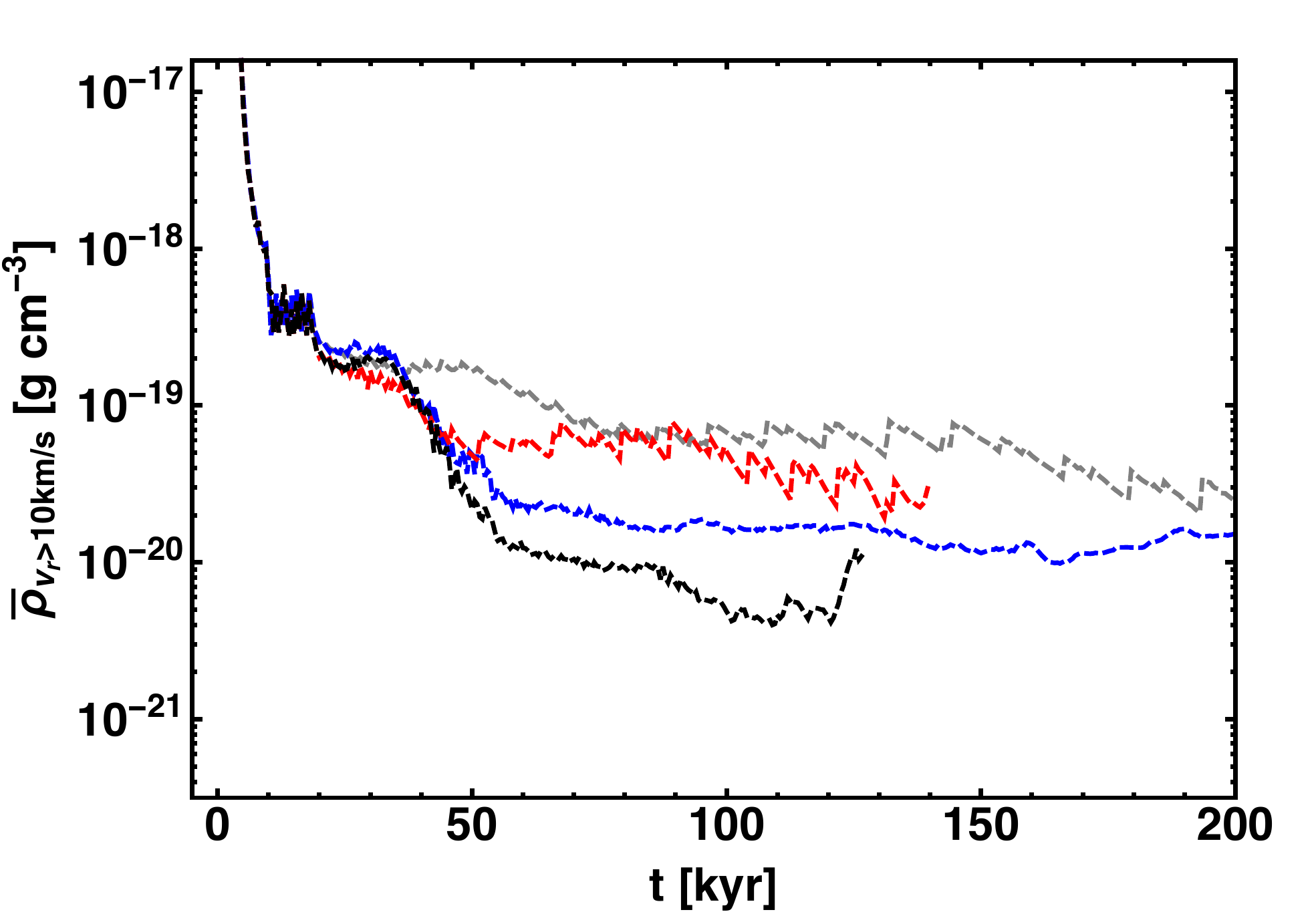}
\includegraphics[width=0.49\textwidth]{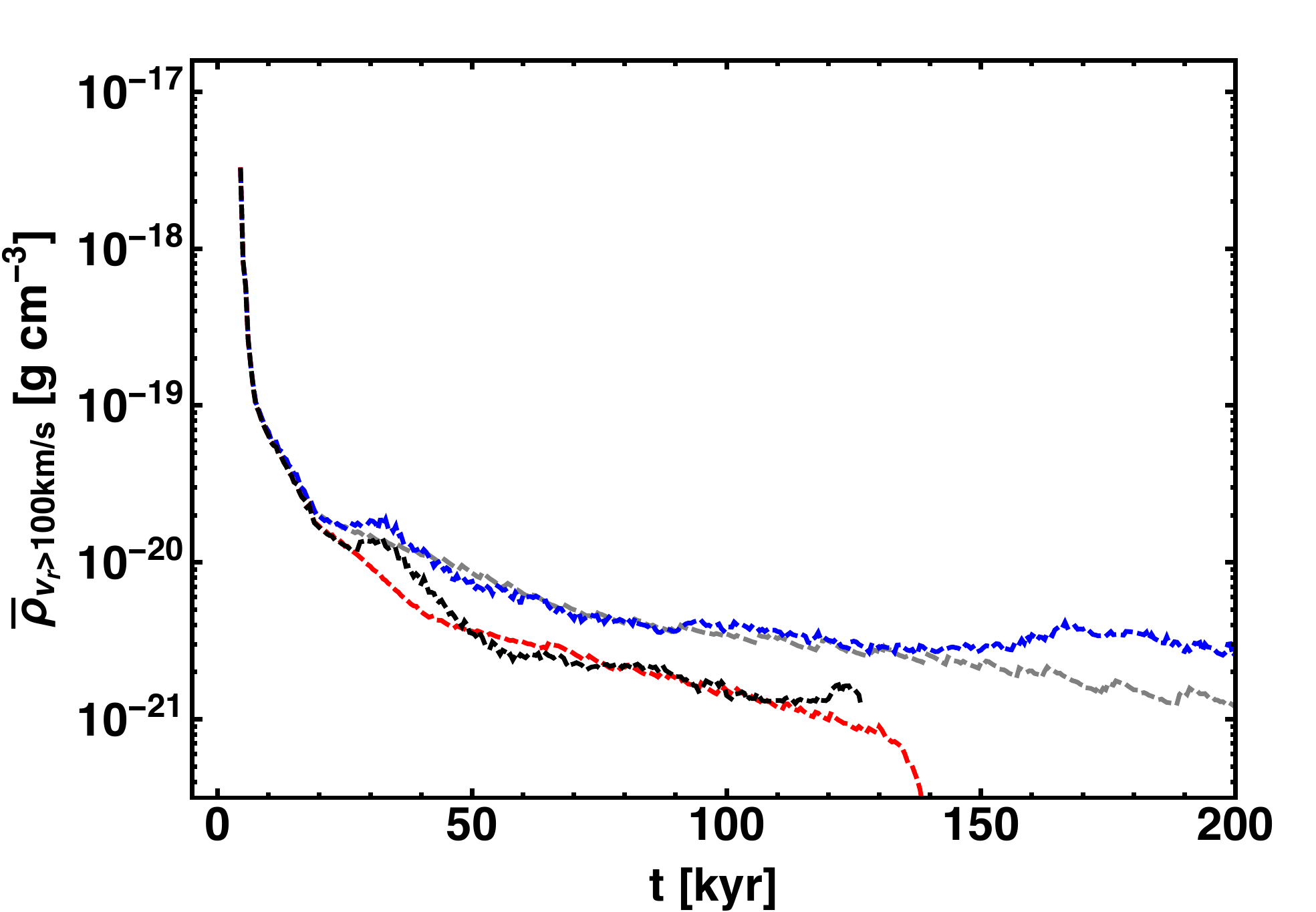}
\caption{
Outflow cavity mean densities:
From top to bottom, the panels show the mean cavity density in regions with radial velocities higher than $1~\kms$, $10~\kms$, and $100~\kms$.
Color and line styles are defined in Table~\ref{tab:sims}.
}
\label{fig:CavityDensity}
\end{figure}

First, we would like to highlight the effect of the ram pressure from the large-scale infall, i.e.~we compare the solid lines (simulations without large-scale infall) with the dashed lines (including large-scale infall) of the same color (same radiative feedback effects included).
On disk scales (bottom panel), the ram pressure from large scales ($> 0.1~\pc$) has only a minor effect; simulations including the ram pressure show higher collimation, but only by a few degrees.
On core scales (middle panel) the collimation effect due to ram pressure becomes significant, especially at later times of evolution.
While all small-scale mass reservoir simulations without ram pressure show an increase of the opening angle down to the disk layer ($\bar{\theta} \ge 80\degr$), the ram pressure from the large-scale infall holds the collimation of the cavity to values below $65\degr$ at the same evolution in time.
The largest scales (top panel) are simply not included in the small-scale mass reservoir simulations, hence, a comparison is obsolete.
In summary, ram pressure is the key process, which counter-balances the broadening of the cavity due to outflow and radiation feedback processes.

Next, we focus only on simulations including the ram pressure (dashed lines) and investigate the effect of radiation force only (red lines) and photoionization feedback only (blue lines) compared to purely protostellar outflow feedback (gray lines).
On disk scales (bottom panel), the opening angle is dominated by radiation forces; photoionization only simulations not show any broadening compared to the protostellar outflow only simulations.
The dominance of radiation pressure still holds on core scales (middle panel), but here photoionization contributes to the broadening as well.
Finally, on the largest spatial scales (top panel), photoionization dominates as the broadening effect; radiation force only simulations do only show a negligible small broadening compared to the outflow only results.

This result, which distinguishes the feedback effects based on spatial scales, is of course expected:
Photoionization yields an increase in thermal pressure, which causes the broadening; its relative importance is determined by the ratio of the sound speed of the ionized gas compared to the escape speed at the spatial scale under consideration (escape speed with respect to the mass of the host star as well as inner cloud mass).
Due to fact that the sound speed of the ionized gas is to first order independent of spatial scale, and the escape speed decreases with distance away from the collapse center, photoionization becomes stronger on larger spatial scales.
On spatial scales larger than those considered herein ($> 1~\pc$), the HII region's expansion is eventually stopped when the ionized gas and the surrounding medium are effectively in pressure balance \cite[see e.g.][]{2015MNRAS.454.4484G, 2015MNRAS.453.1324B}.
Radiation forces instead scale with radiative flux and the radiative flux of the stellar irradiation drops at least as fast as distance to the star squared (in optically thin regions), thereby becoming less effective on large scales.

In addition to the broadening, we want to point out that radiation forces and photoionization not only affect the outflow cavities' opening angle, but impact the bipolar region by sweeping up the gas, i.e.~cavities drop to lower gas density due to feedback, while cavity walls (borders of the HII region) gain in density, see Fig.~\ref{fig:CavityDensity}.
First, we see that higher velocity components have in general lower densities.
The density of the fastest outflow component ($v_\mathrm{r} > 100~\kms$) is not affected by photoionization; in the bottom panel we see that switching the photoionization feedback on or off does not change the mean density of the fast moving gas.
On the other hand, the density of the slow outflow component ($v_\mathrm{r} > 1~\kms$) is only to a minor degree affected by radiation forces; in the top panel we see that switching the radiation force feedback on or off only changes the mean density of the slow moving gas in the case where photoionization is switched on as well. 
The simulation with only protostellar outflows and outflows plus radiation forces yield the same mean densities of the slow outflow component.

Adding up both sources of radiation feedback (radiation forces and photoionization; black lines) yields a combined broadening on large and core scales which is stronger than each of the broadening effects alone.
Only on disk scales, where we already saw that photoionization is a negligible broadening effect, adding up both feedback effects does not lead to an increase in the opening angle compared to radiation pressure only simulations.
The reason that we still see an effect of radiation pressure on the largest scale is that we -- additionally to the direct irradiation of the forming star -- also take into account the thermal (re-)emission by dust.

Last, we want to focus on the simulation with all the feedback effects included (black dashed lines) and the evolution of the opening angle of the cavity in time and space within this most realistic scenario.
The opening angle is largest at its base, in fact its value is controlled by the inner disk rim \citep{2016ApJ...832...40K}, and decreases strongly toward larger and larger spatial scales, i.e.~the cavity region is collimated at this stage in time.
As discussed above, this collimation is the result of counterplay between the broadening feedback effects and the ram pressure from the large-scale gravitational collapse.
Hence, without a sustained infall from larger scales, the cavity broadens quickly and clears the stellar and disk environment.

The determined change in relative importance of radiation forces and photoionization when going from small to large scales is in agreement with observational findings by e.g.~\citet{Lopez:2011cj}, their Fig.~11, although these observations are analyzing these feedback components on (much larger) cluster scales. 
The physical reason for the similarity in the outcome is given by the fact that the radiation force from direct irradiation by a point source decreases geometrically to larger radii $r$ proportional to $r^2$, while photoionization feedback, i.e.~the high thermal pressure of the HII region remains constant to first order.

\section{Limitations and outlook}
\label{sect:limitations}
In this study, we model the formation and long-term evolution of massive (proto)stars including their accretion disks and a variety of feedback effects.
We derive the impact and relative importance of the radiation and photoionization feedback components in terms of final stellar mass and accretion timescale.
Besides the fact that both radiative feedback components -- radiation forces and photoionization -- are taken into account, the simulations are unique in terms of their high spatial (sub-au) resolution and the coverage of extremely long timescales, explicitly the full accretion timescale from free-fall and disk formation to disk destruction.

In order to compensate for the associated high computational effort, simplifications are currently unavoidable to keep the modeling expense doable on modern computing architecture.
The most severe simplifications exploited are 
the assumption of axial and midplane symmetry, which reduces the numerical problem to two dimensions,
as well as neglecting the effect of magnetic fields. 
In reality, the forming accretion disk will evolve non-axial symmetric features such as spiral arms and fragments.
Based on previous three-dimensional cloud collapse simulations \citep{2011ApJ...732...20K, 2017MNRAS.464L..90M, 2018MNRAS.473.3615M} and their comparison with analogous two-dimensional axially symmetric simulations, we can claim that the mean stellar accretion rate during disk formation and the main epoch of disk evolution is not affected by our assumption of axial symmetry.
The underlying physical reason is that the mean accretion rates are governed by the global cloud collapse \citep{Yorke:1999fo} and the gravitational instability in the evolving circumstellar disk, which produces the non-axially symmetric features, gives at the same time rise to an efficient angular momentum transport \citep{2011ApJ...732...20K}, allowing the large-scale infall to be channeled toward the disk's host star.
Nonetheless, no three-dimensional study of high-mass star formation has ever been modeled up to the point in time where stellar accretion ceases.
Hence, we can't directly compare this latest evolutionary phase of disk destruction for now and have to rely on the 2D models presented here.

The change from a relatively smooth accretion history in axially symmetric models compared to accretion bursts in three-dimensional disk evolution will also alter the feedback, see e.g.~\citet{2016ApJ...824..119H} for an example in primordial star formation.
But for the results obtained in this study of present-day star formation, this plays a much more minor (if any) role, because of the following reasoning:
The change of the total luminosity due to accretion bursts is weak once the stellar luminosity is the dominating component, and this happens already when the high-mass protostar has grown to $10~\Msol$ \citep{Hosokawa:2009eu, Hosokawa:2010hs, 2013ApJ...772...61K}, i.e.~long before the onset of either strong radiation force or photoionization feedback.
The impact of accretion bursts on stellar evolution does change the stellar radius, but does not impact the stellar luminosity.
Hence, the radiation force feedback from direct stellar irradiation remains of the same order ($L_*/c$), although emitted in a different part of the spectrum, i.e.~being absorbed at different depths in the environment.
The remaining uncertainty is that strong accretion bursts might lead to a recurrent stellar bloating phase, which can suppress the photoionization feedback \citep{2016ApJ...824..119H}.
In sum, our conclusion that radiation forces are the dominant feedback on accretion scales is still valid in the case of accretion bursts;
the broadening of the outflow cavities on the largest scales and the associated feedback due to photoionization might be altered in the case of accretion bursts.

In terms of neglecting the effect of magnetic fields, we attempt to compensate for this by utilizing the sub-grid module for protostellar outflow feedback as done in \citet{2015ApJ...800...86K} and \citet{2016ApJ...832...40K}.
Although this sub-grid module does not solve self-consistently for the magnetic field evolution during the cloud collapse and disk formation, it handles the injection of jet and outflow momentum into the natal cloud.
As an outlook, we would like to mention that we are currently working on using the Pluto code's capability of solving for the magneto-hydrodynamical evolution of the collapsing mass reservoir including self-consistent jet launching, acceleration, collimation, and jet-cloud interaction.
As of the writing of this paper, a handful of simulation studies have taken into account the effect of magnetic fields during cloud collapse and high-mass star formation, but neglected either radiation forces or photoionization or both
\citep{
Banerjee:2007eh,
Wang:2010ib,
2011A&A...528A..72H,
2011ApJ...742L...9C,
Seifried:2011ef,
Seifried:2012fb,
Seifried:2012ip,
2014MNRAS.439.3420M,
2017MNRAS.470.1026M}.
\vONE{
Additionally, magnetic field are not only important for launching jets and outflows, but the magnetization of the cloud medium will alter their back reaction on stellar feedback to some extent.
For a numerical study on how magnetic fields influence the expansion of HII regions, please see \citet{Krumholz:2007dh}.
}

Last but certainly not least, we would like to mention that radiatively line-driven forces and stellar winds might be a crucial component of the feedback ladder in order to determine the final mass of accreting hot, massive stars.
Pioneering work on the impact of line-driven forces on stellar decretion disks has been presented in \citet{2016MNRAS.458.2323K} and \citet{2018MNRAS.474..847K}.
In future studies, we will include the effect of line-driven forces and stellar winds in our feedback machinery.

%%%%%%%%%%
%%                   %%
%%  Summary  %%
%%                   %%
%%%%%%%%%%
\section{Summary}
\label{sect:summary}
% Highlight:
We modeled -- for the first time -- the formation and evolution of a high-mass star including the feedback due to radiation forces and photoionization simultaneously.
Additionally, the model includes the feedback of protostellar outflows and the ram pressure from large-scale gravitational infall.
We computed the long-term evolution of the system until accretion on the massive stars has either ceased or the star has already gained more than $270~\Msol$ and is still actively accreting.
The latter happens only in cases, where radiation forces were \vONE{on}
%by 
purpose neglected for the evolution of the system.

% Method:
The simulations start from a globally collapsing cloud of gas and dust, 
solve for hydrodynamics, self-gravity, continuum radiation transport, as well as photoionization, and 
include subgrid modules for stellar evolution, shear disk viscosity, and dust evaporation and sublimation.
To enable the computation of the long-term evolution of the system and the parameter study outlined, simulations were performed on two-dimensional grids in spherical coordinates assuming axial and midplane symmetry of the system.
The usage of log-spherical grid coordinates yields sub-au resolution in the vicinity of the forming star.

% Simulations:
To investigate the importance of the individual feedback components by direct comparison, we ran a series of models including and excluding each of the feedback components.
Additionally, the simulation series is carried out for two different initial mass reservoir scenarios, a finite small-scale mass reservoir of $100~\Msol$ in a sphere of $0.1~\pc$ radius and a virtually infinite mass reservoir of $1000~\Msol$ in a sphere of $1~\pc$ radius.
This allowed us to study the importance of ram pressure from the large-scale gravitational infall.

% Basic Evolution of the System - no bullet list, simple paragraph:
\paragraph{Basic Evolution (Sect.~\ref{sect:results_Basics})}
In all simulations, the system goes through epochs of 
global collapse of the mass reservoir, 
the formation of a high-mass (proto)star,
the formation of a circumstellar accretion disk, and
the momentum feedback of protostellar outflows.
Simulations including the feedback of radiation forces yield a broadening of the bipolar low-density cavity regions. 
Eventually radiation forces stop the feeding of the circumstellar accretion disk from larger scales.
Simulations taking into account photoionization feedback include the formation of a large-scale HII region once the protostar has contracted down to the zero-age main sequence.

In models using finite mass reservoir of $0.1~\pc$, accretion onto the star ceases at about $60 - 70~\Msol$.
In models using a virtually infinite mass reservoir of $1~\pc$, the star gains further mass from these large-scales and stops accretion only in simulations including radiation forces.
In these simulations, the accretion phase lasts less than $150~\kyr$ and the final stellar mass is below $100~\Msol$.
In simulations excluding radiation forces, the remaining feedback components are inefficient in stopping the accretion from large scales;
these simulations were stopped after more than $500~\kyr$ of evolution, when the star has accreted more than $270~\Msol$, and was still actively accreting from its circumstellar disk.

The fact that the star stops accretion in simulations including radiation forces in spite of the initial $1000~\Msol$ and $1~\pc$ large-scale reservoir provides strong motivation for the introduction of the concepts of a maximum accretion radius or a region of influence.
Based on a simple free-fall analysis of the initial mass reservoir, we can invert the duration of the stellar accretion phase into a maximum accretion radius of the embedded high-mass star; 
the medium beyond this radius has no influence on the formation process of the massive star.
Applied to the two large-scale simulations, which take into account radiation forces, 
the accretion phase of $t_\mathrm{tot}^\mathrm{acc} = 140~\kyr$ (without photoionization feedback) and $t_\mathrm{tot}^\mathrm{acc} = 125~\kyr$ (with photoionization feedback) yields a radius of the sphere of influence of $0.27~\pc$ and $0.24~\pc$, respectively.
Future studies will especially address the dependence of the maximum accretion radius on the initial density slope, and investigate potential large-scale factors such as cloud fragmentation, filament formation, and initial ram pressure.

In the following, we will briefly overview the most important results ordered by topics.
\paragraph{Stellar Evolution (Sect.~\ref{sect:results_Star})}
The initial collapse phase is dominated by accretion luminosity as the main source of radiation feedback.
The star becomes super-Eddington with respect to dust opacity at about $M_* \approx 10~\Msol$, from then on radiation forces counteract gravity efficiently in the directly irradiated stellar surroundings.
Due to the bloating epoch of the accreting protostar, the photospheric temperature is too low for a substantial amount of photoionizing EUV radiation at that point in time.
The onset of strong EUV photoionization feedback can be estimated to $M_* \approx 30~\Msol$, after the stellar Kelvin-Helmholtz contraction toward the zero-age main sequence. From then on, an HII region will be formed and expand into the small-scale and large-scale stellar surrounding.

\paragraph{Disk Evolution (Sect.~\ref{sect:results_Disk})}
The circumstellar accretion disk forms inside-out due to the angular momentum from larger scales.
The bulk of the disk is in gravito -- centrifugal equilibrium; gravito here includes the stellar mass as well as the disk's self-gravity.
During active accretion from larger scales, the outer disk edge is in (gravito + ram pressure) -- centrifugal equilibrium. As a consequence the azimuthal speed exceeds the Keplerian value by up to 20\%. Question: Is this 20\% of super-Keplerian velocity at the outer edge of accreting disks observationally detectable?
Eventually, the outer disk edge is in gravito -- (centrifugal + radiation pressure) equilibrium. 
As a consequence, the disk becomes strongly sub-Keplerian, decreases in size, and is eventually destroyed from outside-in. 
Once the disk loses its optical thickness, the remnant disk is expelled by radiation forces.
Interestingly, this outside-in destruction of the high-mass accretion disk is opposed to the inside-out disk dispersal in the low-mass regime \citep{2014prpl.conf..475A}.

During phases of strong stellar EUV feedback, the inner disk rim and the upper disk atmosphere layers are photoionized.
Like scissor's handles, the photoionization of the bipolar low-density cavities pushes mass from high latitudes toward the disk. I.e.~the photoionization feedback yields an increase in disk mass and hence stellar accretion rate.

\paragraph{Star Formation and Small-Scale Feedback (Sect.~\ref{sect:results_StarFormation})}
In the scenario of a small-scale finite mass reservoir, such as an isolated pre-stellar core, each feedback component contributes to limiting the final stellar mass via decreasing the mass of the reservoir. 
In the case of photoionization feedback, this negative feedback effect is fully overcompensated for its positive feedback effect mentioned above.
In the scenario of a large-scale virtually infinite mass reservoir, such as a high-density region fed by gravitational collapse from larger scales, even the combined effect of protostellar outflows and photoionization do not prevent stellar accretion at any time.
But the feedback due to radiation forces eventually stop the feeding of the accretion disk, and hence, the disk runs out of material, and stellar accretion is finally stopped. Hence, radiation forces denote an intrinsic limit of the stellar mass growth.
To form a strongly super-Eddington star of a specific mass, an initial mass reservoir of about 2-3 times higher mass is required.

For the chosen initial conditions of our simulations and the forming $95~\Msol$ star, this mass reservoir relates to a size of $\approx 0.24~\pc$ in radius.
In our opinion, these numbers do not allow us to argue conclusively if the underlying accretion physics corresponds to a competitive accretion or a monolithic collapse scenario.
We would e.g.~expect 
a smaller mass reservoir in the case of initially higher-density profiles (due to a faster star formation process) and
a larger mass reservoir in the case of initially shallower density profiles (due to a slower star formation process).

\paragraph{HII Regions and Large-Scale Feedback (Sect.~\ref{sect:results_HII})}
During the contraction phase of the protostar toward the zero-age-main-sequence between $M_* \approx 10~\Msol$ and $M_* \approx 30~\Msol$, the photospheric temperature of the star increases about one order of magnitude up to $T_* \approx 45000~\K$. As a consequence, the stellar spectrum shifts toward the EUV regime and an HII region starts to form and expands into the stellar environment.
The strong dependence of photoionization on density promotes the anisotropy of optical depth caused by the accretion disk. 
The ionized gas of the HII region quickly fills the low-density bipolar cavities. 
Only the inner rim and disk's atmosphere become ionized.
At the inner disk rim, the sound speed of the ionized gas is lower than the escape speed with respect to stellar gravity, hence, photoionization does not stops the disk-to-star accretion flow.
Even more impactfully, the scissor handle like push effect of the HII region expanding away from the bipolar outflow region toward the disk results in a positive feedback on disk mass and stellar accretion rate.
But on larger scales ($> 0.1~\pc$), photoionization and HII regions denote the dominant negative feedback effect, which broadens the outflow cavity and slows down the gravitational collapse of the large-scale environment. 

\paragraph{Outflow Broadening (Sect.~\ref{sect:results_Outflow})}
The opening angle of the low-density bipolar cavity region is both a function of time and spatial scales. The underlying reason for the space- and time-dependence is the multi-physics, which controls the broadening.
The relative contribution of the feedback components of protostellar outflows, radiation forces, and photoionization changes with time and varies with spatial scales.
Without ram pressure from large-scale gravitational infall, the opening angle quickly opens up to the full region not blocked by the (inner rim of the) accretion disk.
Within a globally collapsing large-scale mass reservoir, the ram pressure is key in collimating the expanding outflow and its entrained material.
On top of the protostellar outflow, radiation forces are the dominant broadening of the bipolar region on small-scales comparable to the size of the accretion disk ($\le 2000~\au$).
They are important on pre-stellar core scales ($\le 0.1~\pc$), but rather weak to negligible on largest scales ($> 0.1~\pc$).
Photoionization in turn is negligible on disk scales, relevant on core scales, and dominant on the largest scales.
While radiation forces govern the mean density of the faster outflow component, photoionization causes a significant decrease in mean density for the slower outflow component.

\paragraph{Outlook}
Future studies planned will include
the evolution of magnetized mass reservoirs including self-consistent launching of jets and outflows,
the fragmentation of the massive accretion disk in three-dimensional simulations,
the diffuse EUV radiation field from direct recombination of free electrons into hydrogen's ground state, and
the late-phase feedback from line-driven stellar winds
to further investigate the formation and feedback of high-mass stars, especially in the regime of the most massive stars known in the present-day universe.

\begin{acknowledgements}
This study was conducted within the Emmy Noether research group on ``Accretion Flows and Feedback in Realistic Models of Massive Star Formation'' funded by the German Research Foundation (DFG) under grant no.~KU 2849/3-1.
We thank Harold W.~Yorke for enlightening scientific discussions on star formation, stellar evolution, and stellar feedback.
R.~K.~thanks the StarBench collaboration for instructive discussions on stellar photoionization feedback.
We greatly thank Nathaniel Dylan Kee for proof-reading the manuscript.
The authors acknowledge support by the High Performance and Cloud Computing Group at the ``Zentrum f\"ur Datenverarbeitung'' of the University of T\"ubingen, the state of Baden-W\"urttemberg through bwHPC and the DFG through grant no.~INST 37/935-1 FUGG.
T.~H.~also appreciates financial support in part by MEXT/JSPS Kakenhi Grant Number 16H05996.
\end{acknowledgements}

\bibliographystyle{aa}
\bibliography{Papers3,PapersSubmitted}

\appendix
\section{Convergence studies}
\label{sect:convergence}
In order to check the convergence and robustness of the simulations outcomes, and specifically their dependence on the numerical grid configuration in use, we have performed a series of simulations of the large-scale reservoir scenario ($1000~\Msol \mbox{ in } 1~\pc$ sphere) with different sink cell sizes and varying spatial resolution.

\subsection{Sink cell size check}
\label{sect:Rmin}
\begin{figure}[htbp]
\centering
\includegraphics[width=0.49\textwidth]{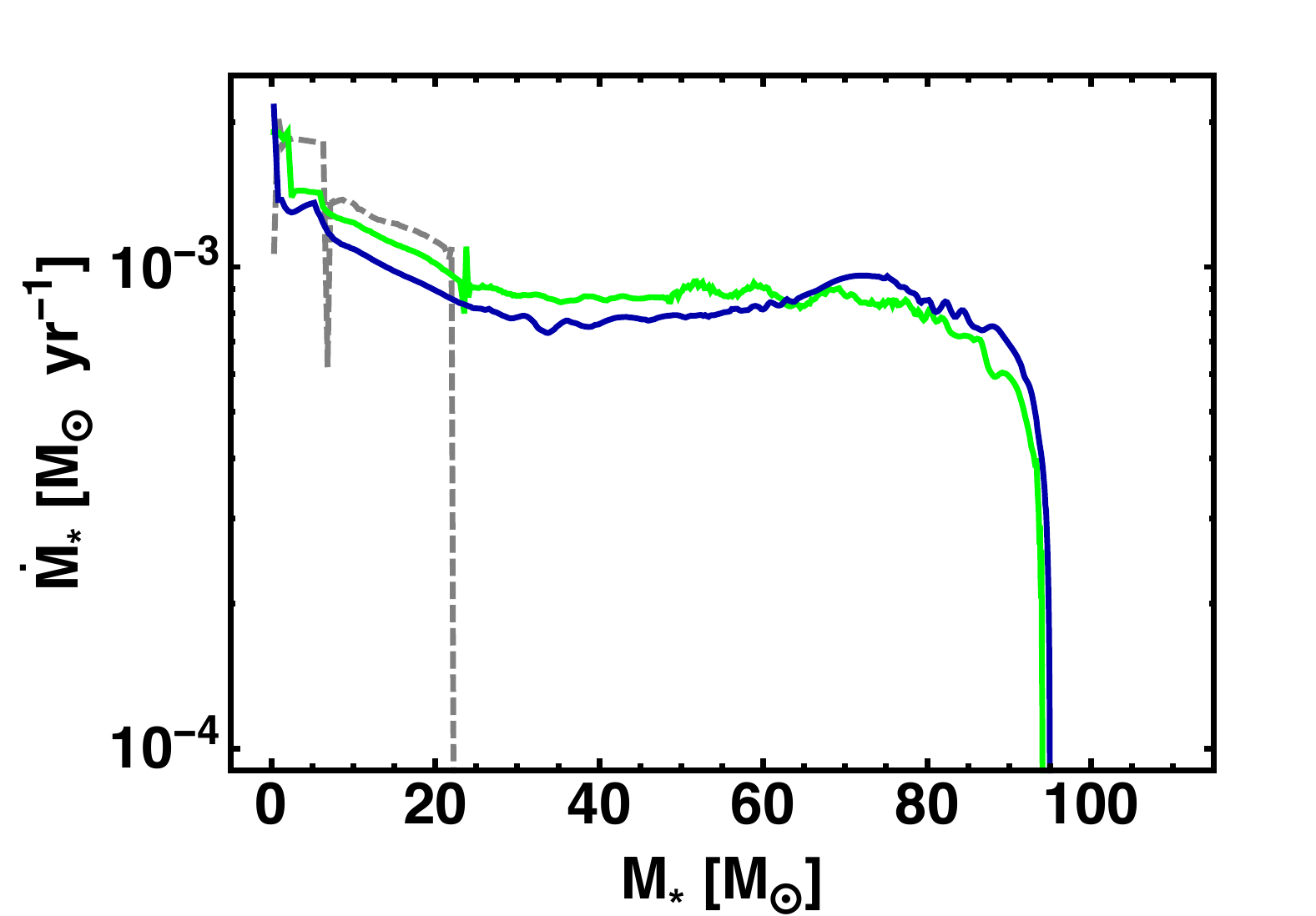}
\caption{
Stellar accretion rate as function of actual stellar mass for the three different sizes of the central sink cell.
Gray-dashed, green, and blue refer to sink cell sizes of $R_\mathrm{min} = 30~\au$, $10~\au$, and $3~\au$, respectively.
}
\label{fig:RMin}
\end{figure}

The size of the innermost sink cell, i.e.~the radially inner boundary of the computational domain, represents the accretion radius of the central star.
Moreover, the region between the stellar photospheric surface and the sink cell radius is assumed to be optically thin for the stellar irradiation.
Hence, the sink cell has to be chosen small enough to correctly account for the optical depth of the forming accretion disk.
We investigated the effect of different sink cell sizes already in \citet{2010ApJ...722.1556K} (see Sect.~5.1 therein), regarding the resulting balance of radiation forces and disk accretion for super-Eddington massive (proto)stars.
Here, we study a similar scenario, but take into account the additional feedback of protostellar outflows and photoionization.
In order to check the robustness of these new simulations as regards their dependence on the size of the central stellar sink cell, we have performed three simulations with different sink cell sizes of $R_\mathrm{min} = 30~\au$, $10~\au$, and $3~\au$.
Fig.~\ref{fig:RMin} depicts the outcome of this parameter scan.

The choice of a fairly large sink cell radius of $R_\mathrm{min} = 30~\au$ yields an early disk destruction due to direct irradiation behind the expected dust sublimation front.
This outcome is in agreement with our earlier sink cell size tests results in \citet{2010ApJ...722.1556K}, see their Fig.~9 for comparison and Sect.~5.1 therein for a detailed discussion. 

The choice of $R_\mathrm{min} = 10~\au$, which was the default sink cell size in our earlier studies, yields slight oscillations of the accretion rate at the epoch where the star has grown up to $M_* \approx 20~\Msol$, but the accretion flow from disk to star remains.
With a choice of a smaller sink cell of $R_\mathrm{min} = 3~\au$, the strength of disk accretion is not impacted anymore.
In general, the simulations with $R_\mathrm{min} = 3~\au$ and $R_\mathrm{min} = 10~\au$ show qualitatively and quantitatively a robust simulation outcome.
The accretion rate from disk to star is slightly smaller in simulations with smaller sink cells due to the fact that the feedback acting on the smaller scales diminishes the accretion flow.
But overall, we see a converged simulation outcome for sink cell sizes smaller than the dust sublimation front.

Explicitly, we find a final stellar mass of 
$M_*^\mathrm{final} = 94~\Msol$ for the $R_\mathrm{min} = 10~\au$ simulation
and 
$M_*^\mathrm{final} = 95~\Msol$ for the $R_\mathrm{min} = 3~\au$ simulation.
The associated stellar accretion timescale and accretion radius, as defined in Table~\ref{tab:sims} are
$t_\mathrm{acc}^\mathrm{tot} = 120~\kyr$ and $R_\mathrm{acc}^\mathrm{max} = 0.23~\pc$ for the $R_\mathrm{min} = 10~\au$ simulation
and 
$t_\mathrm{acc}^\mathrm{tot} = 126~\kyr$ and $R_\mathrm{acc}^\mathrm{max} = 0.24~\pc$ for the $R_\mathrm{min} = 3~\au$ simulation.
This good agreement of the derived values makes us feel certain that the simulation outcome and conclusions drawn are based on robust numerical results.

As an outcome of this sink cell size check, all simulation series analyzed in this study (see Table~\ref{tab:sims}) were performed using a sink cell size of the smaller value of $R_\mathrm{min} = 3~\au$.

\subsection{Resolution check}
\label{sect:resolution}
The spatial resolution can have profound effects on the outcome of numerical simulations, most seriously when certain physical length scales have to be resolved properly.
For the disk accretion and feedback studies presented here, such a scale is e.g.~given by the pressure scale height of the forming accretion disk.
Here, we study the resolution dependence of the simulation outcomes in a simulation series with varying spatial resolution.

Based on the outcome of this parameter check, we have decided to perform all simulations analyzed in this study with a spherical grid with a minimum spatial resolution of $\Delta x_\mathrm{min} = 0.33~\au$ at the innermost cells at a radius of $r = 3~\au$.
Due to the log-radial grid, the spatial resolution coarsens to larger distances $r$ from the forming central star as $\Delta x(r) = f \times r$.
In the default resolution simulation, the factor $f$ is set to $0.1$.
The resolution check includes simulations for half and double this default spatial resolution.
The quantitative results of these simulations are given in the following table:
\begin{table}[h!]
\caption{
Results of resolution check. 
The first column gives the resolution factor $f$ as defined in the main text, a lower factor resembles a higher spatial resolution. 
The last three columns give the final stellar mass, the total duration of the stellar accretion epoch and the maximum size of the associated accretion reservoir.
}
\begin{tabular}{c | c c c}
Resolution factor $f$ & $M_*^\mathrm{final}$ [$\Msol$] & $t_\mathrm{acc}^\mathrm{tot}$ [kyr] & $R_\mathrm{acc}^\mathrm{max}$ [pc] \\
\hline
0.2~~~~	& 82 		& 129 	& 0.25 \\
0.1~~~~	& 95 		& 126 	& 0.24 \\
0.05~~	& 98 		& 132 	& 0.25
\end{tabular}
\end{table}

The simulation with a coarser grid resolution leads to the formation of a less massive final star.
The accretion timescale and radius are not strongly affected by this change due to the fact that a less massive star implies less radiative feedback as well.
In general, an increase in spatial resolution yields an increase in the final stellar mass, but the absolute and relative increase in final stellar mass decreases toward the highest resolution run performed.
The simulation with the default spatial resolution agrees well with the simulation outcomes of the higher resolution run with a maximum deviation in final stellar mass of $\approx 3\%$ compared to the highest resolution run.

All simulations analyzed in this study (see Table~\ref{tab:sims}) were performed used a radially stretched spatial resolution of $\Delta x(r) = 0.1 \times r$, i.e.~with $\Delta x_\mathrm{min} = 0.33~\au$ at the innermost radius of the forming accretion disk.

\end{document}